\documentclass[11pt]{article}
\usepackage[margin=1in]{geometry}
\usepackage{amsmath,amssymb,amsthm,mathtools}
\usepackage{aliascnt}
\usepackage{booktabs,array}
\usepackage{enumitem}
\usepackage{microtype}
\usepackage{xcolor}
\usepackage{tikz}
\usetikzlibrary{decorations.pathreplacing}
\usepackage[hyphens]{url}
\renewcommand{\texttt}[1]{\url{#1}}
\usepackage[colorlinks=true,linkcolor=blue!55!black,citecolor=blue!55!black,urlcolor=blue!55!black]{hyperref}

\usepackage{etoolbox}
\makeatletter
\newif\ifthm@anchor
\patchcmd{\@thm}{\refstepcounter{#2}}{\let\thm@savedrl\Hy@raisedlink\let\Hy@raisedlink\@gobble
  \refstepcounter{#2}\let\Hy@raisedlink\thm@savedrl\global\thm@anchortrue}{}{\errmessage{theorem anchor patch 1 failed}}
\patchcmd{\@begintheorem}{\the\thm@headfont}{\ifthm@anchor\global\thm@anchorfalse
  \hyper@anchorstart{\@currentHref}\hyper@anchorend\fi\the\thm@headfont}{}{\errmessage{theorem anchor patch 2 failed}}
\makeatother

\newtheorem{theorem}{Theorem}[section]
\newaliascnt{lemma}{theorem}
\newtheorem{lemma}[lemma]{Lemma}
\aliascntresetthe{lemma}
\newaliascnt{proposition}{theorem}
\newtheorem{proposition}[proposition]{Proposition}
\aliascntresetthe{proposition}
\newaliascnt{assumption}{theorem}
\newtheorem{assumption}[assumption]{Assumption}
\aliascntresetthe{assumption}
\newaliascnt{corollary}{theorem}
\newtheorem{corollary}[corollary]{Corollary}
\aliascntresetthe{corollary}
\theoremstyle{definition}
\newaliascnt{definition}{theorem}
\newtheorem{definition}[definition]{Definition}
\aliascntresetthe{definition}
\theoremstyle{remark}
\newaliascnt{remark}{theorem}
\newtheorem{remark}[remark]{Remark}
\aliascntresetthe{remark}

\usepackage[nameinlink,noabbrev]{cleveref}
\crefname{lemma}{Lemma}{Lemmas}
\crefname{assumption}{Assumption}{Assumptions}
\Crefname{lemma}{Lemma}{Lemmas}
\crefname{theorem}{Theorem}{Theorems}
\Crefname{theorem}{Theorem}{Theorems}
\crefname{proposition}{Proposition}{Propositions}
\Crefname{proposition}{Proposition}{Propositions}
\crefname{corollary}{Corollary}{Corollaries}
\Crefname{corollary}{Corollary}{Corollaries}
\crefname{definition}{Definition}{Definitions}
\Crefname{definition}{Definition}{Definitions}
\crefname{remark}{Remark}{Remarks}
\Crefname{remark}{Remark}{Remarks}
\crefname{part}{Part}{Parts}
\Crefname{part}{Part}{Parts}
\crefname{figure}{Figure}{Figures}
\Crefname{figure}{Figure}{Figures}

\newcommand{\J}{\mathcal J}
\newcommand{\M}{\mathcal M}
\newcommand{\EB}{\mathcal E_{\mathrm B}}
\newcommand{\ES}{\mathcal E_{\mathrm S}}
\newcommand{\StrongLP}{\textnormal{\textsc{StrongLP}}}
\newcommand{\OPTLP}{\mathrm{OPT}_{\StrongLP}}
\newcommand{\OPT}{\mathrm{OPT}_{\mathrm{mig}}}
\newcommand{\OPTnm}{\mathrm{OPT}_{\mathrm{nonmig}}}
\newcommand{\E}{\mathbb E}
\newcommand{\DPNR}{\hyperref[alg:dpnr]{Algorithm~\textnormal{\textsc{DP-NR}}}}

\title{The Benefit of Migration in Throughput Maximization on Unrelated Machines and under Restricted Assignment}
\author{George Karakostas\thanks{Research partially supported by an NSERC Discovery Grant.}\\
  \small Department of Computing \& Software, McMaster University, Hamilton, ON, Canada\\
  \small\texttt{karakos@mcmaster.ca}
  \and
  Stavros G. Kolliopoulos\\ 
  \small Department of Informatics and Telecommunications,\\
  \small National and Kapodistrian University of Athens, Athens, Greece\\
  \small\texttt{sgk@di.uoa.gr}}
\date{}  

\begin{document}
\maketitle

\begin{abstract}
In preemptive throughput maximization on parallel machines, jobs have
release dates, due dates and weights, and the goal is to maximize the
total weight of the jobs completed by their due dates.  We study the
benefit of migration, i.e., by what factor the optimal weight can
increase when a preempted job is allowed to resume on a different machine.
By a classic result of Kalyanasundaram and Pruhs, on identical machines
the migratory optimum is at most $O(1)$ times the nonmigratory optimum.
We give, for the first time, bounds for unrelated machines and for
restricted assignment.  We show that on unrelated machines, when every
eligible machine can complete a job by itself, the migratory optimum is
at most $O(\sqrt n)$ times the nonmigratory optimum, where $n$ is the
number of jobs.  Under restricted assignment it is at most $O(\log n)$
times the nonmigratory optimum.  We provide instances which show that
both bounds are best possible up to constant factors, even when the jobs
have unit weights.  Since our proofs are constructive, we obtain the
first nontrivial approximation algorithms for migratory throughput
maximization on unrelated machines and under restricted assignment.
\end{abstract}

\tableofcontents

%=====================================================================
\section{Introduction}
%=====================================================================

We are given a set $\M$ of $m$ machines and a set $\J$ of $n$ jobs.
Each job $j$ has a release date $r_j$, a due date $d_j$, a weight
$w_j$, and a set $\Gamma(j)\subseteq\M$ of eligible machines. In the
{\em unrelated machines} setting, on
each $i\in\Gamma(j)$ job $j$ has a processing time $p_{ij}$.  We assume $p_{ij}\le d_j-r_j$, so
that every eligible
machine can complete $j$ by itself. %; \Cref{sec:model} discusses this assumption.
In the 
\emph{restricted assignment} setting, $p_{ij}=p_j$ for all $i\in\Gamma(j)$.
The \emph{window} of job $j$ is $K_j=[r_j,d_j)$, and the \emph{slack} of $j$ on machine $i$ is
$s_{ij}=d_j-r_j-p_{ij}$, the amount by which its window is longer than its processing time on $i$.
Jobs may be \emph{preempted}, that   is, interrupted and resumed later, but a
job is never processed on two machines at the same time.  If
machine $i$ processes job $j$ for a total time $t_{ij}$ within its
window, then $j$ is completed
when $\sum_it_{ij}/p_{ij}\ge1$, and the goal is to maximize the total
weight of completed jobs.  A preemptive schedule is \emph{migratory} if a
preempted job may be resumed on a different machine, and
\emph{nonmigratory} if every completed job runs on one machine only.
Let $\OPT$ and $\OPTnm$ be the largest weights of a migratory and of a
nonmigratory schedule respectively. % (\Cref{sec:model}).
This paper bounds the ratio
$\OPT/\OPTnm$, first on unrelated machines and then under restricted
assignment.

Our study of this ratio is motivated by three reasons.
\begin{enumerate}[label=(\arabic*)]
\item Migration is convenient in analysis but costly in practice:
  moving a partially processed job to another machine transfers its
  state, and nonmigratory schedules are preferred for this reason
  \cite{KP,CLT}; on unrelated machines, which model heterogeneous
  processors, the state of a job may not transfer at all.
\item There is a structural reason to compare the two optima.  Without migration, each job must be assigned
  integrally to a single machine, and given a set $S$ of jobs, even deciding whether all jobs of $S$ can be
  completed is strongly NP-hard \cite{AS}.  With migration, the processing of a job may be divided among its
  eligible machines in arbitrary proportions, so deciding whether $S$ is feasible reduces to solving
  the linear program of Lawler and Labetoulle \cite{LL}.  The ratio $\OPT/\OPTnm$ therefore measures the
  cost of this integrality requirement.  A similar question has been studied for makespan on
  unrelated machines without release dates, where the corresponding
  ratio is at most $4$, and this is tight \cite{CSV}.  For throughput,
  by a classic result of Kalyanasundaram and Pruhs \cite{KP}  the ratio is bounded
  by a constant on identical machines; Megow and Zondervan
  \cite{MZ} recently proved the bound $2 -1/m$, which \cite{KDA} show to be tight.  Since
  the  work of \cite{KP}, related
  questions on the power of migration have been studied extensively,
  cf.~\Cref{sec:related} below.  To our knowledge, no
  bound in terms of $n$ is known on unrelated machines, or even under
  restricted assignment.
\item A nonmigratory schedule is also a migratory one, so a
  constructive proof of a bound $\rho$ on the ratio yields a
  $\rho$-approximation for maximizing throughput in the migratory
  setting.  For that problem, on unrelated machines and under
  restricted assignment, to our knowledge the literature provides no
  nontrivial approximation algorithm.
\end{enumerate}

\subsection{Results and techniques}\label{sec:results}

Both bounds are proved via the $\StrongLP$, a linear relaxation of the migratory problem that we introduce in
\Cref{sec:lp}, strengthened by constraints that bound the rate at which a fractionally accepted job is processed.  Every
migratory schedule gives a $\StrongLP$ solution of the same value, so $\OPT\le\OPTLP$, the optimum value of the
$\StrongLP$.  Our main theorems give polynomial-time algorithms that compute nonmigratory
schedules whose weight is within the stated
factors of the value of an arbitrary $\StrongLP$ solution.  We did not attempt to optimize the constants in these factors.

\begin{theorem}[Unrelated machines]\label{thm:main-unrelated}
There is a deterministic polynomial-time algorithm that, given an
instance on unrelated machines with $n$ jobs and a
$\StrongLP$ solution of value $W$, outputs a nonmigratory schedule of
weight at least
\[
 \frac{W}{246\sqrt n+49152}.
\]
\end{theorem}

\begin{corollary}[Migration on unrelated machines]\label{cor:unrelated}
On unrelated machines,
\[
 \OPT\ \le\ \OPTLP\ \le\ (246\sqrt n+49152)\,\OPTnm .
\]
Moreover, a nonmigratory schedule of weight at least
$\OPT/(246\sqrt n+49152)$ can be computed in deterministic polynomial
time.
\end{corollary}

The following proposition is a corollary of the approach of Kalyanasundaram and Pruhs \cite{KP}, with their factor
$6m-5$ replaced by the bound $2m-1$ of Megow and Zondervan \cite{MZ}.  We include a proof in \Cref{sec:machines} for
completeness.
\begin{proposition}[Dependence on the number of machines]\label{prop:machines}
On unrelated machines with $m$ machines,
\[
 \OPT\ \le\ (2m-1)\,\OPTnm .
\]
Moreover, for every $\varepsilon>0$, a nonmigratory schedule of weight
at least $(1-\varepsilon)\OPT/(2m-1)$ can be computed in time
polynomial in the input size and $1/\varepsilon$, and with unit
weights one of weight at least $\OPT/(2m-1)$ can be computed in
polynomial time.
\end{proposition}

\noindent With \Cref{cor:unrelated}, \Cref{prop:machines} gives
$\OPT\le\min\{246\sqrt n+49152,\,2m-1\}\,\OPTnm$ on unrelated machines.
\Cref{prop:machines} bounds $\OPT$ but not $\OPTLP$, whereas
\Cref{thm:main-unrelated} rounds arbitrary $\StrongLP$ solutions, with a
loss that depends on $n$ and not on $m$.

\begin{theorem}[Restricted assignment]\label{thm:main-ra}
There is a deterministic polynomial-time algorithm that, given a
restricted-assignment instance with $n\ge2$ jobs and a
$\StrongLP$ solution of value $W$, outputs a nonmigratory schedule of
weight at least
\[
 \frac{W}{170\log n+410}.
\]
\end{theorem}

\begin{corollary}[Migration under restricted assignment]\label{cor:ra}
Under restricted assignment with $n\ge2$ jobs,
\[
 \OPT\ \le\ \OPTLP\ \le\ (170\log n+410)\,\OPTnm .
\]
Moreover, a nonmigratory schedule of weight at least
$\OPT/(170\log n+410)$ can be computed in deterministic polynomial
time.
\end{corollary}

These bounds are best possible up to constant factors, already with
unit weights.  On unrelated machines, \Cref{thm:sqrt-tightness} gives instances with
$\OPTLP=\OPT$ and $\OPT/\OPTnm\ge\sqrt n/3=m/6$.  Hence neither the factor $\sqrt n$ of
\Cref{thm:main-unrelated} nor the factor $2m-1$ of
\Cref{prop:machines} can be improved by more than a constant factor.
Under restricted assignment, call a job \emph{rigid} if $p_j=d_j-r_j$.  The factor of \Cref{thm:main-ra} is also best
possible up to a constant factor, already for rigid jobs and unit weights.
Megow and Zondervan \cite{MZ} give rigid instances, with at most two eligible machines per job, on
which eliminating migration requires $\Omega(m\log m)$ machine copies in total.  We observe that their argument also bounds
throughput: in every nonmigratory schedule of these instances each machine completes at most one job, so
$\OPT/\OPTnm=\Omega(\log n)$.  Moreover, all jobs can be completed with migration, so $\OPTLP=\OPT$; \Cref{rem:tower} gives the details.  Our \Cref{thm:restricted-log-gap} gives another family with the same
properties.  
For rigid jobs with unit weights, \Cref{prop:rigid-unit} shows that $\OPT\le(2\log m+6)\,\OPTnm$, which the instances of
\cite{MZ} show to be tight up to a constant factor.

As an easy consequence of our results, we obtain $n$-dependent bounds on the number of copies of machines needed to
eliminate migration, the measure studied in \cite{KP,MZ}: $O(\sqrt n\log n)$ copies of each machine suffice on
unrelated machines, and $O(\log^2n)$ under restricted assignment, and the nonmigratory schedule can be computed in
deterministic polynomial time.  \Cref{app:copies} states and proves this precisely.

\paragraph{Techniques.}  A $\StrongLP$ solution accepts a fraction $z_j\in[0,1]$ of each job $j$ and has value
$\sum_jw_jz_j$.  To round it to a nonmigratory schedule, we must decide which jobs to keep and on which machine each kept job runs.  Our
algorithms use the following ideas.

The first is a matching argument over a common set of times.  Suppose
that, during a common set $X$ of times, every job $j$ of a set
receives LP processing at least a constant fraction of $z_j|X|$.  For
a job $j$ and an eligible machine $i$, let $f_{ij}$ be the LP
processing of $j$ on $i$ during $X$, divided by $|X|$.  Since a machine
processes at total rate at most one, $\sum_jf_{ij}\le1$ for every machine $i$.  Since
a job $j$ is processed at total rate at most $z_j$, $\sum_if_{ij}\le z_j$ for every job $j$, and by
assumption $\sum_if_{ij}$ is at least a constant times $z_j$.  Hence
the numbers $f_{ij}$ form a fractional matching between the jobs and
the machines, of weight at least a constant times $\sum_jw_jz_j$.
Since the bipartite matching polytope is integral, some set of jobs of
at least that weight can be assigned to distinct eligible machines.
We call such an $X$ a \emph{witness} for these jobs.  A witness need
not be long enough to process the jobs; it only certifies that they
can be given distinct machines.  The jobs still have to be scheduled
within their windows, and jobs selected through different witnesses
must not need the same machine at the same time.  Most of this paper
is devoted to these two tasks.

We call a job $j$ \emph{big-covered} if a substantial part of its
accepted fraction $z_j$ comes from processing on machines $i$ on which
its slack $s_{ij}$ is large.
Big-covered jobs need no witnesses: they are rounded by a randomized
greedy algorithm, which \Cref{app:big-derandomization} derandomizes.  This is the easy case.  The algorithm processes the
randomly selected jobs in nondecreasing order of window length, and a
job with large slack on its machine still fits into its window when
the jobs with shorter windows occupy part of it.
For all other jobs,
which we call \emph{small-covered}, witnesses come from a counting
argument.  Most of the accepted fraction of such a job $j$ comes from
machines $i$ with small slack $s_{ij}$, on which $p_{ij}$ is close to
the length $L_j=d_j-r_j$ of its window.  Hence $j$ receives processing
close to $z_jL_j$ inside its window.  Its LP processing never exceeds the rate $z_j$, the
fractional counterpart of never running on two machines at once.  Its
rate is thus close to $z_j$ except on a small part of the window,
whose size we call the \emph{deficit} of $j$; every part of the window
longer than the deficit by a constant factor is a witness for $j$.

On unrelated machines, the small-covered jobs are distributed over the
nodes of a balanced tree of separators on the time axis.  Each
small-covered job is handled at the first node on its path from the root whose separator $t$ lies in
its window.  If its deficit is small compared with the length
$\lambda$ of the shortest window handled at that node, then an
interval of length $\lambda/2$ that starts or ends at $t$ lies in its
window and is a witness for it.  At each node, two such intervals
serve all these jobs.  Every other job has an interval of \emph{usable
distances} $r$ from the separator: for each of them, an interval of
length $r$ that starts or ends at $t$ lies in its window and is a
witness for it.  Jobs with a common usable distance form a group with
a common witness.  Jobs whose intervals of usable distances are
pairwise disjoint form a \emph{ladder}: their windows grow
geometrically, and the deficit of each job yields a machine on which
it has substantial slack, its \emph{reserve machine}, so that a ladder
can be split into four sets, each of which can be scheduled completely
with every job on its reserve machine.

All jobs handled at a node have windows containing its separator $t$.
The jobs of a group are assigned to machines by a matching.  Since the
processing time of a job is known only after the matching has chosen
its machine, the job cannot be confined in advance to a fixed part of
its window.  Two groups at the same node may be matched to a common
machine, and their jobs then compete for it around $t$, so each group
needs a schedule of its own.  A ladder runs each job on its reserve
machine.  Two ladders at the same node may have reserve machines in
common, with the same conflict around $t$, so each ladder needs four
schedules of its own, one per set in the ladder.  Thus a node needs
one schedule for each of its groups and four for each of its ladders,
and the jobs in these schedules have at least half of the LP weight of
its jobs.  If the intervals of usable distances at a node with $N$
jobs are split into at most $N/k$ groups and $k$ ladders, the node
needs about $N/k+4k$ schedules, and $k\approx\sqrt N$ balances the two
terms.  Taking the heaviest of these $O(\sqrt N)$ schedules loses a
factor $O(\sqrt N)$ at that node.  The jobs at different nodes of the
same depth have windows in disjoint time regions, so one candidate
schedule can combine one schedule from each of these nodes, and a
depth whose nodes have at most $N$ jobs each yields $O(\sqrt N)$
candidates.  The number of jobs at a node decreases geometrically with
its depth, so the whole tree yields $O(\sqrt n)$ candidates.  The
final schedule is the heaviest candidate, and it loses a factor
$O(\sqrt n)$.  \Cref{thm:sqrt-tightness} shows that the factor $\sqrt
n$ cannot be improved.

Under restricted assignment, a job has the same processing time $p_j$,
and hence the same slack $s_j$, on all of its machines.  Then any
common part of the windows of a set of jobs that is longer than their
slacks by a constant factor is a witness for them, independently of
the given $\StrongLP$ solution.  Moreover, each job can be confined in
advance to a \emph{support}, a part of its window of length at least
$p_j$, so groups whose supports are disjoint can share machines.  The
difficulty is to decide which groups may share machines.  Jobs of
moderate slack are sorted into classes by window length and grouped at
points of a grid, which yields $O(\log n)$ families of groups whose
supports are disjoint.  Nearly rigid jobs, whose slack is a
polynomially small fraction of their window, are the hardest case.
For them, the length of a witness is no obstacle, since even a short
common interval is longer than their slacks.  A support, however, must
cover almost the whole window, so groups of jobs whose windows overlap
substantially cannot share machines, and choosing the groups to keep
becomes a packing problem.  They are handled by a recursion on
separators with a geometric grid of distances: the grid offsets are
chosen so that the jobs with a witness have at least half of the LP
weight of the nearly rigid jobs, and a dynamic program at each node
assembles the schedule.  At each node, the jobs that stay there are
grouped by the alphabetic code of Gilbert and Moore~\cite{GM}.  A grid cell
whose subtree in the recursion hosts many jobs receives a short code word, so that few
groups compete with those jobs for machines, and the loss adds up
to $O(\log n)$ over the recursion.  The instances of
\Cref{thm:restricted-log-gap} and \cite{MZ} show that the logarithmic
factor is necessary.

\subsection{Related work}\label{sec:related}

\paragraph{Complexity with migration.}  On a single machine, the migratory and nonmigratory problems coincide.  With unit weights, preemptive
throughput maximization with release dates and due dates is solvable
in polynomial time by a dynamic program of Lawler \cite{La90}.  With
arbitrary weights it is NP-hard even without release dates and with a
common due date, since it then contains knapsack \cite{Karp}; Lawler's
program solves it in pseudo-polynomial time \cite{La90}, and it admits
a fully polynomial-time approximation scheme \cite{PW}.  On identical
machines the problem is NP-hard already with unit weights and without
release dates \cite{La83}; a job with $p_j>L_j$ can never be completed
and may be discarded, so this also holds in our model.
Whether it is strongly NP-hard or pseudo-polynomially solvable is
open, and so is the same question under restricted assignment without
release dates \cite{Si}.  With release dates, the problem under
restricted assignment is strongly NP-hard, even with unit weights and
rigid jobs \cite{DL}: in their reduction every job has the same
processing time on all its eligible machines, equal to the length of
its window, so the result holds in our model.  On
unrelated machines the problem is strongly NP-hard, even with unit
weights, also without release dates \cite{Si}.  That reduction relies
on pairs with $p_{ij}> d_j - r_j$, on which a job can make only partial
progress, so it does not establish hardness in our model, where $p_{ij}\le L_j$ for every eligible pair.

\paragraph{Complexity without migration.}  Without migration, the problem is
hard even in two special cases in which
preemption is of no use.  If every job is rigid, a
nonmigratory schedule runs each completed job on a single machine
throughout its window; under restricted assignment the problem is then
interval scheduling with machine eligibility, for which already
deciding whether all jobs can be completed is strongly NP-hard
\cite{AS}.  Without release dates and with a common due date $d$,
every window is $[0,d)$, and a set of jobs can be completed on machine
$i$ without migration if and only if their processing times $p_{ij}$
sum to at most $d$, since they can then run one after another.  On
unrelated machines the problem is then the generalized assignment
problem with bin capacity $d$, which is APX-hard even with unit
weights and only two distinct processing times \cite{CK}.

\paragraph{Offline algorithms.}  Kalyanasundaram and Pruhs
\cite{KP} show that if a migratory schedule on $m$ identical machines
completes a set of jobs, then some nonmigratory schedule on $6m-5$
machines completes the same set; their construction runs in
pseudo-polynomial time, and in polynomial time with $12m-5$ machines.
Keeping the best $m$ of these machines shows that, on identical
machines, the migratory optimum is at most $O(1)$ times the nonmigratory
optimum.
Megow and Zondervan \cite{MZ} show that  given  a set of jobs admitting a
feasible migratory schedule on
$m$ machines a nonmigratory schedule on $2m-1$ machines can be computed in strongly polynomial time.
Hence $\OPT\le\frac{2m-1}{m}\,\OPTnm$ on identical machines, and this is tight \cite{KDA}. In the same paper \cite{MZ} they obtain a polynomial-time
nonmigratory algorithm within a factor $2.542+\delta$ of the migratory optimum, for every $\delta>0$.
Under restricted assignment, \cite{MZ} show that eliminating migration may require $\Omega(m\log m)$ machine copies in
total, even if different machines may receive different numbers of copies, and hence $\Omega(\log m)$ copies of some
machine; they ask whether $O(\log m)$ copies of each machine suffice.
Im and Moseley \cite{IM} eliminate migration on \emph{related machines}, where $p_{ij}=p_j/s_i$ for the speed $s_i$ of
machine $i$, using speed $1+\varepsilon$ and $O(1/\varepsilon)$ copies of each machine, and $O(1/\varepsilon^2)$
copies under restricted assignment.  They also show that additional machines without additional speed do not suffice.
On unrelated machines, Devanur and Kulkarni \cite{DK} show that speed alone suffices: if a time-indexed assignment
relaxation, which allows a job to run on several machines at once, completes every job, then a nonmigratory schedule
that completes every job with speed $16$ can be computed in polynomial time.
Karakostas and Kolliopoulos \cite{KK} give a
polynomial-time $(1-1/e-\delta)$-approximation, for every $\delta>0$, for nonmigratory
weighted throughput on unrelated machines, against a configuration LP
that relaxes the nonmigratory problem.

Sarpatwar, Schieber and Shachnai \cite{SSS} study migratory preemptive throughput on identical machines on which jobs
also have heights, and jobs may share a machine while their heights fit.  For laminar windows on which every job's
slack is at least a constant fraction of $d_j-r_j$, they round an LP with one area constraint per window, at most a constant times $m|\chi|$ inside each laminar
window $\chi$, and obtain a constant factor.  This area constraint is the identical-machine counterpart of the capacity
argument behind \Cref{lem:common-witness}, our common-witness lemma.  For arbitrary windows they require every job to
have slack more than $\frac34(d_j-r_j)$, and first replace every window by a laminar subinterval; such laminarization fails for small slack.

\paragraph{Online algorithms.}  Chan, Lam and To \cite{CLT}
give a nonmigratory online algorithm that completes, on
$m$ identical machines of speed $3+2\sqrt2$, every set of jobs that a
migratory schedule completes on $m$ unit-speed machines.  In the online setting,
forbidding migration can lead to larger losses than offline.  In machine
minimization, a deterministic nonmigratory online algorithm may need
$\Omega(\log n)$ machines on instances that a migratory schedule
completes on a constant number of identical machines \cite{CMS}.  On
identical machines with \emph{slackness} $\eta$, that is, when every
job window is at least $1+\eta$ times the processing time, migration
improves the deterministic competitive ratio for maximizing the total
processing time of completed jobs, when every job must be accepted or
rejected on arrival, from $(1+\eta)/\eta$, attained by the
nonmigratory greedy algorithm \cite{DP}, to a ratio that tends to
$(1+\eta)\ln((1+\eta)/\eta)$ as the number $m$ of machines tends to
infinity \cite{SS}.

%% The smaller the slackness, the harder the problem.
On a single
machine, the best competitive ratio of a deterministic online
algorithm for unweighted throughput grows as the slackness $\eta$ decreases: it is 
$\Theta(1/\eta)$ \cite{CEMSS}.  The lower bound also holds on
unrelated machines, since every job may be eligible on a single
machine.  On unrelated machines, Eberle, Megow and Schewior \cite{EMS}
give a nonmigratory $O(1/\eta)$-competitive algorithm for unit
weights, and Eberle \cite{Eb} one for arbitrary weights, both against
the nonmigratory optimum; on identical machines the guarantees extend
to the migratory optimum, up to a constant factor.  For rigid jobs,
that is, slackness $0$, no deterministic online algorithm has a
bounded competitive ratio for weighted throughput, already on a single
machine \cite{BKM92}, and no randomized one either \cite{CI}.  With
unit weights, the same holds for deterministic algorithms on a single
machine \cite{BHS}, while randomization gives a constant ratio there
\cite{KP03}.  On $m\ge2$ identical machines, in contrast, Moseley,
Pruhs, Stein and Zhou \cite{MPSZ} give a deterministic
$O(1)$-competitive algorithm for unweighted throughput against the
migratory optimum, without any slackness assumption.  Their analysis
first replaces the optimum by a nonmigratory schedule, at the loss of
a constant factor, using \cite{KP}.
%% In the results of \cite{EMS,Eb,MPSZ},
%% eliminating migration turns a guarantee against the nonmigratory optimum into one against the migratory optimum.

\subsection{Organization of the paper}

\Cref{sec:model} defines the model and proves \Cref{prop:machines}.  \Cref{sec:lp} states the $\StrongLP$ and proves
the two lemmas used throughout, \Cref{lem:processing,lem:common-witness}.  \Cref{sec:big-covered} fixes a threshold
$\delta\in(0,1)$ and calls a pair $(i,j)$ \emph{big-slack} if $s_{ij}\ge\delta L_j$.  It rounds the part of a $\StrongLP$ solution on
big-slack pairs, and \Cref{app:big-derandomization} derandomizes this rounding.

\Cref{part:unrelated} proves \Cref{thm:main-unrelated}, with $\delta=1/64$.  \Cref{sec:outline-unrelated} outlines the
proof, \Cref{sec:deficit} introduces the deficit, \Cref{sec:tree,sec:thin,sec:sqrt} round the small-covered jobs with
a balanced tree of separators, \Cref{sec:proof-unrelated} completes the proof, and \Cref{sec:tightness} proves the matching lower
bound.

\Cref{part:ra} proves \Cref{thm:main-ra}, with $\delta=1/8$.  Let $\sigma_j$ be the relative slack of a job $j$, that
is, its slack divided by its window length.  The jobs are split into three classes by the range of $\sigma_j$, each class gives one schedule,
and the heaviest of the three is kept.

\begin{center}
\small
\begin{tabular}{@{}lll@{}}
\toprule
range of $\sigma_j$ & method & loss \\
\midrule
$\sigma_j\ge1/8$ (big-slack) & \Cref{thm:big-rounding} with $\delta=1/8$ & $96$ \\
$\sigma_0\le\sigma_j<1/8$ (moderate-slack) & witnesses and supports, \Cref{thm:moderate} & $O(\log n)$ \\
$\sigma_j<\sigma_0$ (nearly rigid) & recursion tree and dynamic program, \Cref{thm:near} & $O(\log n)$ \\
\bottomrule
\end{tabular}
\end{center}

\noindent Here $\sigma_0$ is a polynomially small threshold, between $2^{-19}n^{-8}$ and $2^{-11}n^{-8}$;
\Cref{def:ra-classes} defines it.  \Cref{sec:ra-prelims} collects preliminaries, \Cref{sec:ws} introduces witnesses
and supports, \Cref{sec:moderate,sec:near} treat moderate-slack and nearly rigid jobs, \Cref{sec:assembly} completes
the proof, and \Cref{sec:restricted-log-gap} gives the logarithmic lower bound.  \Cref{sec:summary} concludes with
open problems.  \Cref{app:fractional-progress} treats the model without \Cref{asm:valid}, \Cref{app:copies} the
number of machine copies, and \Cref{app:big-derandomization} the derandomization of the big-slack rounding.
\Cref{app:hypercube-initial-segment-proof} proves an auxiliary lemma on the boundaries of initial segments of the
hypercube, used in the logarithmic lower bound, \Cref{app:one-machine} proves the classical feasibility criterion of
Horn~\cite{Ho} for one machine, and \Cref{app:rigid-unit} gives a bound in terms of $m$ for rigid jobs with unit
weights.

%=====================================================================
\section{Model}\label{sec:model}
%=====================================================================

\begin{definition}[Instances]\label{def:instance}
An \emph{instance} has a finite set $\M$ of machines and a finite set $\J$ of $n$ jobs.  Each job $j$ has the
following data.
\begin{enumerate}[label=\arabic*.,nosep]
\item A \emph{release date} $r_j\ge0$, a \emph{due date} $d_j>r_j$ and a \emph{weight} $w_j\ge0$.
The job may be processed only within its \emph{window}, that is, the interval $K_j=[r_j,d_j)$ of length $L_j=d_j-r_j>0$.
\item A nonempty set of \emph{eligible machines} $\Gamma(j)\subseteq\M$ on which $j$ may be processed.
\item For every $i\in\Gamma(j)$, a \emph{processing time} $p_{ij}>0$.
\end{enumerate}
A pair $(i,j)$ with $i\in\Gamma(j)$ is an \emph{eligible pair}, and $s_{ij}=L_j-p_{ij}$ is its
\emph{slack}.  Release dates, due dates and processing times are rational numbers, given in binary.
\end{definition}

\noindent Throughout the paper we make the following assumption.

\begin{assumption}\label{asm:valid}
For every job $j$ and every eligible machine $i\in\Gamma(j)$, $p_{ij}\le L_j$.
\end{assumption}

\noindent \Cref{asm:valid} says that every eligible machine can complete $j$ by itself inside $K_j$; equivalently,
every slack $s_{ij}$ is nonnegative.  On unrelated machines it is a genuine
restriction of the model, since pairs with $p_{ij}>L_j$ cannot be discarded without changing $\OPT$, and our proofs need
it: slacks must be nonnegative, and a job
assigned to a machine must be completable on that machine.
\Cref{thm:main-unrelated,cor:unrelated,prop:machines,thm:main-ra,cor:ra,cor:copies} are proved under
\Cref{asm:valid}.  Without it, a machine on which $p_{ij}>L_j$ can still contribute partial progress to a migratory
schedule.  \Cref{app:fractional-progress} shows that the ratio $\OPT/\OPTnm$ can then be as large as the number $n$
of jobs, which is the trivial upper bound.

\begin{definition}[Restricted assignment]\label{def:ra}
An instance is a \emph{restricted-assignment} instance if every job $j$ has a processing time $p_j$ with
$p_{ij}=p_j$ for every $i\in\Gamma(j)$.  Then the \emph{slack} of $j$ is $s_j=L_j-p_j$, its \emph{relative
slack} is $\sigma_j=s_j/L_j$, and $j$ is \emph{rigid} if $s_j=0$.
\end{definition}

Under restricted assignment, \Cref{asm:valid} is without loss of generality.  A job $j$ with $p_j>L_j$ cannot be
completed: it needs processing time $p_j$ inside a window shorter than $p_j$, hence on two machines at the same time,
which \Cref{def:schedule} forbids.  By \Cref{lem:fp-invalid}, it also contributes nothing to the value of any
$\StrongLP$ solution.  Consequently, deleting such jobs
changes neither $\OPT$ nor $\OPTnm$.  The deleted jobs may separate elementary intervals of the reduced instance.
Merging such elementary intervals and adding the corresponding values $y_{ijk}$ turns a $\StrongLP$ solution into one
of the reduced instance with the same value.  Hence
\Cref{thm:main-ra,cor:ra} hold without \Cref{asm:valid}; if fewer
than two jobs remain, running a remaining job alone during the first $p_j$ units of its window gives weight at least
$W$.

By \Cref{asm:valid}, $s_j\ge0$ and $\sigma_j\in[0,1)$.  In the online scheduling literature, an instance has
slackness $\eta$ if every job window is at least $1+\eta$ times the processing time.  Under restricted assignment,
slackness $\eta$ means $s_j\ge\eta p_j$ for every job, or equivalently $\sigma_j\ge\eta/(1+\eta)$.  The slack is
thus measured against the processing time, while the relative slack measures it against the window length.

For a finite union $X$ of intervals, $|X|$ denotes its total
length.  All logarithms are base $2$ unless specified otherwise.

\begin{definition}[Schedules]\label{def:schedule}
\begin{enumerate}[label=\arabic*.,nosep]
\item \emph{Pieces.}  A \emph{piece} is a triple $(i,j,[a,b))$ of a machine $i$, a job $j$ with
$i\in\Gamma(j)$, and a half-open interval $[a,b)\subseteq K_j$ with $a<b$.  It means that machine $i$
processes $j$ during $[a,b)$.
\item \emph{Schedule.}  A \emph{schedule} is a finite set of pieces such that two
pieces with the same machine, or with the same job, have disjoint intervals: every machine processes at most one job
at a time, and no job is processed on two machines at the same time.  Schedules are \emph{preemptive}: a job may have
several pieces, with the same machine or with different machines.
\item \emph{Completion.}  For a job $j$ and a machine $i\in\Gamma(j)$, let $\Pi_{ij}$ be the union of the
intervals of the pieces of $j$ with machine $i$.  A schedule \emph{completes} $j$ if
\[
 \sum_{i\in\Gamma(j)}\frac{|\Pi_{ij}|}{p_{ij}}\ \ge\ 1 .
\]
\item \emph{Nonmigratory.}  A schedule is \emph{nonmigratory} if, for every completed job, all
its pieces have the same machine.
\item \emph{Weight.}  The \emph{weight} $W(A)$ of a schedule $A$ is the total weight of the jobs it
completes.
\end{enumerate}
\end{definition}

\noindent Under restricted assignment, a schedule completes $j$ exactly when the intervals of the pieces of
$j$ have total length at least $p_j$.  A nonmigratory schedule completes $j$ on
machine $i$ exactly when $|\Pi_{ij}|\ge p_{ij}$.  We write $\OPT$ for the largest weight of a
schedule and $\OPTnm$ for the largest weight of a nonmigratory schedule.  Requiring finitely many pieces is no
restriction.  Suppose that machine $i$ processes job $j$ during a measurable subset of $K_j$, where any two such sets with a
common machine, or with a common job, are disjoint.  Then the decomposition of Lawler and Labetoulle \cite{LL} yields a
schedule that completes the same jobs, is nonmigratory if the given processing is, and has a number of pieces bounded
by a polynomial in $n$ and $m$ alone.

% Optional paragraph, if a referee asks whether finitely many pieces is a restriction.
% If it is restored, rename P_{ijk},
% which clashes with P(M), P_j and P_v.
%
% Requiring finitely many pieces loses nothing.  Suppose a schedule in a more general sense spends
% time $P_{ijk}$ of the elementary interval $\Delta_k$ (Section~\ref{sec:lp}) processing job
% $j$ on machine $i$, with $\sum_jP_{ijk}\le|\Delta_k|$ and $\sum_iP_{ijk}\le|\Delta_k|$.  As in
% preemptive scheduling on unrelated machines~\cite{LL}, the matrix $(P_{ijk})_{i,j}$ is a
% nonnegative combination of matchings that use only entries with $P_{ijk}>0$, with total
% coefficient at most $|\Delta_k|$.  Running these matchings one after another in $\Delta_k$ gives
% finitely many pieces with the same amounts $P_{ijk}$.  The result completes the same jobs, and it
% is nonmigratory if the given schedule is.

\begin{definition}[Running a job during a set of times]\label{def:run}
Let $Y$ be a finite union of half-open intervals.
\begin{enumerate}[label=\arabic*.,nosep]
\item \emph{Earliest units.}  For $0<p\le|Y|$, the \emph{earliest $p$ units} of $Y$ are the set $Y\cap(-\infty,x)$, where $x$ is the
least number with $|Y\cap(-\infty,x)|=p$.
\item \emph{Running.}  \emph{Running} a job $j$ on a machine $i\in\Gamma(j)$ during $Y$, where $Y\subseteq K_j$, means using one piece $(i,j,Y')$ for each
maximal interval $Y'$ contained in $Y$; these maximal intervals are finitely many pairwise disjoint half-open intervals.
\end{enumerate}
\end{definition}

\noindent The earliest $p$ units of $Y$ are again a finite union
of half-open intervals, they lie in $Y$, and their total length is $p$.

Both parts of the paper consider windows as seen from a point $t$ of the time axis, a separator, and use the
following two notions.

\begin{definition}[A job seen from a point]\label{def:seen-from}
Fix a point $t\in\mathbb R$.  The two \emph{sides} of $t$ are $+$, the points $x>t$, and $-$, the points $x<t$; the
point $t$ lies on neither side.  For a job $j$, its \emph{reach} is $\rho_j=\max\{|t-r_j|,\,|d_j-t|\}$, and its
\emph{long side} is $\eta_j=+$ if $d_j-t\ge t-r_j$, and $\eta_j=-$ otherwise.
\end{definition}

\noindent Thus $\rho_j$ is the supremum of the distances $|x-t|$ of the points $x\in K_j$ from $t$, and $K_j$ has points
on side $\eta_j$ at distances arbitrarily close to $\rho_j$; the long side is $+$ exactly when $t\le(r_j+d_j)/2$.  The
reach and the long side depend on $t$, which the notation leaves out.

\subsection{Dependence on the number of machines}\label{sec:machines}

This subsection proves \Cref{prop:machines}.  An instance has \emph{identical machines} if $\Gamma(j)=\M$ for every job $j$
and $p_{ij}$ does not depend on $i$.  We use the theorem of Megow and Zondervan \cite{MZ}, which improves the bound $6m-5$ of
Kalyanasundaram and Pruhs \cite{KP}: if a schedule on $m$ identical machines completes a set $C$ of jobs, then some
nonmigratory schedule on $2m-1$ identical machines, with the same windows and processing times, completes every job
of $C$.

\begin{proof}[Proof of \Cref{prop:machines}]
Let $m=|\M|$.  For every job $j$ fix a machine $i^*(j)\in\Gamma(j)$ with $p_{i^*(j)j}=p^*_j:=\min_{i\in\Gamma(j)}p_{ij}$, a
\emph{fastest machine} of $j$; then $p^*_j\le L_j$ by \Cref{asm:valid}.  For $i\in\M$ let $\J_i=\{j\in\J:\ i^*(j)=i\}$, and write
$w(S)=\sum_{j\in S}w_j$ for a set $S$ of jobs.

\noindent\emph{The algorithm.}  For every machine $i$, consider the single-machine instance with the machine $i$ and the jobs of $\J_i$,
with their windows, weights and processing times $p_{ij}$, and compute a schedule $A_i$ of it.  With unit weights, $A_i$ is a
schedule of largest weight, computed in polynomial time by the dynamic program of Lawler \cite{La90}.  With arbitrary weights,
$A_i$ has weight at least $1-\varepsilon$ times the largest, and is computed in time polynomial in the input size and
$1/\varepsilon$ by the approximation scheme of Pruhs and Woeginger \cite{PW}.  Output the union $A'$ of the schedules $A_i$.  Two
pieces of $A'$ with the same machine $i$ belong to $A_i$, and all pieces of a job $j$ belong to $A_{i^*(j)}$, so $A'$ is a
nonmigratory schedule of the given instance, of weight $\sum_{i\in\M}W(A_i)$.

It remains to show that some nonmigratory schedule completes, for every machine $i$, a set $D_i\subseteq\J_i$ of jobs on $i$
alone, with $\sum_{i\in\M}w(D_i)\ge\OPT/(2m-1)$.  This gives $\OPTnm\ge\OPT/(2m-1)$.  The pieces of that schedule with machine $i$
form a schedule of the single-machine instance of $i$ that completes $D_i$, so $W(A_i)\ge(1-\varepsilon)\,w(D_i)$, with
$\varepsilon=0$ for unit weights, and $A'$ has weight at least $(1-\varepsilon)\OPT/(2m-1)$.  The rest of the proof constructs
such a schedule; this construction is used only in the analysis.  Let $A$ be a schedule of weight $\OPT$, and let $C$ be the set
of jobs it completes.

\noindent\emph{Identical machines.}  Let $I'$ be the instance with the machines of $\M$ and the jobs of $C$, with their windows and
weights, in which every job is eligible on every machine with processing time $p^*_j$.  Since $p^*_j\le L_j$, it is an instance
with identical machines.  The pieces of $A$ of the jobs of $C$ form a schedule of $I'$ that completes every job of $C$: the
pieces of $j$ have machines in $\Gamma(j)$, where $p^*_j\le p_{ij}$, so
\[
 \sum_{i\in\M}\frac{|\Pi_{ij}|}{p^*_j}\ \ge\ \sum_{i\in\Gamma(j)}\frac{|\Pi_{ij}|}{p_{ij}}\ \ge\ 1 .
\]

\noindent\emph{Eliminating migration.}  Keeping only the first $p^*_j$ units of processing of every job of $C$ gives a schedule
of $I'$ that processes every job of $C$ for exactly $p^*_j$ units.  By \cite{MZ} there is a nonmigratory schedule $B$ on $2m-1$ identical machines
$1,\dots,2m-1$, with the windows and processing times of $I'$, that completes every job of $C$.  Hence every $j\in C$ has all its
pieces in $B$ on one machine $a(j)$, and these pieces cover at least $p^*_j$ units of $K_j$.

\noindent\emph{Returning to the machines of $\M$.}  For $i\in\M$ and $a\in\{1,\dots,2m-1\}$ let $C_{i,a}=\{j\in C:\ i^*(j)=i,\ a(j)=a\}$, and
choose $a_i$ with $w(C_{i,a_i})$ largest.  For every $i\in\M$, every $j\in C_{i,a_i}$ and every piece of
$j$ in $B$, with interval $[x,y)$, use the piece $(i,j,[x,y))$; here $i=i^*(j)\in\Gamma(j)$.  The pieces with machine $i$ come from
pieces of $B$ with the single machine $a_i$, so their intervals are pairwise disjoint; the pieces of a job $j$ come from its pieces
in $B$, so their intervals are pairwise disjoint, and they all have machine $i^*(j)$.  Therefore these pieces form a nonmigratory
schedule of the given instance.  A job $j\in C_{i,a_i}$ receives at least $p^*_j=p_{ij}$ units of $K_j$ on $i$, so the schedule
completes it.  Thus $D_i=C_{i,a_i}\subseteq\J_i$ is completed on $i$ alone.  The sets $C_{i,a}$ partition $C$, so
\[
 \sum_{i\in\M}w(C_{i,a_i})\ \ge\ \sum_{i\in\M}\frac1{2m-1}\sum_{a=1}^{2m-1}w(C_{i,a})\ =\ \frac{w(C)}{2m-1}\ =\ \frac{\OPT}{2m-1}.
 \qedhere
\]
\end{proof}

%=====================================================================
\section{The \texorpdfstring{\StrongLP}{StrongLP}}\label{sec:lp}
%=====================================================================

This section states the $\StrongLP$, shows that it relaxes migratory scheduling, and proves the two facts
about LP solutions that both parts of the paper use: a job receives most of its LP processing in any large
part of its window, and LP processing on a common set of times yields an assignment of jobs to distinct
machines.

Let $\tau_0<\tau_1<\dots<\tau_\nu$ be the distinct release dates and due dates of the jobs.  Let
$\Delta_k=[\tau_{k-1},\tau_k)$ for $k=1,\dots,\nu$; these are the \emph{elementary intervals}.
Every window is the union of the elementary intervals it contains, and $\nu\le2n-1$.

For every eligible pair $(i,j)$ and every $k$ with $\Delta_k\subseteq K_j$, let $y_{ijk}\ge0$ denote the
amount of processing time assigned to $j$ on $i$ inside $\Delta_k$; set $y_{ijk}=0$ for all other triples
$(i,j,k)$.  Let $z_j\in[0,1]$ be the fractional completion of job $j$.  The strengthened relaxation is
\begin{align}
 \StrongLP:\qquad
 \max\quad &\sum_{j\in\J}w_jz_j \label{lp:strong-objective}\\
 \text{s.t.}\quad
 &\sum_{i\in\Gamma(j)}\sum_k\frac{y_{ijk}}{p_{ij}}=z_j,
 &&j\in\J, \label{eq:lp-coverage}\\
 &\sum_{j}y_{ijk}\le |\Delta_k|,
 &&i\in\M,\ k=1,\dots,\nu, \label{eq:lp-machine}\\
 &\sum_{i\in\Gamma(j)}y_{ijk}\le |\Delta_k|\,z_j,
 &&j\in\J,\ k=1,\dots,\nu, \label{eq:lp-job}\\
 &0\le z_j\le1,\qquad y_{ijk}\ge0. \notag
\end{align}
We call \eqref{eq:lp-coverage}, \eqref{eq:lp-machine} and \eqref{eq:lp-job} the \emph{coverage},
\emph{machine capacity} and \emph{non-parallelism} constraints.  A \emph{$\StrongLP$ solution} is a pair $(y,z)$
that satisfies all constraints of the program; its \emph{value} is $W=\sum_jw_jz_j$, and $\OPTLP$ denotes the optimum value.
The \emph{LP weight} of a set $J$ of jobs is $\sum_{j\in J}w_jz_j$.
Non-parallelism says that a job of which a fraction $z_j$ is accepted never runs faster than $z_j$; for $z_j=1$, that
it never runs on two machines at once.  It is what makes the relaxation strong: without the
non-parallelism constraint the integrality gap of the relaxation is $\Omega(n)$.  For example, on one machine, take
$n$ rigid unit-weight jobs with windows $[0,2^k)$, $k=1,\ldots,n$.  Every two of these windows overlap, so
$\OPT=1$.  Without non-parallelism, job $k$ can be processed at full rate during $[2^{k-1},2^k)$, which gives
$z_k=1/2$ for every $k$ and value $n/2$.  The $\StrongLP$ has
$O(n^2|\M|)$ variables and constraints, and the algorithms below apply to any $\StrongLP$ solution.

\begin{lemma}[Validity of $\StrongLP$]\label{lem:strong-valid}
Every schedule, migratory or not, gives a $\StrongLP$ solution of the same value.  Hence $\OPT\le\OPTLP$.
\end{lemma}

\begin{proof}
Let $A$ be a schedule, and let $\Pi_{ij}$ be as in \Cref{def:schedule}.  For a completed job $j$ and a number
$x$ set
\[
 \pi_j(x)=\sum_{i\in\Gamma(j)}\frac{|\Pi_{ij}\cap(-\infty,x)|}{p_{ij}} .
\]
It is continuous and nondecreasing in $x$, it is $0$ for $x\le r_j$, and it is at least $1$ for $x\ge d_j$,
since $j$ is completed.  Let $x_j$ be the least number with $\pi_j(x_j)=1$.  For every completed job $j$ and
every $i\in\Gamma(j)$ keep the times of $\Pi_{ij}\cap(-\infty,x_j)$ at which $i$ processes $j$, and
drop all other processing.  Set $z_j=1$ for the completed jobs and $z_j=0$ for the others, and let $y_{ijk}$
be the total length of the kept times in $\Delta_k$ at which machine $i$ processes $j$.

The kept times of $j$ lie in $K_j$, which is a union of elementary intervals, so $y_{ijk}>0$ only if
$\Delta_k\subseteq K_j$, and coverage holds because $\pi_j(x_j)=1$.  Machine capacity holds because the
pieces of a machine have disjoint intervals.  Non-parallelism holds because the pieces of a job have
disjoint intervals, so a completed job is processed for a total time of at most $|\Delta_k|=|\Delta_k|z_j$ in
$\Delta_k$.  The value of the point is the weight of $A$.
\end{proof}

For every eligible pair $(i,j)$ define its \emph{normalized coverage}
\begin{equation}\label{eq:xij}
 x_{ij}:=\sum_k\frac{y_{ijk}}{p_{ij}} .
\end{equation}
Then
\begin{equation}\label{eq:x-sums-z}
 \sum_i x_{ij}=z_j,
 \qquad
 p_{ij}x_{ij}=\sum_ky_{ijk}.
\end{equation}
The LP fixes only the amount $y_{ijk}$ of processing of $j$ on $i$ in each elementary interval $\Delta_k$, so we spread
this amount evenly over $\Delta_k$.  A part $X\cap\Delta_k$ of the
interval then receives the fraction $|X\cap\Delta_k|/|\Delta_k|$ of it.  Accordingly, for a finite union $X$ of
intervals, we define the processing of $j$ on $i$ during $X$ and the total processing of $j$ during $X$ as
\begin{equation}\label{eq:yX}
 y_{ij}(X):=\sum_k\frac{|X\cap\Delta_k|}{|\Delta_k|}\,y_{ijk},
 \qquad
 y_j(X):=\sum_{i\in\Gamma(j)}y_{ij}(X).
\end{equation}
In particular, $y_{ij}(\Delta_k)=y_{ijk}$, and $y_{ij}(X)$ is additive over disjoint sets $X$.

\begin{lemma}[Processing in a set of times]\label{lem:processing}
Fix a $\StrongLP$ solution and a finite union $X$ of intervals.
\begin{enumerate}[label=(\alph*),nosep]
\item $y_{ij}(K_j)=p_{ij}x_{ij}$ for every eligible pair $(i,j)$.
\item $\sum_jy_{ij}(X)\le|X|$ for every machine $i$.
\item For every job $j$ and every set $E$ of eligible pairs of $j$,
$\sum_{i:(i,j)\in E}y_{ij}(X)\le z_j\,|X\cap K_j|$.
\item If moreover $X\subseteq K_j$, then
\[
 \sum_{i:(i,j)\in E}y_{ij}(X)\ \ge\ \sum_{i:(i,j)\in E}p_{ij}x_{ij}\ -\ z_j\bigl(L_j-|X|\bigr).
\]
\end{enumerate}
\end{lemma}

\begin{proof}
(a) Every elementary interval lies in $K_j$ or is disjoint from it, and $y_{ijk}>0$ only if $\Delta_k\subseteq K_j$.  Consequently,
$y_{ij}(K_j)=\sum_ky_{ijk}=p_{ij}x_{ij}$ by \eqref{eq:x-sums-z}.

(b) By machine capacity, $\sum_jy_{ij}(X)=\sum_k\frac{|X\cap\Delta_k|}{|\Delta_k|}\sum_jy_{ijk}\le\sum_k|X\cap\Delta_k|\le|X|$, since the
elementary intervals are pairwise disjoint.

(c) Since $y_{ijk}>0$ only if $\Delta_k\subseteq K_j$, non-parallelism gives
\[
 \sum_{i:(i,j)\in E}y_{ij}(X)\ \le\ \sum_{k:\,\Delta_k\subseteq K_j}\frac{|X\cap\Delta_k|}{|\Delta_k|}\sum_{i\in\Gamma(j)}y_{ijk}
 \ \le\ z_j\sum_{k:\,\Delta_k\subseteq K_j}|X\cap\Delta_k|\ =\ z_j|X\cap K_j|,
\]
where the last equality holds because $K_j$ is the disjoint union of the elementary intervals it contains.

(d) $K_j$ is the disjoint union of $X$ and the finite union of intervals $K_j\setminus X$, of length $L_j-|X|$.  By additivity and (a),
$y_{ij}(X)=p_{ij}x_{ij}-y_{ij}(K_j\setminus X)$.  Summing over the pairs of $E$ and applying (c) to $K_j\setminus X$
gives the claim.
\end{proof}

Under restricted assignment with $E$ the set of all eligible pairs of $j$, part (d) becomes
$y_j(X)\ge z_j(|X|-s_j)$; this is \Cref{lem:ra-processing}: a job is busy on most of any large part of its window.
The next lemma is the central operation of both parts of the paper.  We call a set $X$ as in its statement
a \emph{common witness} for $F$.

\begin{lemma}[Common witness]\label{lem:common-witness}
Fix a $\StrongLP$ solution, a set $E$ of eligible pairs, a set $F$ of jobs, a finite union $X$ of intervals with
$|X|>0$, and a number $\rho\in(0,1]$.  Suppose that
\[
 \sum_{i:(i,j)\in E}y_{ij}(X)\ \ge\ \rho\,z_j|X|\qquad(j\in F).
\]
Then one can compute in polynomial time a set $I\subseteq F$ and an assignment of the jobs of $I$ to distinct
machines, such that every $j\in I$ is assigned to a machine $i$ with $(i,j)\in E$ and $y_{ij}(X)>0$, and
\[
 \sum_{j\in I}w_j\ \ge\ \rho\sum_{j\in F}w_jz_j .
\]
\end{lemma}

\begin{proof}
For $j\in F$ and $(i,j)\in E$ set $f_{ij}:=y_{ij}(X)/|X|$.  By \Cref{lem:processing}(b), $\sum_jf_{ij}\le1$ for every
machine $i$.  By \Cref{lem:processing}(c), $\sum_if_{ij}\le z_j\le1$ for every $j\in F$, and by assumption
$\sum_if_{ij}\ge\rho z_j$.  Thus $f$ is a fractional matching of the bipartite graph between $F$ and the
machines whose edges are the pairs $(i,j)\in E$ with $f_{ij}>0$.  Give every edge incident to $j$ the weight
$w_j$; the weight of $f$ is $\sum_{j\in F}w_j\sum_if_{ij}\ge\rho\sum_{j\in F}w_jz_j$.  The bipartite matching
polytope is integral, so a maximum-weight matching of this graph, which can be computed in polynomial time,
has at least this weight.  Let $I$ be the set of its matched jobs, each assigned to its matched machine.
\end{proof}

%=====================================================================
\section{Big slack}\label{sec:big-covered}
%=====================================================================

In this section we round the part of a $\StrongLP$ solution that lies on pairs with large slack.  Fix $\delta\in(0,1)$.  An eligible pair $(i,j)$
is \emph{big-slack} if
\[
 p_{ij}\le(1-\delta)L_j,
\]
and \emph{small-slack} otherwise.  Let $\EB$ and $\ES$ be the sets of big-slack and of small-slack pairs.  For a job
$j$, let
\begin{equation}\label{eq:pair-mass}
 z_j^{\mathrm B}:=\sum_{i:(i,j)\in\EB}x_{ij},
 \qquad
 z_j^{\mathrm S}:=\sum_{i:(i,j)\in\ES}x_{ij}
\end{equation}
be the parts of its coverage on big-slack and on small-slack pairs.  By \eqref{eq:x-sums-z},
\begin{equation}\label{eq:pair-split}
 z_j^{\mathrm B}+z_j^{\mathrm S}=z_j.
\end{equation}
These notions depend on $\delta$; \Cref{part:unrelated} uses $\delta=1/64$ and \Cref{part:ra} uses $\delta=1/8$.

\begin{theorem}[Big-slack rounding]\label{thm:big-rounding}
Let $\delta\in(0,1)$, and fix a $\StrongLP$ solution.  For every subset $\J'\subseteq\J$, one can compute in deterministic
polynomial time a nonmigratory schedule $A_{\J'}$, in which every completed job $j$ runs on a machine $i$ with
$(i,j)\in\EB$ and $x_{ij}>0$, such that
\begin{equation}\label{eq:big-rounding}
 W(A_{\J'})\ \ge\ \frac{\delta}{12}
 \sum_{j\in\J'}w_jz_j^{\mathrm B}.
\end{equation}
\end{theorem}

\noindent The rest of this section gives a randomized rounding algorithm that achieves the bound in expectation.
\Cref{app:big-derandomization} derandomizes it by the method of conditional expectations and proves
\Cref{thm:big-rounding}.

\paragraph{The randomized algorithm.}
Fix a subset $\J'\subseteq\J$, and let $\gamma:=\delta/6$.  Every job $j\in\J'$ chooses, independently of the other jobs, at
most one machine: a machine $i$ with $(i,j)\in\EB$ with probability $\gamma x_{ij}$, and no machine with the remaining
probability $1-\gamma z_j^{\mathrm B}$.  This is a valid distribution, since $z_j^{\mathrm B}\le z_j\le1$ and
$\gamma<1$.  Let $R_{ij}=1$ if $j$ chooses $i$, and $R_{ij}=0$ otherwise.  A job chooses $i$ only if $x_{ij}>0$.

Given the choices, a greedy algorithm builds a nonmigratory schedule $A(R)$, separately on each machine $i$.  It
considers the jobs that chose $i$ in nondecreasing order of window length, with a fixed tie-breaking rule.  When it
considers a job $j$, the \emph{idle time} of $j$ is the set of times of $K_j$ at which no job admitted before $j$ runs on
$i$; it is a finite union of half-open intervals.  If the idle time of $j$ has length at least $p_{ij}$, the algorithm
\emph{admits} $j$ and runs it on $i$ during the earliest $p_{ij}$ units of its idle time; otherwise it discards $j$.  Every
admitted job $j$ is completed on the machine $i$ it chose, so $(i,j)\in\EB$ and $x_{ij}>0$.

A job $j$ that chose $i$ can be discarded only because of the jobs admitted before it on $i$.  These jobs also chose
$i$, their windows are not longer than $K_j$, and those that occupy time in $K_j$ have windows that intersect $K_j$.  For
a big-slack pair $(i,j)$, let
\[
 \mathcal I_{ij}:=\{h\in\J'\setminus\{j\}:(i,h)\in\EB,\ L_h\le L_j,\
 K_h\cap K_j\ne\varnothing\},
 \qquad
 Z_{ij}:=\sum_{h\in\mathcal I_{ij}}p_{ih}R_{ih},
\]
the total processing time of the jobs of $\mathcal I_{ij}$ that choose $i$.

\begin{lemma}[Expected interfering volume]\label{lem:big-volume}
For every big-slack pair $(i,j)$,
\[
 \E[Z_{ij}]\le3\gamma L_j.
\]
\end{lemma}

\begin{proof}
If $h\in\mathcal I_{ij}$, then $K_h$ intersects $K_j$ and $L_h\le L_j$, so
\[
 K_h\subseteq[r_j-L_j,d_j+L_j),
\]
an interval of length $3L_j$, and $y_{ihk}>0$ only if $\Delta_k\subseteq K_h$.  Hence, by \eqref{eq:x-sums-z} and machine
capacity,
\[
 \sum_{h\in\mathcal I_{ij}}p_{ih}x_{ih}
 =\sum_k\sum_{h\in\mathcal I_{ij}}y_{ihk}
 \le\sum\bigl\{|\Delta_k|:\ \Delta_k\subseteq[r_j-L_j,d_j+L_j)\bigr\}
 \le3L_j.
\]
Since $\E[R_{ih}]=\gamma x_{ih}$, linearity of expectation gives the claim.
\end{proof}

\begin{lemma}[Expected weight]\label{lem:big-expected}
The schedule $A(R)$ satisfies
\[
 \E[W(A(R))]\ \ge\ \frac{\delta}{12}\sum_{j\in\J'}w_jz_j^{\mathrm B}.
\]
\end{lemma}

\begin{proof}
Let $(i,j)$ be a big-slack pair, and suppose that $j$ chooses $i$.  The jobs admitted before $j$ on $i$ that occupy time
in $K_j$ lie in $\mathcal I_{ij}$ and chose $i$, so they occupy at most $Z_{ij}$ of $K_j$.  If $Z_{ij}\le\delta L_j$, the
idle time of $j$ therefore has length at least $L_j-\delta L_j\ge p_{ij}$, and $j$ is admitted.  The event $R_{ij}=1$
depends only on the choice of $j$, and $Z_{ij}$ only on the choices of other jobs, so they are independent.  By
\Cref{lem:big-volume} and Markov's inequality, $\Pr[Z_{ij}>\delta L_j]\le3\gamma/\delta=1/2$.  Hence $j$ is admitted on
$i$ with probability at least $\gamma x_{ij}/2$, and
\[
 \E[W(A(R))]\ \ge\ \sum_{j\in\J'}w_j\sum_{i:(i,j)\in\EB}\frac{\gamma x_{ij}}2
 \ =\ \frac\gamma2\sum_{j\in\J'}w_jz_j^{\mathrm B}
 \ =\ \frac\delta{12}\sum_{j\in\J'}w_jz_j^{\mathrm B}.
 \qedhere
\]
\end{proof}

%=====================================================================
\clearpage
\part{Unrelated machines}\label{part:unrelated}
%=====================================================================

%=====================================================================
\section{Outline of the proof}\label{sec:outline-unrelated}
%=====================================================================

This part proves \Cref{thm:main-unrelated}.  The proof rests on one operation, \Cref{lem:common-witness}: if a set of
jobs has a common witness, then a heavy subset of it can be completed on distinct eligible machines.  The proof
partitions the jobs into sets to which this lemma, or the ladder argument of \Cref{sec:thin}, applies, and returns the
heaviest of the resulting schedules.  This section outlines the proof.

Throughout \Cref{part:unrelated} we fix a $\StrongLP$ solution $(y,z)$ of value $W$, and we use the notions of
\Cref{sec:big-covered} with
\[
 \delta=\frac1{64}.
\]
Jobs with $z_j=0$ contribute nothing to $W$ and are ignored: every set of jobs considered below consists of jobs with
$z_j>0$.

\begin{definition}[Big-covered and small-covered jobs]\label{def:covered}
A job $j$ with $z_j>0$ is \emph{big-covered} if
\begin{equation}\label{eq:big-covered}
 z_j^{\mathrm B}\ge \frac{z_j}{64},
\end{equation}
and \emph{small-covered} otherwise.  Let $\J_{\mathrm B}$ and $\J_{\mathrm S}$ be the sets of big-covered and of
small-covered jobs.
\end{definition}

\noindent For $\J'=\J_{\mathrm B}$, \Cref{thm:big-rounding} gives a nonmigratory schedule of weight at least
$\frac{\delta}{768}\sum_{j\in\J_{\mathrm B}}w_jz_j$, since $z_j^{\mathrm B}\ge z_j/64$ for every big-covered job.

A small-covered job has $z_j^{\mathrm B}<z_j/64$, so $z_j^{\mathrm S}>\frac{63}{64}z_j$.  The deficit identity of
\Cref{sec:deficit} then shows that its deficit, the part of its window not filled by its normalized small-slack
processing, is less than $L_j/32$.  Non-parallelism spreads this processing over the window: every part $X$ of the
window receives at least $|X|-\Phi_j$ of it.  We distribute the jobs of $\J_{\mathrm S}$ over the nodes of a balanced tree of separators, the \emph{midpoint tree} of
\Cref{sec:tree}.  Every job is handled at one node $v$ of the tree: the first node on its path from the root whose separator $t_v$ lies
in its window.

At a node, the small-covered jobs split into two types.
\begin{itemize}[leftmargin=2em]
 \item \emph{Thick jobs} have deficit at most one eighth of the shortest window of a job handled at the node.  After
 choosing the long side of each window, all jobs with the same long side have a common witness of the form
 $[t_v,t_v+r)$ or $[t_v-r,t_v)$ next to the separator $t_v$, during which they retain at least three quarters of
 their possible small-slack processing.  \Cref{lem:common-witness} then selects a heavy subset of them and assigns
 its jobs to distinct machines.
 \item \emph{Thin jobs} have larger deficit.  Each such job nevertheless has a whole interval of \emph{usable
 distances} from the separator.  If the usable-distance intervals of several jobs share a point, these jobs again have
 a common witness, and they form a \emph{group}.  If the intervals are pairwise disjoint, the windows grow by a factor
 greater than eight from one job to the next, and the jobs form a \emph{ladder}.  A \emph{reserve machine} for each
 job then allows a ladder to be partitioned into four schedules.
\end{itemize}

The interval partition has two stages.  The \emph{common-point removal stage}
repeatedly extracts a large group of usable-distance intervals sharing one point.  The \emph{coloring stage}
colors the remaining intervals into ladders.  With $k$ of order $\sqrt N$ for a node hosting at most $N$
jobs, the two stages produce $O(\sqrt N)$ candidate schedules at that node.  Nodes of the same depth have disjoint
regions, so for each depth $d$ the candidates of its nodes combine into $O(\sqrt{N_d})$ schedules for the whole depth,
where $N_d$ bounds the number of jobs hosted by a node of depth $d$.  Since $N_d$ decreases geometrically with $d$, the
total number of candidates is
\[
 O\bigl(\sqrt n+\sqrt{n/2}+\sqrt{n/4}+\cdots\bigr)=O(\sqrt n),
\]
and the heaviest of them has at least a $1/O(\sqrt n)$ fraction of the LP weight of the thin jobs.  The factor
$\sqrt n$ of \Cref{thm:main-unrelated} comes from this count alone; big-covered jobs lose only a constant factor, and thick jobs only $O(\log n)$.

%=====================================================================
\section{Small-slack processing and the deficit}\label{sec:deficit}
%=====================================================================

This section defines the deficit of a job, the part of its window not filled by normalized small-slack processing,
and bounds it for small-covered jobs.  It then specializes the common-witness lemma to small-slack pairs.

For a job with $z_j>0$ and a finite union $X$ of intervals, define the \emph{normalized small-slack processing}
\begin{equation}\label{eq:qS}
 q_j^{\mathrm S}(X):=\frac1{z_j}\sum_{i:(i,j)\in\ES}y_{ij}(X),
\end{equation}
with $y_{ij}(X)$ as in \eqref{eq:yX}.  It is additive over disjoint sets $X$, and by \Cref{lem:processing}(c),
\begin{equation}\label{eq:qS-bound}
 0\le q_j^{\mathrm S}(X)\le|X\cap K_j|.
\end{equation}
Set
\begin{equation}\label{eq:deficit-def}
 T_j^{\mathrm S}:=q_j^{\mathrm S}(K_j),
 \qquad
 \Phi_j:=L_j-T_j^{\mathrm S}.
\end{equation}
Thus $\Phi_j\ge0$ is the \emph{deficit} of normalized small-slack processing.  Recall that $s_{ij}=L_j-p_{ij}\ge0$
is the slack of an eligible pair.

\begin{lemma}[Exact deficit identity]\label{lem:deficit}
For every job with $z_j>0$,
\begin{equation}\label{eq:deficit}
 \Phi_j=
 \frac{L_jz_j^{\mathrm B}
       +\sum_{i:(i,j)\in\ES}s_{ij}x_{ij}}{z_j}.
\end{equation}
\end{lemma}

\begin{proof}
By \Cref{lem:processing}(a),
\[
 T_j^{\mathrm S}
 =\frac1{z_j}\sum_{i:(i,j)\in\ES}p_{ij}x_{ij}
 =\frac1{z_j}\Bigl(
 L_jz_j^{\mathrm S}-\sum_{i:(i,j)\in\ES}s_{ij}x_{ij}
 \Bigr).
\]
Substituting $z_j^{\mathrm S}=z_j-z_j^{\mathrm B}$, which holds by \eqref{eq:pair-split}, and subtracting from $L_j$
gives \eqref{eq:deficit}, since $\Phi_j=L_j-T_j^{\mathrm S}$.
\end{proof}

\begin{remark}[Coverage values $z_j$ are kept]\label{rem:no-projection}
Every job keeps its value $z_j$ from the given $\StrongLP$ solution.  The coverage masses $z_j^{\mathrm B}$ and
$z_j^{\mathrm S}$ of \eqref{eq:pair-mass} are related to $z_j$ only by $z_j=z_j^{\mathrm B}+z_j^{\mathrm S}$, from
\eqref{eq:pair-split}, and by the threshold $z_j/64$ of \eqref{eq:big-covered}.
\end{remark}

\begin{lemma}[Deficit of small-covered jobs]\label{lem:sc-deficit}
If $j$ is small-covered, then
\begin{equation}\label{eq:sc-deficit}
 \Phi_j<\frac{L_j}{32}.
\end{equation}
\end{lemma}

\begin{proof}
For a small-slack pair, $s_{ij}<L_j/64$.  Since $z_j^{\mathrm B}<z_j/64$ and $z_j^{\mathrm S}\le z_j$,
\Cref{lem:deficit} gives
\[
 \Phi_j
 <\frac{L_jz_j/64+(L_j/64)z_j^{\mathrm S}}{z_j}
 \le\frac{L_j}{32}.
\]
\end{proof}

\begin{lemma}[Common witness for small-slack processing]\label{lem:small-witness}
Let $\J'\subseteq\J$ be a set of jobs with $z_j>0$, let $X$ be a finite union of intervals with $|X|>0$ and $X\subseteq K_j$ for
every $j\in \J'$, and let $\rho\in(0,1]$.  Suppose that
\begin{equation}\label{eq:small-witness-assumption}
 \Phi_j\le(1-\rho)|X|
 \qquad(j\in \J').
\end{equation}
Then one can compute in polynomial time a set $I\subseteq \J'$ with
\begin{equation}\label{eq:small-witness-conclusion}
 \sum_{j\in I}w_j\ \ge\ \rho\sum_{j\in \J'}w_jz_j
\end{equation}
and an assignment of the jobs of $I$ to distinct machines, in which every $j\in I$ is assigned to a machine $i$
with $(i,j)\in\ES$, $y_{ij}(X)>0$ and $x_{ij}>0$.  Running each $j\in I$ on its machine $i$ during the
earliest $p_{ij}$ units of $K_j$ gives a nonmigratory schedule that completes every job of $I$.
\end{lemma}

\begin{proof}
The set $X$ serves only to find a machine for each selected job; each selected job then runs during the earliest
$p_{ij}$ units of its window, which need not lie in $X$.

Fix $j\in \J'$.  By \Cref{lem:processing}(d), with $E$ the set of small-slack pairs of $j$, and by
\Cref{lem:processing}(a),
\[
 z_j\,q_j^{\mathrm S}(X)
 \ \ge\ \sum_{i:(i,j)\in\ES}p_{ij}x_{ij}-z_j(L_j-|X|)
 \ =\ z_j\bigl(T_j^{\mathrm S}-L_j+|X|\bigr)
 \ =\ z_j\bigl(|X|-\Phi_j\bigr)
 \ \ge\ \rho\,z_j|X| .
\]
Thus the hypothesis of \Cref{lem:common-witness} holds with $E=\ES$, which gives $I$ and the assignment.
Since $y_{ij}(X)>0$, some $y_{ijk}$ is positive, so $x_{ij}>0$.  The pair $(i,j)$ is eligible and
$p_{ij}\le L_j=|K_j|$, so the earliest $p_{ij}$ units of $K_j$ are defined.  The pieces of a job
of $I$ are on its machine, have pairwise disjoint intervals, lie in $K_j$ and have total length $p_{ij}$, and
pieces of different jobs of $I$ are on different machines.  Therefore these pieces form a nonmigratory schedule that
completes every job of $I$.
\end{proof}

%=====================================================================
\section{The midpoint tree and thick jobs}\label{sec:tree}
%=====================================================================

This section organizes the small-covered jobs by a balanced recursion on window midpoints, the midpoint tree.  Nodes of the
same depth have pairwise disjoint time regions, so schedules computed independently at these nodes combine into one
schedule.  Thick jobs are then handled directly by \Cref{lem:small-witness}.

For a job $j$ let $m_j=(r_j+d_j)/2$ be the midpoint of its window.  For a nonempty set $J$ of jobs, let $t(J)$ be the
$\lceil|J|/2\rceil$-th smallest of the midpoints $m_j$, $j\in J$, counted with multiplicity, and set
\[
 F(J)=\{j\in J:\ t(J)\in K_j\},\qquad J^{\mathrm L}=\{j\in J\setminus F(J):\ d_j\le t(J)\},
\]
\[
 J^{\mathrm R}=\{j\in J\setminus F(J):\ r_j\ge t(J)\}.
\]
A job $j\in J$ with $m_j=t(J)$ satisfies $r_j<t(J)<d_j$, so $F(J)\ne\varnothing$, and $J^{\mathrm L}$ and $J^{\mathrm R}$
are proper subsets of $J$.

\begin{definition}[The midpoint tree]\label{def:midpoint-tree}
Let $\J_0$ be a nonempty set of jobs.  The \emph{midpoint tree} on $\J_0$ is the rooted binary tree defined
recursively as follows.  Its root \emph{hosts} $\J_0$ and has \emph{region} $\mathbb R$.  A node $v$ that hosts $J_v$
and has region $\mathcal T_v$ has \emph{separator} $t_v=t(J_v)$ and \emph{crossing set} $F_v=F(J_v)$.  If
$J_v^{\mathrm L}\ne\varnothing$, then $v$ has a left child $v_L$ that hosts $J_v^{\mathrm L}$ and has region
$\mathcal T_v\cap(-\infty,t_v)$.  If $J_v^{\mathrm R}\ne\varnothing$, then $v$ has a right child $v_R$ that hosts
$J_v^{\mathrm R}$ and has region $\mathcal T_v\cap[t_v,\infty)$.  The node $v$ has no other children.  The \emph{depth}
of a node is its distance from the root.
\end{definition}

\noindent Since every child hosts a proper subset of the jobs of its parent, the recursion ends, and the tree is
finite.  We set
\begin{equation}\label{eq:H}
 H:=1+\lceil\log n\rceil,
 \qquad
 N_d:=\left\lceil\frac{n}{2^d}\right\rceil\quad(d\ge0).
\end{equation}

\begin{lemma}[Properties of the midpoint tree]\label{lem:midpoint-tree}
Consider the midpoint tree on a nonempty set $\J_0$ of jobs.
\begin{enumerate}[label=(\alph*),nosep]
\item Every job of $J_v\setminus F_v$ is hosted by exactly one child of $v$, and every child of $v$ hosts at most
$\lfloor|J_v|/2\rfloor$ jobs.
\item A node of depth $d$ hosts at most $\lfloor n/2^d\rfloor\le N_d$ jobs, and every node has depth at most $H-1$.
\item $K_j\subseteq\mathcal T_v$ for every node $v$ and every $j\in J_v$.
\item Distinct nodes of the same depth have disjoint regions.
\item Every job of $\J_0$ lies in the crossing set of exactly one node.
\item If $n\ge2$, then $H\le2\sqrt n$.
\end{enumerate}
\end{lemma}

\begin{proof}
(a) Let $N=|J_v|$.  A job $j\in J_v\setminus F_v$ has $t_v\notin[r_j,d_j)$, so exactly one
of $d_j\le t_v$ and $r_j\ge t_v$ holds, and $j$ is hosted by exactly one child.  A job of the left child has
$m_j<d_j\le t_v$, and at most $\lceil N/2\rceil-1\le\lfloor N/2\rfloor$ midpoints are smaller than $t_v$.  A job of the
right child has $m_j>r_j\ge t_v$, and at most $N-\lceil N/2\rceil=\lfloor N/2\rfloor$ midpoints are larger than
$t_v$.

(b) The root hosts at most $|\J_0|\le n$ jobs.  By (a) and induction, a node of depth $d$ hosts at most
$\lfloor n/2^d\rfloor$ jobs, since $\lfloor\lfloor n/2^d\rfloor/2\rfloor=\lfloor n/2^{d+1}\rfloor$.  A node hosts at
least one job, so $2^d\le n$ and $d\le\lfloor\log n\rfloor\le H-1$.

(c) For the root this is trivial.  If $j$ is hosted by the left child $v_L$ of $v$, then $K_j=[r_j,d_j)\subseteq
(-\infty,t_v)$, since $d_j\le t_v$, and $K_j\subseteq\mathcal T_v$ by induction; so $K_j\subseteq\mathcal T_{v_L}$.  The right
child is symmetric, since $r_j\ge t_v$ gives $K_j\subseteq[t_v,\infty)$.

(d) The two children of a node have disjoint regions, and the regions of the children lie in the region of
their parent.  Thus, by induction on the depth, distinct nodes of the same depth have disjoint regions.

(e) The root hosts every job of $\J_0$.  By (a), a job hosted by $v$ lies in $F_v$ or is hosted by exactly one child
of $v$, which hosts fewer jobs than $v$.  Consequently, the nodes hosting the job form a path from the root, which ends at the
unique node whose crossing set contains it.

(f) For $n=2,3$ this is immediate.  For $n\ge4$, $H\le\log n+2\le2\sqrt n$: the last inequality is equality at
$n=4$, and its right minus left side is increasing for $n\ge4$.
\end{proof}

\begin{definition}[The midpoint tree on $\J_{\mathrm S}$]\label{def:mt-part1}
Assume that $\J_{\mathrm S}\ne\varnothing$.  From now on in \Cref{part:unrelated}, \emph{the midpoint tree} is the midpoint
tree on $\J_{\mathrm S}$, and nodes, depths, separators $t_v$, crossing sets $F_v$ and regions are those of this tree.
\end{definition}

\noindent If $\J_{\mathrm S}=\varnothing$, there is nothing to do in this and the following two sections.  For every node
$v$ of the midpoint tree let
\begin{equation}\label{eq:thick-def}
 L_v^{\min}:=\min_{j\in F_v}L_j,
 \qquad
 D_v:=\{j\in F_v:\Phi_j\le L_v^{\min}/8\}.
\end{equation}
The jobs of $D_v$ are the \emph{thick} jobs at $v$, and the jobs of $F_v\setminus D_v$ the \emph{thin} jobs.

For a node $v$ of the midpoint tree and a job $j\in F_v$, the reach $\rho_j$ and the long side $\eta_j$ are those of \Cref{def:seen-from},
seen from $t_v$; they depend on the node $v$, which is determined by $j$, by \Cref{lem:midpoint-tree}(e).  Write
\[
 \lambda_j^-:=t_v-r_j,
 \qquad
 \lambda_j^+:=d_j-t_v .
\]
Since $r_j\le t_v<d_j$, these are the lengths of $K_j\cap(-\infty,t_v)$ and $K_j\cap[t_v,\infty)$, and
$\lambda_j^-+\lambda_j^+=L_j$.  The points of $K_j$ on side $+$ have distances in $(0,\lambda_j^+)$, and those on side $-$
have distances in $(0,\lambda_j^-]$.  Hence $\eta_j=+$ exactly when $\lambda_j^+\ge\lambda_j^-$, and
\begin{equation}\label{eq:reach}
 \rho_j=\max\{\lambda_j^-,\lambda_j^+\}=\lambda_j^{\eta_j}\ge\frac{L_j}2 .
\end{equation}
For $r>0$ let
\[
 I_{v,+}(r):=[t_v,t_v+r),
 \qquad
 I_{v,-}(r):=[t_v-r,t_v).
\]
If $0<r\le\rho_j$, then $I_{v,\eta_j}(r)\subseteq K_j$: for $\eta_j=+$, $[t_v,t_v+r)\subseteq[t_v,d_j)\subseteq K_j$, since
$r\le d_j-t_v$ and $r_j\le t_v$; for $\eta_j=-$, $[t_v-r,t_v)\subseteq[r_j,t_v)\subseteq K_j$, since $r\le t_v-r_j$ and
$t_v<d_j$.

\begin{theorem}[Thick jobs]\label{thm:thick-direct}
Consider the midpoint tree on $\J_{\mathrm S}$.  One can compute in polynomial time a nonmigratory schedule of weight at
least
\[
 \frac{3}{8H}\sum_v\sum_{j\in D_v}w_jz_j .
\]
\end{theorem}

\begin{proof}
Fix a node $v$ of the midpoint tree and a side $\pm$, and let $D_v^\pm$ be the set of thick jobs at $v$ with long side $\pm$.  Every
$j\in D_v^\pm$ has $\rho_j\ge L_j/2\ge L_v^{\min}/2$, so $X=I_{v,\pm}(L_v^{\min}/2)$ lies in $K_j$.  Moreover
$\Phi_j\le L_v^{\min}/8=|X|/4$.  Hence \Cref{lem:small-witness} with $\rho=3/4$ gives a nonmigratory schedule of
weight at least $\frac34\sum_{j\in D_v^\pm}w_jz_j$, whose pieces lie in the windows of jobs of $J_v$.

Now fix a depth $d$ and a side $\pm$, and take the union of these schedules over all nodes $v$ of depth $d$.  Their
pieces lie in the windows of the jobs hosted by these nodes, hence in their regions, by \Cref{lem:midpoint-tree}(c),
which are pairwise disjoint, by \Cref{lem:midpoint-tree}(d).  Hence pieces from different nodes have disjoint
intervals, and the union is a nonmigratory schedule.  There are at most $H$ depths, by \Cref{lem:midpoint-tree}(b), and two sides, so these at most $2H$ schedules have total
weight at least $\frac34\sum_v\sum_{j\in D_v}w_jz_j$.  The heaviest of them has weight at least this total divided by
$2H$.
\end{proof}

%=====================================================================
\section{Usable distances and thin jobs}\label{sec:thin}
%=====================================================================

This section associates a usable-distance interval with every small-covered job.  At every distance $r$ in that
interval, the interval of length $r$ next to the separator on the long side of the job lies in its window and is at
least twice its deficit.  We then specialize to thin jobs.  For them,
pairwise disjoint usable-distance intervals force geometric growth; the resulting ladders are scheduled with reserve
machines and a four-coloring.

Fix a node $v$ of the midpoint tree and a job $j\in F_v$.  Its \emph{usable-distance interval} is
\begin{equation}\label{eq:usable-distance}
 U_j:=[2\Phi_j,\rho_j].
\end{equation}
The interval has positive
length, because \Cref{lem:sc-deficit} and \eqref{eq:reach} give $2\Phi_j<L_j/16<\rho_j$.

If $r\in U_j$ and $r>0$, then $I_{v,\eta_j}(r)\subseteq K_j$ and $\Phi_j\le r/2$.  The distance $r$ is measured from the
separator $t_v$ of the node of $j$, on the long side of $j$.  Consequently, if several jobs of $F_v$ with the same long
side have a common positive point $r$ in their usable-distance intervals, then the \emph{same} set
$I_{v,\pm}(r)$ lies in every one of their windows and is long enough compared with every deficit: it is a common
witness.

\begin{lemma}[Common usable distance]\label{lem:common-distance}
Let $v$ be a node of the midpoint tree, let $G\subseteq F_v$ be a set of jobs with a common long side $\pm$, and let $r>0$
lie in $U_j$ for every $j\in G$.
Then one can compute in polynomial time a set $I\subseteq G$ with
\begin{equation}\label{eq:common-distance-weight}
 \sum_{j\in I}w_j\ \ge\ \frac12\sum_{j\in G}w_jz_j
\end{equation}
and a nonmigratory schedule that completes every job of $I$ on a machine $i$ with $(i,j)\in\ES$ and $x_{ij}>0$, and
whose pieces lie in the windows of the jobs of $I$.
\end{lemma}

\begin{proof}
The set $X=I_{v,\pm}(r)$ lies in every $K_j$, $j\in G$, and $\Phi_j\le r/2=|X|/2$.  Therefore \Cref{lem:small-witness},
applied with $\rho=1/2$, gives the claim.
\end{proof}

A thin job $j\in F_v\setminus D_v$ satisfies, by \eqref{eq:thick-def} and \Cref{lem:sc-deficit},
\begin{equation}\label{eq:thin-deficit-range}
 \frac{L_v^{\min}}{8}<\Phi_j<\frac{L_j}{32}.
\end{equation}

\subsection{Reserve machines}

The next observation applies to every job with $z_j>0$, although we need it only for thin jobs.  It supplies an
LP-supported machine on which the job has enough slack, using all eligible pairs rather than only the
small-slack pairs.

\begin{lemma}[Reserve machine]\label{lem:slack-machine}
Every job $j$ with $z_j>0$ has an eligible machine $i$ with $x_{ij}>0$ and
\begin{equation}\label{eq:slack-machine}
 s_{ij}=L_j-p_{ij}\ge\frac{\Phi_j}{64}.
\end{equation}
\end{lemma}

\begin{proof}
On a big-slack pair, $s_{ij}\ge L_j/64$.  Therefore
\begin{align*}
 \sum_i s_{ij}x_{ij}
 &\ge \frac{L_j}{64}z_j^{\mathrm B}
      +\sum_{i:(i,j)\in\ES}s_{ij}x_{ij}\\
 &\ge \frac1{64}\Bigl(
       L_jz_j^{\mathrm B}
       +\sum_{i:(i,j)\in\ES}s_{ij}x_{ij}\Bigr)
 =\frac{z_j\Phi_j}{64},
\end{align*}
where the last equality is \Cref{lem:deficit}.  Since $\sum_i x_{ij}=z_j>0$, the weighted average of $s_{ij}$ over the
pairs with $x_{ij}>0$, with weights $x_{ij}$, is at least $\Phi_j/64$, and one such pair attains at least the
average.
\end{proof}

\noindent For every job $j$ with $z_j>0$ we fix such a machine and call it the \emph{reserve machine} $i(j)$ of $j$.  It
is chosen among \emph{all} LP-supported pairs, including big-slack pairs; the conclusion of \Cref{lem:slack-machine}
does not in general hold when the choice is restricted to $\ES$.  For example, if $0<z_j^{\mathrm B}<z_j/64$ and every
small-slack pair of $j$ has slack $0$, then $\Phi_j=L_jz_j^{\mathrm B}/z_j>0$ by \Cref{lem:deficit}, while every
small-slack pair has slack $0<\Phi_j/64$.

\subsection{Four colors for ladders}

We first record the classical feasibility test for jobs that have been assigned to one machine.  We then show that
the jobs of a ladder can be split into four sets, each of which can be scheduled on the reserve machines of its jobs.  A schedule \emph{on machine $i$} is a schedule all
of whose pieces have machine $i$.

The test follows from the optimality of the earliest-due-date algorithm for preemptive scheduling with release dates on
one machine \cite{Ho}.  For the interested reader, we include a proof in \Cref{app:one-machine}.

\begin{lemma}[Feasibility on one machine]\label{lem:processor-demand}
Let $i$ be a machine and $\J'$ a finite set of jobs with $i\in\Gamma(j)$ for every $j\in \J'$.  There is a preemptive
schedule on machine $i$ that completes every job of $\J'$ if and only if, for every interval $I$,
\begin{equation}\label{eq:processor-demand}
 \sum_{j\in \J':\,K_j\subseteq I}p_{ij}\le |I|.
\end{equation}
If so, the schedule can be chosen with every piece inside the window of its job, and the earliest-due-date algorithm (EDD)
computes such a schedule in polynomial time.
\end{lemma}

\begin{lemma}[Four-color feasibility for eightfold ladders]\label{lem:four-color-ladder}
Let $i$ be a machine, and let $j_1,\ldots,j_q$ be jobs with $i\in\Gamma(j_r)$ for every $r$, such that
\begin{align}
 L_{j_{r+1}}&>8L_{j_r} &&(r<q), \label{eq:eight-growth}\\
 L_{j_r}-p_{ij_r}&>\frac{L_{j_{r-1}}}{256} &&(r\ge2). \label{eq:ladder-slack}
\end{align}
Color the jobs cyclically with four colors according to their index.  Then for each color there is a schedule on
machine $i$ that completes every job of that color, with every piece inside the window of its job.
\end{lemma}

\noindent Besides eligibility on $i$ and the displayed inequalities, the proof uses only the inequalities
$p_{ij_r}\le L_{j_r}$ of \Cref{asm:valid}; it requires neither LP support nor membership in either pair class.

\begin{proof}
Fix one color and an interval $I$.  By \Cref{lem:processor-demand}, the only jobs relevant to $I$ are the jobs of
the chosen color whose \emph{entire windows} are contained in $I$.  Set
\[
 \mathcal C(I):=\{j\in\{j_1,\ldots,j_q\}:\ K_j\subseteq I\text{ and }j\text{ has the chosen color}\}.
\]
If $\mathcal C(I)=\varnothing$, its demand is zero.  Otherwise let $j_r$ be the member of $\mathcal C(I)$ with
largest index.  Every other member is an
earlier job of the same color, and hence has index among $r-4,r-8,\ldots$.

Therefore the total processing of all members of $\mathcal C(I)$ other than $j_r$ is at most the total window
length of \emph{all} earlier jobs of that color, whether or not their windows are contained in $I$.  By
\eqref{eq:eight-growth}, this is at most
\[
 L_{j_{r-4}}\sum_{h\ge0}8^{-4h}
 \le \frac{L_{j_{r-1}}}{8^3(1-8^{-4})}
 <\frac{L_{j_{r-1}}}{256},
\]
whenever $\mathcal C(I)$ has a member other than $j_r$.  In that case \eqref{eq:ladder-slack} gives
\[
 p_{ij_r}<L_{j_r}-\frac{L_{j_{r-1}}}{256},
\]
so the total demand of $\mathcal C(I)$ is less than $L_{j_r}$.  If $j_r$ is the only member of $\mathcal C(I)$, its
demand is $p_{ij_r}\le L_{j_r}$.  Finally, $K_{j_r}\subseteq I$ implies $L_{j_r}\le |I|$.  Thus in all cases
\[
 \sum_{j\in\mathcal C(I)}p_{ij}\le |I|.
\]
Since $I$ was arbitrary, \Cref{lem:processor-demand} gives the schedule for the color class.
\end{proof}

\noindent A \emph{ladder} at a node $v$ of the midpoint tree is a set of thin jobs at $v$ with a common long side whose
usable-distance intervals $U_j$ are pairwise disjoint.

\begin{lemma}[Ladder schedules]\label{lem:separated-four}
Let $G$ be a ladder at a node $v$ of the midpoint tree.  Then $G$ can be partitioned into four sets, each completed by a
nonmigratory schedule in which every job runs on its reserve machine, and every piece lies in the window of its job.
The partition and the four schedules can be computed in polynomial time.
\end{lemma}

\noindent The reserve machines may belong to pairs of $\EB$ or of $\ES$.

\begin{proof}
Let $U_{j_1},U_{j_2},\ldots$ be the intervals in order from left to right.  Since they are closed and pairwise disjoint,
\[
 \rho_{j_r}<2\Phi_{j_{r+1}}.
\]
Using \eqref{eq:sc-deficit} and $\rho_{j_r}\ge L_{j_r}/2$ gives
\[
 L_{j_{r+1}}>32\Phi_{j_{r+1}}>16\rho_{j_r}\ge8L_{j_r}.
\]
The reserve machine of $j_{r+1}$ satisfies
\[
 L_{j_{r+1}}-p_{i(j_{r+1})j_{r+1}}
 \ge\frac{\Phi_{j_{r+1}}}{64}
 >\frac{\rho_{j_r}}{128}
 \ge\frac{L_{j_r}}{256}.
\]

Every job is assigned to its reserve machine.  On a fixed machine, consider the subsequence of the jobs assigned to
it, in the order inherited from the left-to-right order.  Consecutive jobs of this subsequence still grow by a factor
greater than eight.  Moreover, if $j_{r'}$ precedes $j_r$ in the subsequence, then $r'\le r-1$ and
$L_{j_{r-1}}\ge L_{j_{r'}}$, so the slack of $j_r$ on its machine is greater than $L_{j_{r'}}/256$.  On every
machine, color its jobs cyclically along its subsequence with the colors $1,2,3,4$, and apply
\Cref{lem:four-color-ladder}.  For each color $c$, the union over all machines of the schedules of color $c$ is a
nonmigratory schedule: pieces on different machines never conflict, and all pieces of a job are on its reserve
machine.  The four colors partition $G$.  The reserve machines, the coloring and the schedules, by
\Cref{lem:processor-demand}, are computed in polynomial time.
\end{proof}

\subsection{The common-point removal and coloring stages}

The groups and ladders of thin jobs are formed from their usable-distance intervals.  For a node $v$ of the midpoint
tree, the window of every job of $F_v$ contains the separator $t_v$, but this alone gives no common witness of positive
length on which a thin job has small enough deficit.  A common point $r>0$ of the intervals $U_j$ does:
$I_{v,\pm}(r)$ is then a common witness, and $\Phi_j\le r/2$ for every job of the group.  If those $U_j$ were already pairwise disjoint, \Cref{lem:separated-four} would
apply to all of them directly.

\begin{lemma}[Interval partition]\label{lem:interval-partition}
Let $\mathcal A$ be a family of at most $N$ closed intervals on a line, and let $k\ge1$.  It can be partitioned into
\begin{enumerate}[label=(\roman*),leftmargin=2em]
 \item at most $\lceil N/k\rceil$ parts whose intervals have a common point, and
 \item at most $k$ parts whose intervals are pairwise disjoint.
\end{enumerate}
The partition can be found in polynomial time.
\end{lemma}

\begin{proof}
In the \emph{common-point removal stage}, while some point belongs to at least $k$ remaining intervals, remove all
remaining intervals containing that point and make them one part of type~(i).  Every such part has at least $k$
intervals, so there are at most $\lfloor N/k\rfloor$ of them.

If $k=1$, the common-point removal stage leaves no interval.  Assume $k\ge2$.  In the remaining family no point
belongs to $k$ intervals.  In the \emph{coloring stage}, sort these intervals by left endpoint and color them
greedily, always using a color not carried by an earlier interval that contains the current left endpoint.  At
most $k-2$ earlier intervals contain it, so $k-1$ colors suffice.  Each color class is a part of type~(ii).  The
common points and the coloring are found by sorting interval endpoints.
\end{proof}

\noindent Larger $k$ gives fewer parts of type~(i) and more parts of type~(ii); \Cref{sec:sqrt} chooses $k$ of order
$\sqrt N$ to balance the two.

%=====================================================================
\section{Square-root aggregation for thin jobs}\label{sec:sqrt}
%=====================================================================

This section combines the two mechanisms for thin jobs.  At one depth, the common-point removal stage and the
coloring stage, with $k=\lceil\sqrt N\rceil$, create only $O(\sqrt N)$ schedule candidates.  Since nodes of the same
depth have disjoint regions, the schedules computed at these nodes combine into schedules for the whole depth, and
summing their number over the depths gives $O(\sqrt n)$.

By \Cref{lem:midpoint-tree}(b), a node of depth $d$ of the midpoint tree hosts at most $N_d$
jobs.  Set
\[
 k_d:=\left\lceil\sqrt{N_d}\right\rceil.
\]

\begin{theorem}[Thin jobs]\label{thm:thin-candidates}
Consider the midpoint tree on $\J_{\mathrm S}$, and let $n\ge2$.  One can compute in polynomial time a nonmigratory
schedule of weight at least
\[
 \frac{1}{240\sqrt n}\sum_v\sum_{j\in F_v\setminus D_v}w_jz_j .
\]
\end{theorem}

\begin{proof}
We construct a set $\mathcal S_{\mathrm{thin}}$ of nonmigratory schedules, which we call \emph{candidates}, separately
for every depth $d$ and every side $\pm$, and output the heaviest of them.

\noindent\emph{Groups and ladders at one node.}  Fix a depth $d$ and a side $\pm$.  At a node $v$ of depth $d$, the thin jobs
with long side $\pm$ have at most $N_d$ usable-distance intervals, by \Cref{lem:midpoint-tree}(b).
\Cref{lem:interval-partition} with $k=k_d$ partitions these intervals into at most $\lceil N_d/k_d\rceil$ parts whose
intervals have a common point, the \emph{groups}, and at most $k_d$ parts whose intervals are pairwise disjoint, the
\emph{ladders}.  Every common point $r$ of a group is positive: for every member $j$, $r\ge2\Phi_j>L_v^{\min}/4>0$, by
\eqref{eq:thin-deficit-range}.  Number the groups and the ladders at each node.

\noindent\emph{Candidates of one depth.}  Distinct nodes of depth $d$ have disjoint regions, and the windows of the jobs at a
node lie in its region, by \Cref{lem:midpoint-tree}(c),(d).  Hence a union of schedules, one for each node of depth
$d$, whose pieces lie in the windows of the jobs of that node, is a nonmigratory schedule.  For a group index, take
at every node the group with that index, if it exists; \Cref{lem:common-distance} gives for it a schedule of weight
at least half of its LP weight whose pieces lie in the windows of its jobs.  The union of these schedules over the
nodes is one candidate.  For a ladder index, \Cref{lem:separated-four} partitions the ladder with that index at every
node into four sets, each completed by a schedule whose pieces lie in the windows of its jobs.  Number the four sets
at each node; for each of the four numbers, the union over the nodes is one candidate.  Thus depth $d$ and side
$\pm$ give at most $\lceil N_d/k_d\rceil+4k_d$ candidates.  Every thin job at a node of depth $d$ with long side $\pm$
lies in exactly one group or ladder.  The candidates of groups weigh at least half of the LP weight of the groups,
and the four candidates of a ladder index complete all jobs of these ladders, so their weight is at least the LP
weight of the ladders.  The candidates of depth $d$ and side $\pm$ therefore weigh at least half of the LP weight of
these thin jobs.

\noindent\emph{Counting.}  Since $k_d\le\sqrt{N_d}+1\le2\sqrt{N_d}$ and $\lceil N_d/k_d\rceil\le\sqrt{N_d}+1\le2\sqrt{N_d}$,
depth $d$ gives, over both sides, at most $2(\lceil N_d/k_d\rceil+4k_d)\le20\sqrt{N_d}$ candidates.  Every node has
depth at most $H-1$, by \Cref{lem:midpoint-tree}(b), and every thin job lies in the crossing set of exactly one node,
by \Cref{lem:midpoint-tree}(e).  Summing over depths, $\mathcal S_{\mathrm{thin}}$ has at most
$20\sum_{d=0}^{H-1}\sqrt{N_d}$ schedules, of total weight at least $\frac12\sum_v\sum_{j\in F_v\setminus D_v}w_jz_j$.
Since $\sqrt{\lceil x\rceil}\le\sqrt x+1$ for every $x\ge0$, we have $\sqrt{N_d}\le\sqrt n\,2^{-d/2}+1$, and
$H\le2\sqrt n$ by \Cref{lem:midpoint-tree}(f).  Hence
\[
 \sum_{d=0}^{H-1}\sqrt{N_d}
 \ \le\ \sqrt n\sum_{d\ge0}2^{-d/2}+H
 \ \le\ \Bigl(\frac1{1-2^{-1/2}}+2\Bigr)\sqrt n
 \ =\ (4+\sqrt2)\sqrt n\ <\ 6\sqrt n ,
\]
and $\mathcal S_{\mathrm{thin}}$ has fewer than $120\sqrt n$ schedules.  Finally, the heaviest schedule of a set weighs
at least the total weight of the set divided by the number of its schedules, so the heaviest candidate has weight at
least $\frac1{240\sqrt n}\sum_v\sum_{j\in F_v\setminus D_v}w_jz_j$.  The construction uses sorting, \Cref{lem:common-distance} and
\Cref{lem:separated-four}, so it runs in polynomial time.
\end{proof}

%=====================================================================
\section{Proof of the unrelated-machine theorem}\label{sec:proof-unrelated}
%=====================================================================

This section combines the three classes of jobs.  Big-covered jobs are rounded at a constant-factor loss,
thick jobs at a loss $O(\log n)$, and thin jobs at a loss $O(\sqrt n)$.  Returning the heaviest of the three schedules
proves \Cref{thm:main-unrelated}.  Write
\[
 W_{\mathrm{big}}:=\sum_{j\in\J_{\mathrm B}}w_jz_j,
 \qquad
 W_{\mathrm{thick}}:=\sum_v\sum_{j\in D_v}w_jz_j,
 \qquad
 W_{\mathrm{thin}}:=\sum_v\sum_{j\in F_v\setminus D_v}w_jz_j,
\]
where $v$ ranges over the nodes of the midpoint tree on $\J_{\mathrm S}$; if $\J_{\mathrm S}=\varnothing$, then
$W_{\mathrm{thick}}=W_{\mathrm{thin}}=0$.  Every job with $z_j>0$ lies in $\J_{\mathrm B}$ or in $\J_{\mathrm S}$, and every job of $\J_{\mathrm S}$ lies in exactly one crossing set, by \Cref{lem:midpoint-tree}(e).  Hence
\begin{equation}\label{eq:three-way}
 W_{\mathrm{big}}+W_{\mathrm{thick}}+W_{\mathrm{thin}}=\sum_jw_jz_j=W .
\end{equation}

\begin{proof}[Proof of \Cref{thm:main-unrelated}]
If $n=1$, run the single job $j$ on any eligible machine $i$ during the earliest $p_{ij}$ units of $K_j$; this
schedule has weight $w_j\ge w_jz_j=W$.  Assume $n\ge2$.

Compute the schedule $A_{\mathrm{big}}$ of \Cref{thm:big-rounding}, with $\delta=1/64$ and $\J'=\J_{\mathrm B}$, and, if
$\J_{\mathrm S}\ne\varnothing$, the schedules $A_{\mathrm{thick}}$ and $A_{\mathrm{thin}}$ of
\Cref{thm:thick-direct,thm:thin-candidates}; if $\J_{\mathrm B}$ or $\J_{\mathrm S}$ is empty, use the empty schedule
instead.  Output the heaviest of these three schedules.

For a big-covered job, $z_j^{\mathrm B}\ge z_j/64$, so
\[
 W(A_{\mathrm{big}})\ \ge\ \frac1{768}\sum_{j\in\J_{\mathrm B}}w_jz_j^{\mathrm B}\ \ge\ \frac{W_{\mathrm{big}}}{49152}.
\]
By \Cref{thm:thick-direct} and \Cref{lem:midpoint-tree}(f),
\[
 W(A_{\mathrm{thick}})\ \ge\ \frac{3W_{\mathrm{thick}}}{8H}
 \ \ge\ \frac{3W_{\mathrm{thick}}}{16\sqrt n},
\]
and by \Cref{thm:thin-candidates},
\[
 W(A_{\mathrm{thin}})\ \ge\ \frac{W_{\mathrm{thin}}}{240\sqrt n}.
\]
Since $\max\{a/\alpha,b/\beta,c/\gamma\}\ge(a+b+c)/(\alpha+\beta+\gamma)$ for nonnegative $a,b,c$ and positive
$\alpha,\beta,\gamma$, by \eqref{eq:three-way} the output has weight at least
\[
 \frac{W_{\mathrm{big}}+W_{\mathrm{thick}}+W_{\mathrm{thin}}}{49152+\frac{16}{3}\sqrt n+240\sqrt n}
 \ \ge\ \frac{W}{246\sqrt n+49152}.
\]

\noindent\emph{Running time.}  The numbers $y_{ij}(X)$, $x_{ij}$, $z_j^{\mathrm B}$ and $\Phi_j$ are finite sums of rationals
computed from the LP solution.  The schedule $A_{\mathrm{big}}$ is computed in deterministic polynomial time, by \Cref{thm:big-rounding}.  The midpoint tree is built by repeated median selection, and it has at most $n$ nodes,
since their crossing sets are nonempty and pairwise disjoint, by \Cref{lem:midpoint-tree}(a),(e).  Every application of \Cref{lem:small-witness} is a maximum-weight bipartite matching
computation.  Reserve machines are found by scanning the pairs with $x_{ij}>0$.  The interval partition is found by
sorting endpoints, and the ladder schedules by coloring and EDD on each machine.  The schedules $A_{\mathrm{thick}}$ and
$A_{\mathrm{thin}}$ are computed in polynomial time, by \Cref{thm:thick-direct,thm:thin-candidates}.  Therefore the algorithm runs in deterministic polynomial time.
\end{proof}

\begin{proof}[Proof of \Cref{cor:unrelated}]
The $\StrongLP$ has $O(n^2|\M|)$ variables and constraints, with rational coefficients of polynomial encoding length,
so an optimal solution can be computed in polynomial time.  For it, $W=\OPTLP$, and the schedule of
\Cref{thm:main-unrelated} is nonmigratory, so $\OPTLP\le(246\sqrt n+49152)\,W(A)\le(246\sqrt n+49152)\,\OPTnm$.  By
\Cref{lem:strong-valid}, $\OPT\le\OPTLP$, so the same schedule has weight at least $\OPT/(246\sqrt n+49152)$.
\end{proof}

\begin{remark}[Which pairs the three branches use]\label{rem:pair-requirements}
The three schedules use different kinds of pairs.  The big-covered rounding uses only LP-supported
big-slack pairs.  Thick jobs and groups of thin jobs use only small-slack pairs with positive LP processing on
their common witness.  Ladders use reserve machines from \emph{either} pair class.  In particular, small-covered jobs are not
restricted to small-slack pairs in the final schedule, since a small-covered job need not have a reserve machine on a
small-slack pair.
\end{remark}

%=====================================================================
\section{A matching \texorpdfstring{$\Omega(\sqrt n)$}{Omega(sqrt(n))} lower bound}\label{sec:tightness}
%=====================================================================

The $O(\sqrt n)$ dependence in \Cref{thm:main-unrelated} is asymptotically unavoidable if the output is required
to be nonmigratory.  The instance that shows this has two families of machines, of sizes $q$ and $T$, and every job is
eligible on exactly one machine of each family.

\begin{definition}[The crossbar instance]\label{def:crossbar}
Let $q,T\ge1$ be integers, and set
\begin{equation}\label{eq:tight-lengths}
 \beta:=2q,
 \qquad
 L_k:=\beta^{k-1}\quad(k=1,\ldots,T).
\end{equation}
The \emph{$(q,T)$-crossbar instance} has $q$ \emph{vertical} machines $a_1,\ldots,a_q$, one \emph{level} machine $b_k$ for each
level $k=1,\ldots,T$, and one job $j_{k,\ell}$ for every level $k$ and every $\ell=1,\ldots,q$.  All weights are one.
Job $j_{k,\ell}$ has the window
\[
 K_k=[-L_k/2,L_k/2),
\]
it is eligible precisely on $a_\ell$ and $b_k$, and its processing times are
\begin{equation}\label{eq:tight-processing}
 p_{a_\ell,j_{k,\ell}}=L_k,
 \qquad
 p_{b_k,j_{k,\ell}}=L_k/2.
\end{equation}
\end{definition}

\noindent The jobs and their eligible machines form the following pattern, with $qT$ jobs but only $q+T$ machines:
\[
\begin{array}{c|cccc}
 & a_1 & a_2 & \cdots & a_q \\ \hline
 b_1 & j_{1,1} & j_{1,2} & \cdots & j_{1,q}\\
 b_2 & j_{2,1} & j_{2,2} & \cdots & j_{2,q}\\
 \vdots & \vdots & \vdots & \ddots & \vdots\\
 b_T & j_{T,1} & j_{T,2} & \cdots & j_{T,q}
\end{array}
\]
All $q$ jobs of level $k$ have the same centered window $K_k$, so $K_1\subset K_2\subset\cdots\subset K_T$.  These
windows have negative release dates; we center them at $0$ only for convenience.  Shifting all times by $L_T/2$ makes
every release date nonnegative and changes no schedule weight.  Both processing times of a job of level $k$ are at most
$L_k$, so the instance satisfies \Cref{asm:valid}.

\begin{theorem}[A matching $\Omega(\sqrt n)$ lower bound]\label{thm:sqrt-tightness}
For every perfect square $n$, the $(\sqrt n,\sqrt n)$-crossbar instance has $n$ jobs and $m=2\sqrt n$ machines, and
\[
 \OPTLP\ =\ \OPT\ \ge\ \frac{\sqrt n}3\,\OPTnm\ =\ \frac m6\,\OPTnm .
\]
\end{theorem}

\begin{proof}
Let $s=\sqrt n$ and $q=T=s$, so that the instance has $qT=n$ jobs and $q+T=2s=m$ machines.

\noindent\emph{A migratory schedule completing every job.}
The nested windows define pairwise disjoint \emph{shells}
\begin{equation}\label{eq:tight-shells}
 S_1:=K_1,
 \qquad
 S_k:=K_k\setminus K_{k-1}\quad(k\ge2),
\end{equation}
each a union of at most two half-open intervals.  Their lengths are
\begin{equation}\label{eq:tight-shell-length}
 |S_1|=L_1,
 \qquad
 |S_k|=L_k-L_{k-1}=L_k(1-1/\beta)\quad(k\ge2).
\end{equation}
Every level-$k$ job has all of $S_k$ in its window.  For each job $j_{k,\ell}$ we prescribe, entirely during $S_k$,
\begin{align}
 A_k&:=L_k(1-1/q)
 &&\text{units on its vertical machine }a_\ell, \label{eq:tight-work-a}\\
 B_k&:=\frac{L_k}{2q}
 &&\text{units on the shared level machine }b_k. \label{eq:tight-work-b}
\end{align}
It is important to distinguish the load per shell from the total load of a vertical machine.  Fix a level $k$.
Among the level-$k$ jobs, machine $a_\ell$ is eligible for exactly one job, $j_{k,\ell}$, so its load during $S_k$ is
$A_k$, not $qA_k$.  In contrast, all $q$ level-$k$ jobs use the single machine $b_k$, whose load during $S_k$ is
$qB_k$.  For $k\ge2$ we have
\begin{align}
 A_k&=L_k(1-1/q)\le L_k(1-1/\beta)=|S_k|, \label{eq:tight-a-load}\\
 qB_k&=L_k/2\le |S_k|, \label{eq:tight-b-load}\\
 A_k+B_k&=L_k(1-1/(2q))\le |S_k|, \label{eq:tight-job-load}
\end{align}
where the first inequality uses $\beta\ge q$, the second $\beta\ge2$, and the third $\beta\ge2q$.  For $k=1$ the same
three inequalities hold with $|S_1|=L_1$.

We realize these loads by explicit pieces in the shell $S_k$.  Let $Y_{k,1}$ be the earliest $B_k$ units of $S_k$,
and for $\ell=2,\dots,q$ let $Y_{k,\ell}$ be the earliest $\ell B_k$ units of $S_k$ minus the earliest $(\ell-1)B_k$ units; these are pairwise disjoint finite
unions of half-open intervals in $S_k$, with $|Y_{k,\ell}|=B_k$, and they exist because $qB_k\le|S_k|$ by
\eqref{eq:tight-b-load}.  Run $j_{k,\ell}$ on $b_k$ during $Y_{k,\ell}$, and on $a_\ell$ during the earliest $A_k$ units of
$S_k\setminus Y_{k,\ell}$; these exist because $|S_k\setminus Y_{k,\ell}|=|S_k|-B_k\ge A_k$ by \eqref{eq:tight-job-load}.
Different jobs use different vertical machines, so their vertical pieces may overlap in time.

These pieces satisfy \Cref{def:schedule} at level $k$.  Pieces on $b_k$ have disjoint intervals because the
$Y_{k,\ell}$ are disjoint.  Each vertical machine $a_\ell$ processes only $j_{k,\ell}$ during $S_k$.  The pieces of
$j_{k,\ell}$ lie in $Y_{k,\ell}$ and in its complement in $S_k$, so they have disjoint intervals: the job never runs on
two machines at once.

The pieces of different levels lie in the pairwise disjoint shells, so all pieces together form a schedule.  The
\emph{total} load of a vertical machine $a_\ell$ is $\sum_{k=1}^T A_k$, but its level-$k$ portion lies in $S_k$; indeed
\begin{equation}\label{eq:tight-total-a-load}
 \sum_{k=1}^T A_k
 \le \sum_{k=1}^T |S_k|
 =|K_T|=L_T.
\end{equation}
Each level machine $b_k$ is used only by its own level.

Every job $j_{k,\ell}$ satisfies the completion condition of \Cref{def:schedule}:
\[
 \frac{A_k}{L_k}+\frac{B_k}{L_k/2}
 =1-\frac1q+\frac1q=1.
\]
Thus all $qT$ jobs are completed by a migratory schedule, so $\OPT=qT$.  By \Cref{lem:strong-valid}, $\OPTLP\ge qT$,
and no $\StrongLP$ solution has value larger than the total weight $qT$, since $z_j\le1$.  Hence $\OPTLP=\OPT=qT$.

\noindent\emph{Why a nonmigratory schedule completes at most $q+2T$ jobs.}
In a nonmigratory schedule, the pieces of the jobs completed on a machine $i$ form a schedule on $i$ that completes
them, so \Cref{lem:processor-demand} applies to them.  Fix a vertical machine $a_\ell$.  Its eligible jobs are $j_{1,\ell},\ldots,j_{T,\ell}$.  Their windows are nested,
and job $j_{k,\ell}$ requires $p_{a_\ell,j_{k,\ell}}=L_k=|K_k|$, its entire window.  If $k<k'$, then $K_k\subset
K_{k'}$, while the two processing times sum to
\[
 L_k+L_{k'}>L_{k'}=|K_{k'}|.
\]
By \Cref{lem:processor-demand}, the two jobs cannot both be completed on $a_\ell$.  Hence every vertical machine
completes at most one job in a nonmigratory schedule.

Fix instead a level machine $b_k$.  Its $q$ eligible jobs all have the window $K_k$ and each requires $L_k/2$ on
$b_k$.  By \Cref{lem:processor-demand}, at most two of them can be completed on $b_k$.  Since every completed job of
a nonmigratory schedule is completed on one machine,
\[
 \OPTnm\le q\cdot1+T\cdot2=q+2T.
\]
With $q=T=s$, this gives $\OPTnm\le3s$, while $\OPTLP=\OPT=s^2$.  Hence $\OPT\ge(s/3)\,\OPTnm$, and
$s/3=\sqrt n/3=m/6$.
\end{proof}

%=====================================================================
\clearpage
\part{Restricted assignment}\label{part:ra}
%=====================================================================

%=====================================================================
\section{Preliminaries for restricted assignment}\label{sec:ra-prelims}
%=====================================================================

This part proves \Cref{thm:main-ra}.  The jobs are split by relative slack into the three classes of
\Cref{def:ra-classes}, which are rounded separately.  Big-slack jobs have only big-slack pairs and are rounded by
\Cref{thm:big-rounding}.  Moderate-slack jobs are rounded through witness systems in \Cref{sec:moderate}.  Nearly rigid
jobs are rounded by a recursion with a dynamic program at every node in \Cref{sec:near}.  The algorithms for the
last two classes rest on \Cref{sec:ws}: a set of jobs whose windows contain a common witness, longer than their slacks by a constant factor, has a
heavy subset that runs on distinct machines.  \Cref{sec:assembly} combines the three schedules, and
\Cref{sec:restricted-log-gap} shows that the logarithmic factor cannot be avoided.

Throughout \Cref{part:ra} the instance is a restricted-assignment instance
 with $n\ge2$ jobs, and we fix a $\StrongLP$ solution $(y,z)$ of value $W$.  We use the notions of
\Cref{sec:big-covered} with
\[
 \delta=\frac18 .
\]
Under restricted assignment the coverage constraint reads $\sum_i\sum_ky_{ijk}=p_jz_j$, and $p_{ij}x_{ij}=\sum_ky_{ijk}$
gives $x_{ij}=\frac1{p_j}\sum_ky_{ijk}$.

\begin{definition}[Parameters and job classes]\label{def:ra-classes}
Set
\[
 \ell=\lfloor\log n\rfloor+1,\qquad K'=4(\ell+1),\qquad Q=4^{K'},\qquad \sigma_0=\frac{1}{8Q}.
\]
A job $j$ is \emph{big-slack} if $\sigma_j\ge1/8$, \emph{moderate-slack} if $\sigma_0\le\sigma_j<1/8$, and \emph{nearly
rigid} if $\sigma_j<\sigma_0$.  Let $\J_{\mathrm{big}}$, $\J_{\mathrm{mod}}$ and $\J_{\mathrm{nr}}$ be the sets of these jobs,
and let $W_{\mathrm{big}}$, $W_{\mathrm{mod}}$ and $W_{\mathrm{nr}}$ be their LP weights $\sum_jw_jz_j$, so that
$W=W_{\mathrm{big}}+W_{\mathrm{mod}}+W_{\mathrm{nr}}$.
\end{definition}

\noindent \Cref{sec:params} explains the choice of $\ell$, $K'$, $Q$ and $\sigma_0$.

\begin{lemma}[Pairs of a job]\label{lem:ra-split}
Let $j$ be a job.  If $\sigma_j\ge\delta$, then every eligible pair of $j$ is big-slack and $z_j^{\mathrm B}=z_j$.  If
$\sigma_j<\delta$, then every eligible pair of $j$ is small-slack and $z_j^{\mathrm B}=0$.
\end{lemma}

\begin{proof}
A pair $(i,j)$ is big-slack exactly when $p_{ij}\le(1-\delta)L_j$.  Since $p_{ij}=p_j$, this means $s_j\ge\delta L_j$,
i.e., $\sigma_j\ge\delta$, which does not depend on $i$.  The values of $z_j^{\mathrm B}$ follow from
\eqref{eq:pair-mass} and \eqref{eq:x-sums-z}.
\end{proof}

\begin{lemma}[Busy time in the window]\label{lem:ra-processing}
For every job $j$ and every finite union $X\subseteq K_j$ of intervals,
\[
 y_j(X)\ \ge\ z_j\bigl(|X|-s_j\bigr).
\]
\end{lemma}

\begin{proof}
We apply \Cref{lem:processing}(d) with $E$ the set of all eligible pairs of $j$.  By \eqref{eq:x-sums-z},
$\sum_ip_{ij}x_{ij}=p_jz_j$, so $y_j(X)\ge p_jz_j-z_j(L_j-|X|)=z_j(|X|-s_j)$.
\end{proof}

\subsection{The transversal matroid}

At any moment, the jobs that a schedule processes run on distinct machines, each eligible for its job.  The sets of
jobs that can be assigned to distinct eligible machines are the independent sets of a transversal matroid.  The
algorithms for moderate-slack and nearly rigid jobs choose such sets with the greedy algorithm of \Cref{lem:greedy}.

\begin{definition}[Transversal matroid]\label{def:matroid}
\begin{enumerate}[label=\arabic*.,nosep]
\item \emph{Eligibility graph.}  The bipartite graph between jobs and machines with an edge $\{j,i\}$ whenever
$i\in\Gamma(j)$.
\item \emph{Independent sets.}  A set $T$ of jobs is \emph{independent} if the eligibility graph has a matching that
covers $T$, that is, if the jobs of $T$ can be assigned to distinct eligible machines.
\item \emph{Transversal matroid.}  The matroid $M$ on $\J$ whose independent sets are the independent sets of~2.
\end{enumerate}
\end{definition}

\noindent The independent sets of~2 form a matroid, the transversal matroid of the eligibility graph.

\begin{lemma}[Greedy algorithm]\label{lem:greedy}
Let $G$ be a set of jobs with weights $w_j\ge0$.  The greedy algorithm, which considers the jobs of $G$ in
nonincreasing order of weight and keeps a job if it is independent together with the jobs kept before, finds an
independent subset of $G$ of maximum weight.  An independence test is a bipartite matching computation.
\end{lemma}

\noindent This is the greedy algorithm for matroids \cite{Ed}, applied to the restriction of $M$ to $G$.

%=====================================================================
\section{Witnesses and supports}\label{sec:ws}
%=====================================================================

This section gives the tool that the algorithms for moderate-slack and nearly rigid jobs use to assign jobs to distinct machines.  A
\emph{witness} for a set of jobs is a common part of their windows that is longer than their slacks by a constant
factor; \Cref{def:witness} makes this precise.  If a set of jobs has a witness, then a subset of it with a
constant fraction of its LP weight can be assigned to distinct eligible machines, by \Cref{lem:group}, and each chosen
job then runs in a \emph{support}, a part of its window long enough for its processing.  Jobs chosen from different
sets may still need the same machine at the same time.  \Cref{sec:moderate} prevents this by putting the sets into
families in which jobs of different groups have disjoint supports, and \Cref{sec:near} by a dynamic program over a recursion tree.  The
results of this section hold for every set of jobs; \Cref{sec:moderate} applies them to $\J_{\mathrm{mod}}$ and
\Cref{sec:near} to $\J_{\mathrm{nr}}$.

\begin{definition}[Witnesses and supports]\label{def:witness}
\begin{enumerate}[label=\arabic*.,nosep]
\item \emph{Witness.}  Let $\varepsilon>0$, and let $G$ be a nonempty set of jobs.  An \emph{$\varepsilon$-witness} for $G$ is a finite union $X$ of intervals with
$|X|>0$ such that $X\subseteq K_j$ and $|X|\ge(1+\varepsilon)s_j$ for every $j\in G$.
\item \emph{Support.}  A \emph{support} of a job $j$ is a finite union $Y_j\subseteq K_j$ of
half-open intervals with $|Y_j|\ge p_j$.
\end{enumerate}
\end{definition}

The jobs of $G$ may have very different lengths, and supports may have holes.  The definition involves only windows and
slacks, not machines or LP values.  Since a support $Y_j$ is a finite union of half-open intervals with $|Y_j|\ge p_j$, its
earliest $p_j$ units are defined.

\begin{lemma}[A group with a common witness]\label{lem:group}
Let $X$ be an $\varepsilon$-witness for a nonempty set $G$ of jobs.  The greedy algorithm, run on
$G$ with the weights $w_j$, finds an independent set $I\subseteq G$ with
\[
 \sum_{j\in I}w_j\ \ge\ \frac{\varepsilon}{1+\varepsilon}\sum_{j\in G}w_jz_j .
\]
If every job $j\in I$ has a support $Y_j$, then matching $I$ to distinct eligible machines
and running each $j\in I$ on its machine during the earliest $p_j$ units of $Y_j$ gives a
nonmigratory schedule that completes every job of $I$.
\end{lemma}

\begin{proof}
Set $\rho=\varepsilon/(1+\varepsilon)\in(0,1)$.  For $j\in G$ we have $X\subseteq K_j$ and $s_j\le|X|/(1+\varepsilon)$, so
$|X|-s_j\ge\rho|X|$, and \Cref{lem:ra-processing} gives $y_j(X)\ge\rho z_j|X|$.  Thus \Cref{lem:common-witness}, with $E$ the
set of all eligible pairs of the jobs of $G$ and $F=G$, gives a set $I'\subseteq G$ of jobs assigned to distinct
eligible machines with $\sum_{j\in I'}w_j\ge\rho\sum_{j\in G}w_jz_j$.  The set $I'$ is independent, so the independent set $I$
of maximum weight found by the greedy algorithm of \Cref{lem:greedy} weighs at least as much.  Since $I$ is independent,
it can be matched to distinct eligible machines.  By \Cref{def:run}, the pieces of each $j\in I$ are on its machine, have
pairwise disjoint intervals, lie in $Y_j\subseteq K_j$, and have total length $p_j$.  Pieces of different jobs of $I$ are on
different machines.  Consequently, these pieces form a nonmigratory schedule that completes every job of $I$.
\end{proof}

The greedy algorithm uses only the weights and the eligibility sets; the LP values $z_j$ enter only the bound.
Supports and witnesses need not follow the distribution of the LP processing over time.  The $\StrongLP$ solution
enters the proof of \Cref{lem:group} only through \Cref{lem:ra-processing}, which holds for every finite union of
intervals $X\subseteq K_j$ regardless of the processing outside $X$, and through \Cref{lem:common-witness}; the schedule
is built from scratch.  Hence a support or a witness may omit parts of $K_j$ on which $j$ receives much LP processing;
the supports of \Cref{sec:near}, for example, omit a hole around a separator point.

%=====================================================================
\section{Moderate slack}\label{sec:moderate}
%=====================================================================

Moderate-slack jobs have relative slack between $\sigma_0$ and $1/8$.  Their slack is small compared with their windows,
so a group of them with a common witness can be rounded by \Cref{lem:group}.  The difficulty is to combine many groups.
The next subsection introduces witness systems, which put the groups into families in which jobs of different
groups never have overlapping supports, and shows how to round from them.  \Cref{sec:mod-construction} constructs a witness system of
logarithmic size for the moderate-slack jobs.

\subsection{Witness systems}

Jobs chosen by \Cref{lem:group} from different groups may be assigned to the same machine.  A witness system rules out such
conflicts by putting the groups into \emph{families} in which jobs of different groups never have overlapping supports.  All
chosen jobs of one family can then run together, and if there are $\phi$ families, the best one has at least a $1/\phi$ fraction
of the LP weight, up to the factor $\varepsilon/(1+\varepsilon)$ of \Cref{lem:group}.  \Cref{def:ws,thm:ws} apply to every set
$\J'$ of jobs; \Cref{sec:mod-construction} applies them to $\J_{\mathrm{mod}}$.

\begin{definition}[Witness system]\label{def:ws}
Fix $\varepsilon>0$ and a subset $\J'\subseteq\J$.  An \emph{$\varepsilon$-witness system} for $\J'$ consists of
\begin{enumerate}[label=(W\arabic*),nosep]
\item a partition of $\J'$ into \emph{families}, and of each family into \emph{groups};
\item for every group $G$ an $\varepsilon$-witness $X_G$ for $G$;
\item for every job $j\in\J'$ a support $Y_j$;
\end{enumerate}
such that
\begin{enumerate}[label=(W\arabic*),nosep,start=4]
\item if $j,j'$ lie in the same family but in different groups, then $Y_j\cap Y_{j'}=\varnothing$.
\end{enumerate}
The \emph{size} of the system is its number of families.
\end{definition}

Supports within one group may overlap.  Like \Cref{def:witness}, this definition involves only windows and slacks.

\begin{theorem}[Rounding from witness systems]\label{thm:ws}
Let an $\varepsilon$-witness system of size at most $\phi$ for $\J'$ be given, where $\phi\ge1$.  In time polynomial in the
input size and in the total number of intervals of the supports, one can compute a nonmigratory schedule of weight at least
\[
 \frac{\varepsilon}{1+\varepsilon}\cdot\frac{1}{\phi}\sum_{j\in\J'}w_jz_j .
\]
Each completed job $j$ is run on one machine during the earliest $p_j$ units of its support.
\end{theorem}

\begin{proof}
If $\J'$ is empty, the empty schedule suffices.

\noindent\emph{One group.}  For a group $G$, with the witness $X_G$ of (W2), \Cref{lem:group} finds an independent set $T_G\subseteq G$ with
$\sum_{j\in T_G}w_j\ge\frac{\varepsilon}{1+\varepsilon}\sum_{j\in G}w_jz_j$, and schedules it on distinct eligible machines, each
$j\in T_G$ during the earliest $p_j$ units of $Y_j$.

\noindent\emph{One family.}  Take the union of the schedules of all groups of a family $F$.  Two jobs on the same
machine come from different groups, because within a group machines are distinct.  Their pieces lie in their supports, which are
disjoint by (W4), so the intervals of their pieces are disjoint.  Each job runs inside its window on one machine.  The family's schedule is
therefore a feasible nonmigratory schedule of weight at least $\frac{\varepsilon}{1+\varepsilon}\sum_{j\in F}w_jz_j$.  There are at most $\phi$ families, so the best family schedule has at least a
$1/\phi$ fraction of their total weight.

\noindent\emph{Polynomial time.}  The algorithm uses the partition into families and groups, the supports and the weights, but not the
witnesses.  For each group it runs the greedy algorithm of \Cref{lem:greedy}, whose independence tests are bipartite matching
computations, and it computes one matching.  If a support consists of $k$ pairwise disjoint half-open intervals, its earliest $p_j$
units consist of the first of these intervals in increasing order, up to the one at which their total length reaches $p_j$, with
that last interval shortened; they are therefore found with $O(k\log k)$ arithmetic operations.
\end{proof}

\subsection{A witness system for moderate slack}\label{sec:mod-construction}

Jobs are sorted into length classes, in powers of two.  Within a length class, a grid whose spacing is the class length gives each
job a nearby grid point, its \emph{pivot}; the half-cell next to the pivot, on one side, lies in the window and is long compared with
the slack.  The jobs of one length class with the same pivot and side form a group, with this half-cell as its witness, and
coloring the pivots with three colors splits the groups of a length class into a constant number of families in which different
groups have disjoint windows.  To include several length classes in the same family, the support of each job must avoid the windows of
the shorter jobs whose length classes are congruent to its own modulo a number $q$.  This is possible when these windows are much shorter than the slack of the job; such windows are
called \emph{dust}.  All shorter windows of a family are dust if its length classes differ by multiples of $q$ and
$q=O(\log(n/\sigma_0))$ is large enough, and this gives $O(\log n)$ families when $\sigma_0$ is polynomially small.

\begin{definition}[Length classes, grids, pivots, dust]\label{def:grid}
\begin{enumerate}[label=\arabic*.,nosep]
\item \emph{Length class.}  The \emph{length class} of a job $j$ is the integer $\lambda_j$ with $2^{\lambda_j}\le L_j<2^{\lambda_j+1}$.
\item \emph{Grid and cells.}  For a length class $\lambda$ let $g=2^\lambda$.  The \emph{grid} of the length class is $g\mathbb Z$, and its \emph{cells} are the
intervals $[kg,(k+1)g)$, $k\in\mathbb Z$.
\item \emph{Pivot, index, color.}  The \emph{pivot} $\theta_j$ of a job $j$ of length class $\lambda$ is the least point of $g\mathbb Z$ with $\theta_j\ge r_j$;
its \emph{index} is $\theta_j/g$, and its \emph{color} is the index modulo $3$.
\item \emph{Three-cell box.}  The \emph{three-cell box} of a job $j$ of length class $\lambda$ is the open interval $(\theta_j-g,\ \theta_j+2g)$, where $g=2^\lambda$.
\item \emph{Dust.}  A window $K_{j'}$ is \emph{dust} for a job $j$ if $L_{j'}<s_j/(4n)$.
\end{enumerate}
\end{definition}

\begin{theorem}[Moderate slack]\label{thm:moderate}
The set $\J_{\mathrm{mod}}$ of moderate-slack jobs has a $1$-witness system of size at most $54\log n+138$, computable in
polynomial time.
\end{theorem}

\begin{proof}
Let $q=\lceil\log (8n/\sigma_0)\rceil$.  Fix a job $j\in\J_{\mathrm{mod}}$, let $\lambda=\lambda_j$ be its length class, and let $g=2^\lambda$.  The window
of $j$ is at least one cell and less than two cells long.

\noindent\emph{Pivots and the three-cell box.}  Since $\theta_j$ is the least multiple of $g$ with $\theta_j\ge r_j$, the
grid point before it lies before $r_j$: $\theta_j-g<r_j\le\theta_j$.  Since the window is shorter than two cells,
$d_j=r_j+L_j\le\theta_j+L_j<\theta_j+2g$.  Therefore the window lies in its three-cell box:
\begin{equation}\label{eq:window-box}
 K_j\subseteq(\theta_j-g,\ \theta_j+2g).
\end{equation}
Also $\theta_j<r_j+g\le d_j$, so the pivot lies in the window.

\noindent\emph{Sides and witnesses: the half-cell next to the pivot.}  If $d_j\ge\theta_j+g/2$, give $j$ side $+$ and the interval
$X_j=[\theta_j,\theta_j+g/2)$; it lies in $K_j$ because $r_j\le\theta_j$.  Otherwise give $j$ side $-$ and
$X_j=[\theta_j-g/2,\theta_j)$.  It lies in $K_j$ because $d_j>\theta_j$ (as $d_j\ge r_j+g>\theta_j$) and
$r_j=d_j-L_j<\theta_j+g/2-g$.  In both cases $|X_j|=g/2>2s_j$, since $s_j\le L_j/8<g/4$.

\noindent\emph{Groups.}  A group consists of the jobs with the same length class, pivot and side.  The half-cell $X_j$ depends only on
these three, so it is the same for all members of a group, and it is the witness of the group.  Thus condition (W2) of
\Cref{def:ws} holds with $\varepsilon=1$.

\noindent\emph{Families.}  A family consists of the jobs $j$ with the same
\[
 (\lambda_j\bmod q,\ \ \theta_j/2^{\lambda_j}\bmod3,\ \ \text{side of }j),
\]
where $\theta_j/2^{\lambda_j}\bmod3$ is the color of $j$, so there are at most $6q$ nonempty families.  The length class, pivot and side of a job determine this triple, so every
group lies in one family, and the groups partition each family; this is condition (W1).

\noindent\emph{Supports.}  Let $Y_j$ be $K_j$ minus the windows of all jobs $j''\in\J_{\mathrm{mod}}$ with
$\lambda_{j''}\equiv\lambda_j\pmod q$ and $\lambda_{j''}<\lambda_j$.  Such a job has $\lambda_{j''}\le\lambda_j-q$, and $2^q\ge8n/\sigma_0$, so its
window is dust for $j$:
\[
 L_{j''}<2^{\lambda_{j''}+1}\le2^{\lambda_j+1}2^{-q}\le2^{\lambda_j+1}\frac{\sigma_0}{8n}\le\frac{\sigma_0L_j}{4n}\le\frac{s_j}{4n}.
\]
At most $n$ windows are removed, so $|K_j\setminus Y_j|\le s_j/4$ and $|Y_j|\ge p_j$; also $Y_j\subseteq K_j$.  Since $K_j$ and the
removed windows are half-open intervals, $Y_j$ is a finite union of half-open intervals.  This is (W3).

\noindent\emph{Disjoint supports.}  We now establish condition (W4).  Let $j$ and $j'$ lie in the same family but in different
groups, and assume that $\lambda_j\ge\lambda_{j'}$.
\begin{itemize}[nosep]
\item If $\lambda_j=\lambda_{j'}$ (coloring), they have the same side, so $\theta_j\ne\theta_{j'}$, and
the same color, $\theta_j/g\equiv\theta_{j'}/g\pmod3$.  Their indices $\theta_j/g$ and $\theta_{j'}/g$ are therefore distinct and
congruent modulo $3$, so $|\theta_j-\theta_{j'}|\ge3g$, and by \eqref{eq:window-box} their windows are disjoint, and so are
their supports, which lie in the windows.
\item If $\lambda_j>\lambda_{j'}$ (dust), then $\lambda_{j'}\equiv\lambda_j\pmod q$, so $K_{j'}$ was removed from $Y_j$, and
$Y_j\cap Y_{j'}\subseteq Y_j\cap K_{j'}=\varnothing$.
\end{itemize}

\noindent\emph{Size.}  Since $\log (1/\sigma_0)=2K'+3=8\ell+11\le8\log n+19$, we have $q\le9\log n+23$, so the at most $6q$
nonempty families number at most $54\log n+138$.

\noindent\emph{Polynomial time.}  The length class, pivot and side of each job are computed from its release date, due date and
length by comparisons with powers of $2$ and their multiples.  If $L_j=P/R$ with integers $1\le P,R\le2^b$, then
$2^{-b}\le L_j\le2^b$, so $|\lambda_j|\le b$; hence $g$ and the pivot $\theta_j$ have polynomial bit length.  Each witness is a half-cell, and each support is a window minus
at most $n$ windows, so a union of at most $n+1$ intervals with endpoints among the input numbers.
\end{proof}

%=====================================================================
\section{Nearly rigid jobs}\label{sec:near}
%=====================================================================

Nearly rigid jobs have relative slack less than $\sigma_0$, which is polynomially small in $n$.  This section rounds them
and proves \Cref{thm:near}.  As a warm-up, \Cref{sec:rigid} treats instances in which every job is rigid.  The proof of
\Cref{thm:main-ra} does not use this case: in a general instance, rigid jobs are nearly rigid, and \Cref{thm:near} rounds them
together with the others.  \Cref{sec:near-overview} explains why the general case needs a different method.

\subsection{Rigid jobs}\label{sec:rigid}

We begin with instances in which every job is rigid.  For them, the tools of \Cref{part:unrelated} suffice: the midpoint tree of
\Cref{sec:tree} and \Cref{lem:common-witness}.  At every node of the tree, the rigid jobs whose windows contain the
separator have a common witness: the interval from the separator to the earliest due date among them.  The tree has
logarithmic depth, and taking the heaviest of the depth schedules loses only a logarithmic factor.

\begin{lemma}[Depth schedules for rigid jobs]\label{lem:rigid-depth}
Assume that every job is rigid and that some job has $z_j>0$, and consider the midpoint tree on the set of jobs with
$z_j>0$.  For every depth $d$, one can compute in polynomial time a nonmigratory schedule that completes, for every node
$v$ of depth $d$, a set $I_v\subseteq F_v$ of jobs on pairwise distinct machines with
\[
 \sum_{j\in I_v}w_j\ \ge\ \sum_{j\in F_v}w_jz_j ,
\]
and completes no other jobs.
\end{lemma}

\begin{proof}
At a node $v$ set
\[
 d_v^{\min}:=\min_{j\in F_v}d_j>t_v,
 \qquad
 X_v:=[t_v,d_v^{\min}).
\]
Then $X_v\subseteq K_j$ for every $j\in F_v$, since $r_j\le t_v$, and $s_j=0$, so \Cref{lem:ra-processing} gives
$y_j(X_v)\ge z_j|X_v|$.  \Cref{lem:common-witness} with $\rho=1$ and $E$ the set of all eligible pairs gives a set
$I_v\subseteq F_v$ of weight at least $\sum_{j\in F_v}w_jz_j$, assigned to distinct eligible machines.  Run each $j\in I_v$
on its machine during $K_j$; since $p_j=L_j$, this completes $j$.  For a depth $d$, take the union of these schedules
over the nodes of depth $d$.  Jobs of the same node are on distinct machines, and the windows of jobs of distinct
nodes of depth $d$ lie in their pairwise disjoint regions, by \Cref{lem:midpoint-tree}(c),(d); hence this union is a
nonmigratory schedule, and it completes exactly the jobs of the sets $I_v$.  The midpoint tree and the matchings of
\Cref{lem:common-witness} are computed in polynomial time.
\end{proof}

\begin{proposition}[Rigid restricted assignment]\label{prop:ra-rigid}
Assume that every job is rigid.  Given a $\StrongLP$ solution of value $W$, one can compute in deterministic
polynomial time a nonmigratory schedule of weight at least $W/(1+\lceil\log n\rceil)$.  In particular,
\[
 \OPTLP\ \le\ (1+\lceil\log n\rceil)\,\OPTnm .
\]
\end{proposition}

\begin{proof}
If no job has $z_j>0$, then $W=0$ and there is nothing to show.  Otherwise \Cref{lem:rigid-depth} gives a nonmigratory
schedule for each depth of the midpoint tree on the set of jobs with $z_j>0$.  There are at most $1+\lceil\log n\rceil$ depths, by \Cref{lem:midpoint-tree}(b), and the depth
schedules have total weight at least $\sum_v\sum_{j\in F_v}w_jz_j=W$, by \Cref{lem:midpoint-tree}(e), so the heaviest has
weight at least $W/(1+\lceil\log n\rceil)$.  Applied to an optimal $\StrongLP$ solution, this gives the last
statement.
\end{proof}

With unit weights, a bound in terms of the number $m$ of machines alone also holds: \Cref{app:rigid-unit} shows that
$\OPT\le(2\log m+6)\,\OPTnm$, by an  argument that compares with $\OPT$ rather than with the $\StrongLP$.
The rest of this section treats all nearly rigid jobs, rigid ones included.  Once some of them have positive slack, the
jobs whose windows contain a separator no longer share a witness.

\subsection{Overview of the proof}\label{sec:near-overview}

\paragraph{Why a different method is needed.}  The groups of \Cref{thm:moderate} remain valid for nearly rigid jobs, but its families
do not: they keep different length classes apart only through dust, which gives $O(\log n)$
families only if all relative slacks are at least a polynomially small $\sigma_0$.  For rigid jobs, \Cref{sec:rigid} uses a tree of
separator points: the jobs whose windows contain a separator $t$ form one group, whatever their lengths, and the $O(\log n)$
levels of the tree give the families.  With slack this fails too, since a witness must be longer than the slack of every member; the jobs whose windows contain $t$ must be sorted further, by their distances from $t$.  This
section measures distances from $t$ on a geometric grid.  The recursion differs from the midpoint tree in one
respect.  In the midpoint tree, a node passes to its two children the jobs on the two sides of $t$.  Here, a node has
one child for each cell of the grid and each side of $t$, so that every job passed to a child is busy at distances in
one cell only.  A job whose window crosses a grid point on one side of $t$ therefore stays at the node, together
with the jobs whose windows contain $t$.  Instead of a witness system, the schedule is computed by a dynamic program; groups appear only
in its analysis.

\paragraph{The recursion.}  The nearly rigid jobs, the jobs of $\J_{\mathrm{nr}}$, are
processed by a recursion whose nodes form a rooted tree $R$.  Each node $v$ of $R$ hosts a
set $J_v$ of jobs and a point $t=t_v$, its \emph{separator}, which is a median of the endpoints of the windows of
the jobs of $J_v$.  Everything at $v$ is measured by distance from $t$.

\paragraph{Two radii and a geometric grid.}  The jobs that stay at a node form its \emph{chain}.  Seen from the separator $t$, a
job is described by two distances, its \emph{inner radius} and its \emph{reach}, and the schedule runs
it only at distances between them.  A chain job is \emph{covered} if its inner radius is $0$ or a grid point
lies strictly between eight times its inner radius and its reach; it then has a witness ending at that grid point.  The
radii of the jobs at a node can be of any scale, and a grid of fixed spacing would be too coarse at small distances and too fine at
large ones.  The grid is therefore geometric, evenly spaced on a logarithmic scale, with a ratio $Q$ polynomial in $n$.  With a smaller ratio, a larger fraction of the offsets would leave a chain job uncovered.  A larger ratio would lower the threshold $\sigma_0=1/(8Q)$ of near-rigidity, and the jobs above it are
handled in \Cref{sec:moderate} with $O(\log(n/\sigma_0))$ families.

\paragraph{The dynamic program.}  Two jobs whose ranges of busy distances from $t$ meet in at most one point never run at the same
time, on either side of $t$, because the intervals of their pieces are half-open.  The \emph{busy-distance interval} of a chain job is the
interval between its two radii, rounded outward to grid points or to $0$; it contains at least two cells.  Every job in the
subtree of a child of $v$ in cell $m$ is busy only at distances between $\gamma_m$ and $\gamma_{m+1}$, the two grid points that bound
cell $m$.  Consequently, a nonmigratory schedule can be assembled at $v$ from a \emph{packing}: a set of \emph{blocks}, which are intervals of
distances whose endpoints are grid points or $0$, any two of which meet in at most one point.  A block $[\gamma_m,\gamma_{m+1}]$ for a cell $m$ in which
$v$ has children contributes the schedules already computed for these children.  Any other block carries a set of chain jobs whose
busy-distance intervals lie in it and which is independent in the transversal matroid; the greedy
algorithm finds the heaviest such set, and its jobs run on distinct machines.  Given the values already computed for the
children, choosing a packing of maximum value is weighted interval scheduling, see \Cref{fig:dp}.  The dynamic program runs from the
leaves of $R$ to the root.

\paragraph{Why the loss stays logarithmic.}  The values computed for the children of a node are themselves approximate; the
question is why the losses do not accumulate from the leaves to the root.  For a node $v$, call the LP weight of the covered chain jobs of the nodes in
the subtree of $v$ the \emph{covered weight} of $v$.  The analysis shows, by induction from the leaves, that the schedule computed at every
node $v$ has weight at least the covered weight of $v$ divided by $\Psi(|J_v|)$, where $\Psi(N)=\Theta(\log N)$; call
$\Psi(|J_v|)$ the \emph{loss factor} of $v$.  The loss factor depends only on the number $|J_v|$ of jobs hosted by $v$.  Every
child $u$ of $v$ hosts at most half of the jobs hosted by $v$, so from $u$ to $v$ the loss factor grows from
$\Psi(|J_u|)$ to $\Psi(|J_v|)$, i.e., by $\Theta(\log(|J_v|/|J_u|))$, and this growth suffices to account for the covered chain
jobs of $v$; the next paragraph explains how.  Along a path from the root to a leaf these growths telescope: their
sum is the loss factor of the root minus that of the leaf, which is less than $\Psi(n)=O(\log n)$, whatever the length of the path.

\paragraph{A fractional packing used only in the analysis.}  The analysis makes no use of the packing that the dynamic program
finds.  It constructs a \emph{fractional packing} of blocks at $v$, which the algorithm never computes; interval packing has an
integral linear relaxation, so the dynamic program does at least as well, and it
suffices that the fractional packing have value at least the covered weight of $v$ divided by $\Psi(|J_v|)$.  The covered weight of
$v$ is the covered weight of its children plus the LP weight of the covered chain jobs of $v$, and the fractional packing accounts
for the two parts separately.  The covered chain jobs of $v$ are partitioned into groups with a common witness.  The block of a group has value at least a constant fraction of the LP weight of the group, so every group receives the same multiplicity, however many jobs it contains, and what limits the fractional
packing at a distance is the number of groups whose blocks contain it.  The block of a cell in which $v$ has children needs as its
multiplicity the ratio of their larger loss factor to that of $v$, which is larger the more jobs they host.  Since the multiplicities
at a distance add up to at most $1$, fewer groups may then have blocks at the distances of that cell.  The groups are formed with
the alphabetic code of Gilbert and Moore~\cite{GM}, which determines only, for each covered chain job of positive inner radius, which of the grid points that
make it covered ends the common witness of its group.  The jobs of inner radius $0$ form at most two more
groups.  The code gives each item, one of the cells relevant at $v$, a code word of length approximately the logarithm of the total
frequency divided by the frequency of the item, where the frequency of a cell $m$ counts the jobs of the children of $v$ in the cells
$m$ and $m-1$.  A cell whose children host many jobs therefore has a short code word, and the number of groups whose blocks contain
its distances is at most a constant times its length.

\paragraph{Outline of the presentation.}  \Cref{sec:params} sets the parameters, the radii and the grid.
\Cref{sec:recursion} defines the recursion tree and proves the two facts about it that the rest uses:
covered chain jobs have witnesses ending at grid points, and the offsets can be chosen
so that the LP weight of the covered jobs is at least half of the LP weight $\sum_{j\in\J_{\mathrm{nr}}}w_jz_j$ of $\J_{\mathrm{nr}}$.  \Cref{sec:dpalg} gives the
dynamic program, and \Cref{sec:guarantee} proves a lower bound on the weight of the schedule it produces.

\subsection{Parameters, radii and the grid}\label{sec:params}

This subsection explains the parameters, describes a job by its distances from a point $t$, and introduces
the grid of distances that cuts them into cells.  \Cref{lem:radii} collects the facts about these
distances that the rest of the section uses.

We use the parameters $\ell$, $K'$, $Q$ and $\sigma_0=1/(8Q)$ of \Cref{def:ra-classes}.  All jobs considered in this
section belong to $\J_{\mathrm{nr}}$, so they have $\sigma_j<\sigma_0$, in particular $8s_j<L_j/Q$.  The recursion will have at most $\ell$ nodes on
every root-to-leaf path.  The number $K'$ of offsets is chosen so that the probability that a job
is uncovered at a node, at most $2/K'$, summed over the at most $\ell$ nodes of a
path, stays below $1/2$.  The threshold $\sigma_0$ is chosen so that a job whose window comes within $s_j/2$ of
the separator always has a grid point strictly between $8\iota_j$ and $\rho_j$.

\begin{definition}[Gap, support and inner radius]\label{def:job-from-t}
Fix a point $t$ and a job $j\in\J_{\mathrm{nr}}$, with reach $\rho_j$ and long side $\eta_j$ seen from $t$.  Its
\emph{gap} $\delta_j$ is the smallest distance from $t$ of a point of $K_j$, in the sense of infimum.  Its
\emph{support} and its \emph{inner radius} are
\[
 Y_j=K_j\setminus\bigl[t-\tfrac{s_j}2,\ t+\tfrac{s_j}2\bigr),\qquad \iota_j=\max\{\delta_j,\ s_j/2\}.
\]
\end{definition}

The gap, reach, long side, support and inner radius depend on $t$, which the notation leaves out.  At a node $v$ of the recursion
below, $t$ is the separator $t_v$ of $v$, so the same job can have different values of them at different nodes.

\begin{definition}[Grid and cells]\label{def:grid-cells}
Fix an \emph{offset} $o\in\{0,\dots,K'-1\}$.  The \emph{grid} of $o$ is the set of distances
$\mathcal G(o)=\{\gamma_m:m\in\mathbb Z\}$ with $\gamma_m=4^{o}Q^m$.  The \emph{cells} are the integers
$m$, in their usual order, and the cell $m$ \emph{covers} the distances in
$[\gamma_m,\gamma_{m+1})$.  For a point $t$, the \emph{annuli} of the cell $m$ around $t$ are
\[
 A^+_m=[t+\gamma_m,\ t+\gamma_{m+1}),\qquad A^-_m=(t-\gamma_{m+1},\ t-\gamma_m].
\]
For $\xi>0$ we write $\mathrm{cell}(\xi)$ for the $m$ with $\gamma_m\le\xi<\gamma_{m+1}$, the cell
that covers $\xi$, and $\mathrm{cell}^-(\xi)$ for the $m$ with $\gamma_m<\xi\le\gamma_{m+1}$.
\end{definition}

The grid points are the powers of $4$ whose exponent is congruent to $o$ modulo $K'$, so each power of $4$ is
a grid point for exactly one offset.

\begin{lemma}[Radii]\label{lem:radii}
Let $j\in\J_{\mathrm{nr}}$ be seen from $t$, and let $\mathcal G$ be the grid of any offset, with annuli around $t$.
\begin{enumerate}[label=(\alph*),nosep]
\item $Y_j$ is a support of $j$: $Y_j\subseteq K_j$ and $|Y_j|\ge p_j$.  Every point of $Y_j$ is at distance between $\iota_j$ and $\rho_j$ from $t$, and $Y_j$ contains every
point on side $\eta_j$ at distance strictly between $\iota_j$ and $\rho_j$.  Also $\iota_j=0$ exactly when $j$ is rigid and $K_j$
contains $t$ or ends at $t$.
\item If $\delta_j\le s_j/2$, in particular if $K_j$ contains $t$ or ends at $t$, then $\mathcal G\cap(8\iota_j,\rho_j)\ne\varnothing$.
\item If $\mathcal G\cap(\iota_j,\rho_j)=\varnothing$, then $\delta_j=\iota_j>0$, $K_j$ lies on one side $\pm$ of $t$, and
$K_j\subseteq A^\pm_m\cup\{r_j\}$, where $m=\mathrm{cell}(\iota_j)$; if $\pm$ is $+$, then $K_j\subseteq A^+_m$.
\end{enumerate}
\end{lemma}

\begin{proof}
(a) The support $Y_j$ is $K_j$ minus the interval $[t-s_j/2,\,t+s_j/2)$, which has length $s_j$.
Hence $Y_j\subseteq K_j$ and $|Y_j|\ge L_j-s_j=p_j$.  Moreover $Y_j$ is a union of at most two
half-open intervals, so it is a support.  Points of $K_j$ are at distance at least $\delta_j$ and
at most $\rho_j$, and points of $Y_j$ are moreover at distance at least $s_j/2$.  On its long side,
the interval $K_j$ contains every point at distance strictly between $\delta_j$ and $\rho_j$, and
removing $[t-s_j/2,\,t+s_j/2)$ keeps those at distance greater than $s_j/2$.  Finally $\iota_j=0$ if and only if $s_j=0$ and $\delta_j=0$, and $\delta_j=0$ means that $r_j\le t\le d_j$.

(b) If $K_j$ contains $t$ or ends at $t$, then $\delta_j=0\le s_j/2$, so it suffices to assume $\delta_j\le s_j/2$.

\noindent\emph{Step 1: $\rho_j\ge L_j/2>0$.}  Every point of $K_j$ is at distance at most $\rho_j$ from $t$, so
$K_j\subseteq[t-\rho_j,\,t+\rho_j]$ and $L_j\le2\rho_j$.  Also $L_j>0$.

\noindent\emph{Step 2: the case $s_j=0$.}  The assumption gives $\delta_j=0$, so $\iota_j=\max\{\delta_j,s_j/2\}=0$, and the
interval in question is $(0,\rho_j)$.  It is nonempty by Step~1.  Since $\gamma_m=4^oQ^m\to0$ as $m\to-\infty$, some grid
point lies in it.

\noindent\emph{Step 3: the case $s_j>0$.}  The assumption gives $\iota_j=\max\{\delta_j,s_j/2\}=s_j/2$, so $8\iota_j=4s_j$.  Since
$j\in\J_{\mathrm{nr}}$, we have $\sigma_j=s_j/L_j<\sigma_0=1/(8Q)$, i.e., $L_j>8Q\,s_j$.  With Step~1,
\[
 \frac{\rho_j}{8\iota_j}\ \ge\ \frac{L_j/2}{4s_j}\ =\ \frac{L_j}{8s_j}\ >\ Q\ =\ 4^{K'} .
\]
Set $a=\log_4(8\iota_j)$ and $b=\log_4\rho_j$, so that $b-a>K'$.  The $K'$ consecutive integers
$\lfloor a\rfloor+1,\dots,\lfloor a\rfloor+K'$ all lie in the open interval $(a,b)$, because $\lfloor a\rfloor+1>a$ and
$\lfloor a\rfloor+K'\le a+K'<b$.  One of them is an integer $e$ with $e\equiv o\pmod{K'}$, say $e=o+K'm$ with $m\in\mathbb Z$.
Then $4^e=4^oQ^m=\gamma_m$ is a grid point, and $a<e<b$ gives $8\iota_j<\gamma_m<\rho_j$.

(c) By (b), $\delta_j>s_j/2$, so $\iota_j=\delta_j>0$ and $K_j$ lies on one side of $t$.  Since $m=\mathrm{cell}(\iota_j)$ and no grid point
lies strictly between $\iota_j$ and $\rho_j$, we have $\gamma_m\le\iota_j<\rho_j\le\gamma_{m+1}$.  On side $+$, the points of $K_j=[r_j,d_j)$ are at
distances in $[\iota_j,\rho_j)$ from $t$; on side $-$, its points other than $r_j$ are at distances in $(\iota_j,\rho_j)$.  These distances lie in
$[\gamma_m,\gamma_{m+1})$, the distances covered by the cell $m$, so these points lie in $A^\pm_m$.
\end{proof}

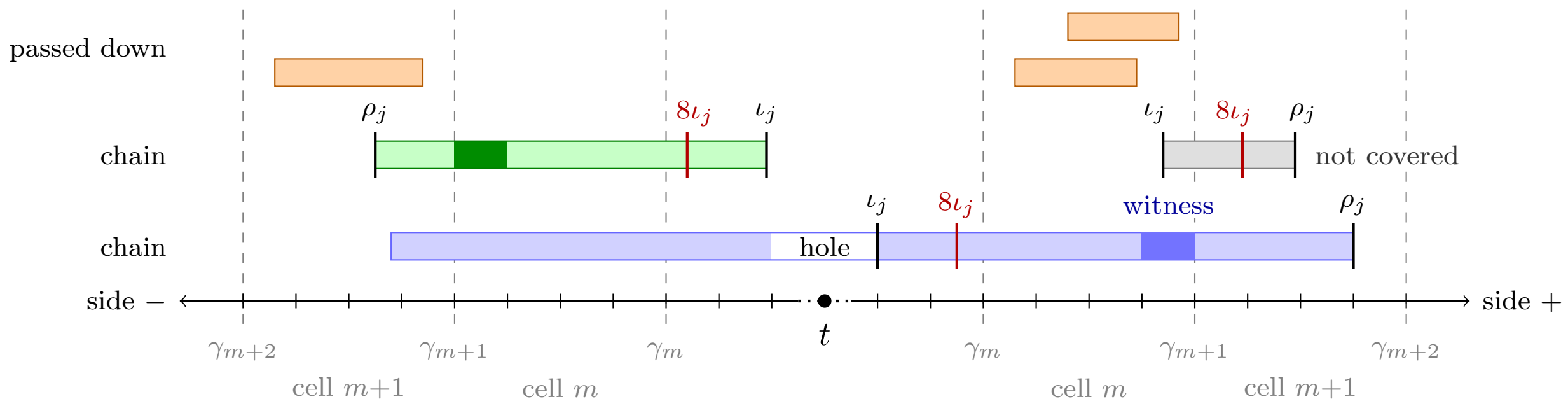
\begin{figure}[t]
\centering
\begin{tikzpicture}[x=0.55cm,y=0.95cm,font=\small]
 % grid lines (cells)
 \foreach \g/\lab in {3/{\gamma_m},7/{\gamma_{m+1}},11/{\gamma_{m+2}}} {
   \draw[dashed,gray] (\g,-0.25) -- (\g,3.3);
   \draw[dashed,gray] (-\g,-0.25) -- (-\g,3.3);
   \node[font=\scriptsize,gray] at (\g,-0.55) {$\lab$};
   \node[font=\scriptsize,gray] at (-\g,-0.55) {$\lab$};
 }
 % axis: logarithmic in the distance on each side; the dotted segment stands for all smaller distances
 \draw[->] (0.5,0) -- (12.2,0) node[right,font=\scriptsize] {side $+$};
 \draw[->] (-0.5,0) -- (-12.2,0) node[left,font=\scriptsize] {side $-$};
 \draw[dotted,thick] (-0.5,0) -- (0.5,0);
 \foreach \y in {1,...,11} {\draw (\y,-0.08) -- (\y,0.08); \draw (-\y,-0.08) -- (-\y,0.08);}
 \fill (0,0) circle (2pt) node[below=3pt] {$t$};
 \foreach \c/\lab in {5/{cell $m$},9/{cell $m{+}1$},-5/{cell $m$},-9/{cell $m{+}1$}} {\node[font=\scriptsize,gray] at (\c,-0.95) {\lab};}
 % row 1: two-sided chain job, long side +
 \fill[blue!18] (-8.2,0.45) rectangle (10,0.75);
 \fill[white] (-1,0.45) rectangle (1,0.75);
 \draw[blue!60] (-8.2,0.45) rectangle (10,0.75);
 \node[font=\scriptsize] at (0,0.6) {hole};
 \fill[blue!55] (6,0.45) rectangle (7,0.75);
 \draw[thick] (1,0.35) -- (1,0.85); \node[font=\scriptsize] at (1,1.05) {$\iota_j$};
 \draw[thick,red!70!black] (2.5,0.35) -- (2.5,0.85); \node[font=\scriptsize,red!70!black] at (2.5,1.05) {$8\iota_j$};
 \draw[thick] (10,0.35) -- (10,0.85); \node[font=\scriptsize] at (10,1.05) {$\rho_j$};
 \node[font=\scriptsize,blue!60!black,fill=white,inner sep=1pt] at (6.5,1.05) {witness};
 \node[anchor=east,font=\scriptsize] at (-12.2,0.6) {chain};
 % row 2: covered one-sided chain job on side - (green), uncovered chain job on side + (gray)
 \fill[green!22] (-8.5,1.45) rectangle (-1.1,1.75);
 \fill[green!55!black] (-7,1.45) rectangle (-6,1.75);
 \draw[green!50!black] (-8.5,1.45) rectangle (-1.1,1.75);
 \draw[thick] (-1.1,1.35) -- (-1.1,1.85); \node[font=\scriptsize] at (-1.1,2.05) {$\iota_j$};
 \draw[thick,red!70!black] (-2.6,1.35) -- (-2.6,1.85); \node[font=\scriptsize,red!70!black] at (-2.45,2.05) {$8\iota_j$};
 \draw[thick] (-8.5,1.35) -- (-8.5,1.85); \node[font=\scriptsize] at (-8.5,2.05) {$\rho_j$};
 \fill[gray!25] (6.4,1.45) rectangle (8.9,1.75);
 \draw[gray] (6.4,1.45) rectangle (8.9,1.75);
 \draw[thick] (6.4,1.35) -- (6.4,1.85); \node[font=\scriptsize] at (6.25,2.05) {$\iota_j$};
 \draw[thick,red!70!black] (7.9,1.35) -- (7.9,1.85); \node[font=\scriptsize,red!70!black] at (7.75,2.05) {$8\iota_j$};
 \draw[thick] (8.9,1.35) -- (8.9,1.85); \node[font=\scriptsize] at (9.05,2.05) {$\rho_j$};
 \node[font=\scriptsize,gray!40!black,anchor=west,fill=white,inner sep=1pt] at (9.2,1.6) {not covered};
 \node[anchor=east,font=\scriptsize] at (-12.2,1.6) {chain};
 % row 3: jobs passed to children
 \fill[orange!35] (3.6,2.35) rectangle (5.9,2.65); \draw[orange!70!black] (3.6,2.35) rectangle (5.9,2.65);
 \fill[orange!35] (4.6,2.85) rectangle (6.7,3.15); \draw[orange!70!black] (4.6,2.85) rectangle (6.7,3.15);
 \fill[orange!35] (-10.4,2.35) rectangle (-7.6,2.65); \draw[orange!70!black] (-10.4,2.35) rectangle (-7.6,2.65);
 \node[anchor=east,font=\scriptsize] at (-12.2,2.75) {passed down};
\end{tikzpicture}
\caption{A node seen from its separator $t$.  Distances from $t$ are drawn on a logarithmic scale on each side: each tick is a
factor $4$, the dashed lines are grid points, drawn far closer together than the true ratio $Q=4^{K'}$, and the dotted segment next to
$t$ stands for all smaller distances.  Each bar is the window of a job.  A job with no grid point strictly between $\iota_j$ and
$\rho_j$ lies, except possibly for its release date, in one cell on one side (orange) and is passed to the child of that cell and side; the other jobs form the chain.  A chain
job is \emph{covered} if a grid point $B$ lies strictly between $8\iota_j$ (red) and $\rho_j$; the distances in $[B/4,B)$ on its long
side (dark) then form a witness for it.  The blue job contains $t$, and its support leaves out the interval
$[t-s_j/2,\,t+s_j/2)$ (``hole''), so $\iota_j=s_j/2$.  The blue and green jobs are covered.  The gray job is not: the
grid point $\gamma_{m+1}$ lies between $\iota_j$ and $8\iota_j$, and no grid point lies strictly between $8\iota_j$ and $\rho_j$.}
\label{fig:node}
\end{figure}

%---------------------------------------------------------------------
\subsection{The recursion tree}\label{sec:recursion}

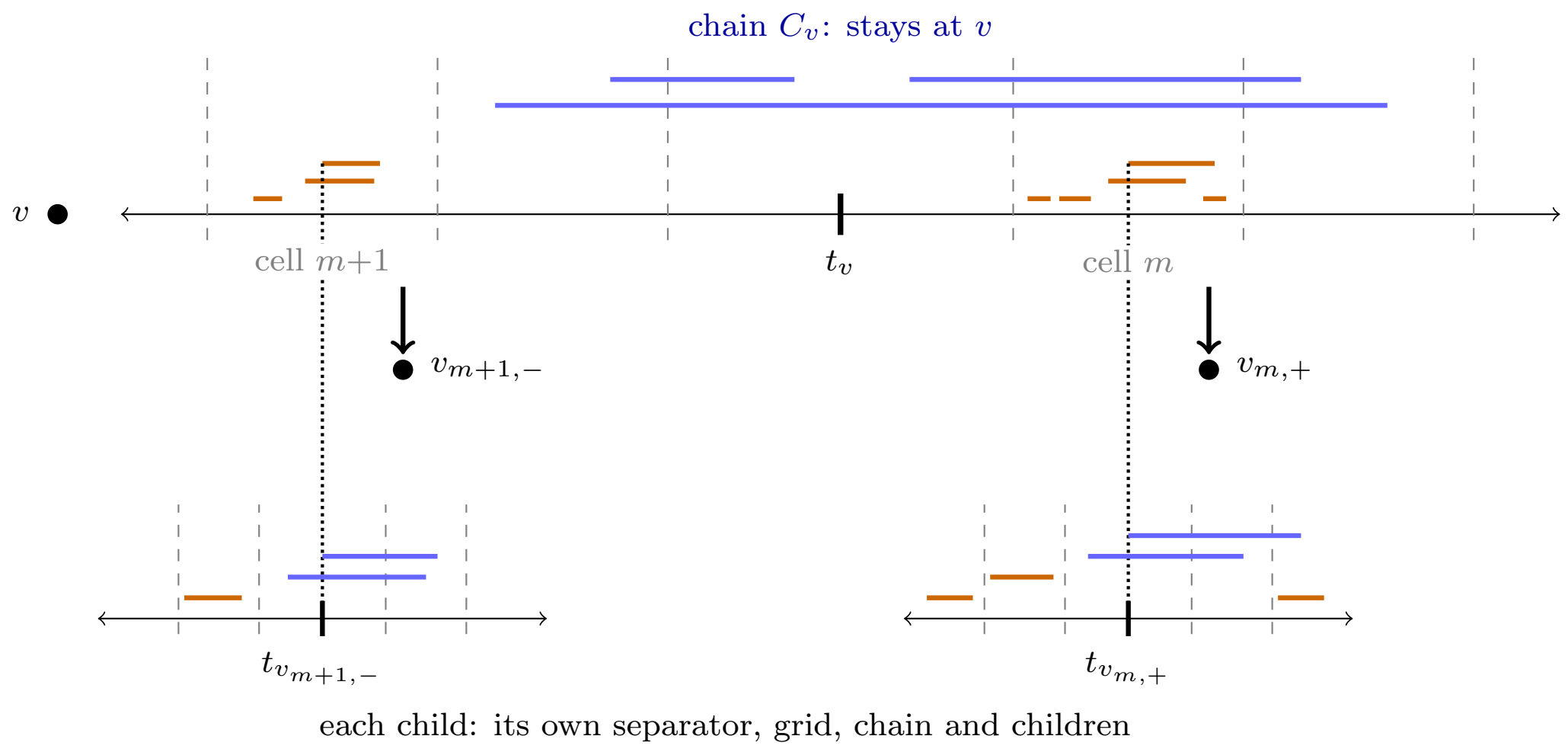
\begin{figure}[t]
\centering
\begin{tikzpicture}[x=0.5cm,y=0.9cm,font=\small]
 % node v: axis, grid, chain (blue), jobs passed to two of its children (orange)
 \fill (-13.6,0) circle (2.5pt) node[left=3pt,font=\scriptsize] {$v$};
 \draw[->] (0,0) -- (12.5,0); \draw[->] (0,0) -- (-12.5,0);
 \draw[line width=1.4pt] (0,-0.2) -- (0,0.2); \node[font=\scriptsize,below] at (0,-0.2) {$t_v$};
 \foreach \g in {3,7,11} {\draw[dashed,gray] (\g,-0.25) -- (\g,1.6); \draw[dashed,gray] (-\g,-0.25) -- (-\g,1.6);}
 % the same jobs as in the children below, with their endpoints in the same order
 \foreach \a/\b/\y in {3.25/3.65/0.15,3.8/4.35/0.15,4.65/6.0/0.32,5.0/6.5/0.49,6.3/6.7/0.15,-10.2/-9.7/0.15,-9.3/-8.1/0.32,-9.0/-8.0/0.49}
   {\draw[orange!80!black,very thick] (\a,\y) -- (\b,\y);}
 % dotted: the endpoint that becomes the separator of the child
 \draw[densely dotted,thick] (5,0.49) -- (5,-3.9);
 \draw[densely dotted,thick] (-9,0.49) -- (-9,-3.9);
 \draw[blue!60,very thick] (-6,1.05) -- (9.5,1.05); \draw[blue!60,very thick] (-4,1.3) -- (-0.8,1.3); \draw[blue!60,very thick] (1.2,1.3) -- (8,1.3);
 \node[font=\scriptsize,blue!60!black] at (0,1.8) {chain $C_v$: stays at $v$};
 \node[font=\scriptsize,gray,fill=white,inner sep=1pt] at (5,-0.45) {cell $m$}; \node[font=\scriptsize,gray,fill=white,inner sep=1pt] at (-9,-0.45) {cell $m{+}1$};
 % two of the children
 \fill (6.4,-1.5) circle (2.5pt) node[right=3pt,font=\scriptsize] {$v_{m,+}$};
 \fill (-7.6,-1.5) circle (2.5pt) node[right=3pt,font=\scriptsize] {$v_{m+1,-}$};
 \draw[->,very thick] (6.4,-0.7) -- (6.4,-1.35);
 \draw[->,very thick] (-7.6,-0.7) -- (-7.6,-1.35);
 % the children's own axes, grids and jobs; each separator is the median endpoint of its jobs
 \begin{scope}[shift={(5,-3.9)}]
  \draw[->] (0,0) -- (3.9,0); \draw[->] (0,0) -- (-3.9,0);
  \draw[line width=1.2pt] (0,-0.17) -- (0,0.17); \node[font=\scriptsize,below] at (0,-0.17) {$t_{v_{m,+}}$};
  \foreach \g in {1.1,2.5} {\draw[dashed,gray] (\g,-0.15) -- (\g,1.1); \draw[dashed,gray] (-\g,-0.15) -- (-\g,1.1);}
  \draw[orange!80!black,very thick] (-3.5,0.2) -- (-2.7,0.2);
  \draw[orange!80!black,very thick] (-2.4,0.4) -- (-1.3,0.4);
  \draw[blue!60,very thick] (-0.7,0.6) -- (2.0,0.6);
  \draw[blue!60,very thick] (0,0.8) -- (3.0,0.8);
  \draw[orange!80!black,very thick] (2.6,0.2) -- (3.4,0.2);
 \end{scope}
 \begin{scope}[shift={(-9,-3.9)}]
  \draw[->] (0,0) -- (3.9,0); \draw[->] (0,0) -- (-3.9,0);
  \draw[line width=1.2pt] (0,-0.17) -- (0,0.17); \node[font=\scriptsize,below] at (0,-0.17) {$t_{v_{m+1,-}}$};
  \foreach \g in {1.1,2.5} {\draw[dashed,gray] (\g,-0.15) -- (\g,1.1); \draw[dashed,gray] (-\g,-0.15) -- (-\g,1.1);}
  \draw[orange!80!black,very thick] (-2.4,0.2) -- (-1.4,0.2);
  \draw[blue!60,very thick] (-0.6,0.4) -- (1.8,0.4);
  \draw[blue!60,very thick] (0,0.6) -- (2.0,0.6);
 \end{scope}
 \node[font=\scriptsize] at (-2,-4.95) {each child: its own separator, grid, chain and children};
\end{tikzpicture}
\caption{The recursion tree, with two of the children of a node $v$.  The node $v$ keeps its chain (blue) and assigns the jobs of each
nonempty cell and side (orange) to a child.  The child measures the distances of its jobs from its own separator, a median of their endpoints, and
uses its own grid (dotted: the endpoint that becomes the separator of the child); again some of its jobs form its chain and the others are
passed further down.  All jobs of the subtree of $v_{m,+}$ have their
windows in the annulus of cell $m$ on side $+$ of $t_v$, by \Cref{lem:shape}(a), and each child hosts at most half of the jobs of $v$, by \Cref{lem:shape}(b).}
\label{fig:tree}
\end{figure}

This subsection builds the recursion tree $R$, see \Cref{fig:tree}.  Each node takes a median endpoint of its jobs as its separator, keeps as its chain the jobs
whose two radii are separated by a grid point, and passes every other job to a child, one child for each cell and side.  \Cref{lem:shape} bounds the depth of $R$ and locates the jobs of each child.  A chain job is \emph{covered} if its
inner radius is $0$ or a grid point lies strictly between eight times its inner radius and its reach.
Covered chain jobs with a common long side and a common such grid point have a common witness, so by
\Cref{lem:witness} they have an independent subset with at least two thirds of their LP weight, and \Cref{lem:coverage}
shows how to choose the offsets so
that the covered chain jobs have at least half of the LP weight of $\J_{\mathrm{nr}}$.

For a nonempty set $J\subseteq\J_{\mathrm{nr}}$, let $t(J)$ be the entry of index $|J|$, counted from $0$, in the sorted list
of the $2|J|$ endpoints of the windows of the jobs of $J$, with multiplicity.  For an offset $o$, with its grid
$\mathcal G$ and annuli around $t(J)$, let $C(J,o)$ be the set of jobs $j\in J$ with $\mathcal G\cap(\iota_j,\rho_j)\ne
\varnothing$, seen from $t(J)$.  For a cell $m$ and a side $\pm$, let $J_{m,\pm}(J,o)$ be the set of jobs $j\in J\setminus
C(J,o)$ with $\mathrm{cell}(\iota_j)=m$ whose window lies on side $\pm$ of $t(J)$; every job of $J\setminus C(J,o)$ has
its window on one side of $t(J)$, by \Cref{lem:radii}(c).  \Cref{fig:node} shows these sets.

\begin{definition}[The recursion tree $R$]\label{def:R}
Assume that $\J_{\mathrm{nr}}\ne\varnothing$.  The \emph{recursion tree} $R$ is the rooted tree defined recursively as
follows.  Its root \emph{hosts} $\J_{\mathrm{nr}}$.  A node $v$ that hosts $J_v$ has \emph{separator} $t_v=t(J_v)$ and
an \emph{offset} $o_v\in\{0,\dots,K'-1\}$, with grid $\mathcal G_v=\mathcal G(o_v)$.  Its \emph{chain} is $C_v=C(J_v,o_v)$, and its
\emph{cell sets} are $J_{m,\pm}=J_{m,\pm}(J_v,o_v)$.  For every $(m,\pm)$ with $J_{m,\pm}\ne\varnothing$, the node $v$
has a child $v_{m,\pm}$ that hosts $J_{m,\pm}$.  The node $v$ has no other children.  The \emph{depth} of a node is
its distance from the root.
\end{definition}

\noindent The offsets may be chosen arbitrarily, and each choice gives a tree.  Every child hosts at most half of the
jobs of its parent, by \Cref{lem:shape}(b), whose proof concerns a single node, so the recursion ends and $R$ is
finite.

For a node $v$ of $R$, the set $\operatorname{Occ}(v)$ of \emph{occupied cells} of $v$ consists of the cells $m$ for which $v$ has a child $v_{m,+}$ or $v_{m,-}$, or both; for
$m\in\operatorname{Occ}(v)$, the \emph{children of $v$ in cell $m$} are those of $v_{m,+}$ and $v_{m,-}$ that exist.  Cells, cell sets $J_{m,\pm}$ and
$\operatorname{Occ}(v)$ always refer to the grid of the node $v$ at hand.

\begin{remark}[Which jobs stay at a node]\label{rem:chain}
A job $j\in J_v$ joins the chain $C_v$ if and only if $\mathcal G_v\cap(\delta_j,\rho_j)\ne\varnothing$.
Indeed, $\iota_j\ge\delta_j$, so every chain job satisfies this condition.  Conversely, let $j$
satisfy it.  If $\delta_j\ge s_j/2$, then $\iota_j=\delta_j$, and $j$ joins the chain by the definition of the
chain.  If $\delta_j<s_j/2$, then $j$ joins the chain by
\Cref{lem:radii}(b).  Informally, the chain consists of the jobs whose windows contain $t_v$
or cross a grid point, i.e., a point $t_v\pm\gamma$ with $\gamma\in\mathcal G_v$.
\end{remark}

The tree $R$ depends on the offsets $o_v$; \DPNR{} chooses them in Step~1.  Unless a statement says
otherwise, the offsets are arbitrary, $R$ is the tree they produce, and the statement holds for every choice; Steps~2 and~3 of
\DPNR{} apply to this tree.

Let $v$ be a node of $R$.  Every job of $J_v$ joins either the chain or a cell set, so no job is dropped.  A chain job $j\in C_v$ is \emph{covered} if $\iota_j=0$ or $\mathcal G_v\cap(8\iota_j,\rho_j)\ne\varnothing$.  The algorithm does
not use this notion except to choose the offsets.  The analysis has two parts: it bounds the weight of the schedule from below by
a fraction of the LP weight of the covered jobs in \Cref{sec:guarantee}, and it bounds the LP weight of the covered jobs from below
by half of the LP weight of $\J_{\mathrm{nr}}$, by \Cref{lem:coverage}.

\begin{lemma}[Shape of $R$]\phantomsection\label{lem:shape}
Let $v$ be a node of $R$.
\begin{enumerate}[label=(\alph*),nosep]
\item Every job $j$ hosted by a node in the subtree of a child $v_{m,\pm}$ of $v$ has $K_j\subseteq A^\pm_m\cup\{r_j\}$, where $A^\pm_m$ is
the annulus of $v$.
\item Every child of $v$ hosts at most $|J_v|/2$ jobs.
\item Nodes of the same depth host pairwise disjoint job sets, every node
has depth at most $\ell-1$, and $R$ has at most $n\ell$ nodes.
\item Every job of $\J_{\mathrm{nr}}$ joins the chain of exactly one node, its \emph{home}; the nodes that host the job are those on the path
from the root of $R$ to its home.
\end{enumerate}
\end{lemma}

\begin{proof}
(a) By \Cref{lem:radii}(c), and because descendants of $v_{m,\pm}$ host subsets of $J_{m,\pm}$.

(b) By \Cref{lem:radii}(b) and~(c), a job passed to a child has $\delta_j>0$, so both of its
endpoints lie strictly on one side of $t_v$.  In the sorted list of endpoints that defines
$t_v=t(J_v)$, at most $|J_v|-1$ entries exceed $t_v$ and at most $|J_v|$ are less than
$t_v$.  A job passed to a child on side $+$ has both endpoints greater than $t_v$, and one passed
to a child on side $-$ has both less than $t_v$.  Therefore the children on side $+$ together host at most
$(|J_v|-1)/2$ jobs, and those on side $-$ at most $|J_v|/2$; in particular, every child hosts at most $|J_v|/2$ jobs.

(c) The nodes of depth $k+1$ are the children of the nodes of depth $k$.  The children of one node host its cell sets, which are
pairwise disjoint subsets of its job set; so, by induction on the depth, the nodes of one depth host pairwise disjoint job sets.  By (b), a node of depth $\ell'$ hosts at most $n/2^{\ell'}$ jobs, and at least one, so $\ell'\le\lfloor\log n\rfloor=\ell-1$.  Each
depth has at most $n$ nodes, since their job sets are disjoint and nonempty.

(d) The root hosts every job of $\J_{\mathrm{nr}}$.  At each node that hosts it, a job joins the chain or moves to a child, which hosts fewer jobs.
Thus the nodes that host the job form a path from the root, which ends at the node whose chain it joins.
\end{proof}

\begin{lemma}[A common witness]\label{lem:witness}
Let $G$ be a nonempty set of jobs of $\J_{\mathrm{nr}}$, seen from the same point $t$, with a common long side $\pm$, and
let $B$ be a number with $8\iota_j<B<\rho_j$ for every $j\in G$.  Then $G$ has an independent subset of weight at
least $\frac23\sum_{j\in G}w_jz_j$.
\end{lemma}

\begin{proof}
Let $X=\{x\text{ on side }\pm\text{ of }t:\ B/4\le|x-t|<B\}$, of length $3B/4$.  For $j\in G$, the points of $X$ have
distances $d$ with $\iota_j<B/8<B/4\le d<B<\rho_j$, so $X\subseteq Y_j\subseteq K_j$ by \Cref{lem:radii}(a).  If
$s_j>0$, then $\iota_j\ge s_j/2$, so $B>4s_j$ and $|X|>3s_j$.  Hence $X$ is a $2$-witness for $G$, and
\Cref{lem:group} gives the independent subset.
\end{proof}

\paragraph{Random offsets.}  Label the child $v_{m,\pm}$ of a node by $(m,\pm)$, and give every node of $R$ its \emph{address}, the
sequence of labels on the path from the root to it; the root has the empty address, and the length of the address of a node is its
depth in $R$.  The children of a node have different labels, so different nodes have different addresses.  Let $\Lambda$ be the set of pairs
$(m,\pm)$ for which there are a job $j\in\J_{\mathrm{nr}}$, an endpoint $t$ of a job of $\J_{\mathrm{nr}}$ and an offset $o$ such that, seen from $t$, $\iota_j>0$ and
$m=\mathrm{cell}(\iota_j)$ for the grid of $o$; $\Lambda$ is finite, since there are finitely many triples $(j,t,o)$ and each gives one $m$.  Every
label that occurs lies in $\Lambda$.  Indeed, if a node $v$ has the child $v_{m,\pm}$, then $J_{m,\pm}$ contains a job $j$, and, by the definition of the cell sets
and \Cref{lem:radii}(c), $\iota_j>0$ and $m=\mathrm{cell}(\iota_j)$, seen from $t_v$ and with the grid of $o_v$.
Since $t_v=t(J_v)$ is an endpoint of the window of a job of $\J_{\mathrm{nr}}$, the triple $(j,t_v,o_v)$ shows that $(m,\pm)\in\Lambda$.  Let $\mathcal A$ be the finite set of sequences of elements of $\Lambda$ of length
at most $\ell-1$; by \Cref{lem:shape}(c), $\mathcal A$ contains the address of every node, for every choice of the offsets.  An \emph{outcome}
is a vector $\xi=(\xi_a)_{a\in\mathcal A}$ whose entries are independent and each is chosen uniformly from $\{0,\dots,K'-1\}$.  With \emph{random offsets}, the node with address $a$, if it exists, uses the offset $\xi_a$.  Every
outcome is thus one choice of the offsets.  Random offsets are the random shifts familiar from shifted quadtrees \cite{Ar}
and from tree embeddings of metrics \cite{Ba,FRT}; Fakcharoenphol, Rao and Talwar draw one random scale for all levels
\cite{FRT}, while here every node draws its own offset.  The randomness serves only the exposition, in
\Cref{lem:coverage}(b) and in \Cref{rem:lp} at the end of this section.  The
algorithm itself is deterministic: Step~1 of \DPNR{} chooses at each node the offset that leaves the least LP weight
uncovered, and \Cref{lem:coverage}(c) shows that this choice keeps the bound that random offsets give.

For a point $t$ and a job $j\in\J_{\mathrm{nr}}$ seen from $t$, an offset is \emph{bad} for $(t,j)$ if its grid $\mathcal G(o)$
satisfies $\iota_j>0$, $\mathcal G(o)\cap(\iota_j,\rho_j)\ne\varnothing$ and $\mathcal G(o)\cap(8\iota_j,\rho_j)=\varnothing$; let
$\mathrm{Bad}(t,j)$ be the set of these offsets.  By the definitions of the chain and of covered chain jobs, a job
$j\in J_v$ is an uncovered chain job of a node $v$ exactly when $o_v\in\mathrm{Bad}(t_v,j)$.

\begin{lemma}[Coverage]\label{lem:coverage}
\begin{enumerate}[label=(\alph*),nosep]
\item $|\mathrm{Bad}(t,j)|\le2$ for every point $t$ and every job $j\in\J_{\mathrm{nr}}$.
\item With random offsets, every job of $\J_{\mathrm{nr}}$ is covered at its home with probability more than $1/2$.
\item With the offsets chosen in Step~1 of \DPNR{}, the LP weight of the jobs covered at their homes is at least half of
the LP weight $W_{\mathrm{nr}}$ of $\J_{\mathrm{nr}}$.
\end{enumerate}
\end{lemma}

\begin{proof}
(a) Let $o$ be bad for $(t,j)$.  Some grid point $g$ of $o$ lies in $(\iota_j,\rho_j)$, and it does not lie in $(8\iota_j,\rho_j)$, so
$\iota_j<g\le8\iota_j$.  The grid points of $o$ are the numbers $\gamma_m=4^oQ^m=4^{o+K'm}$, i.e.,
the powers $4^c$ with $c\equiv o\pmod{K'}$, so $g=4^c$ for an integer $c\equiv o\pmod{K'}$ in the interval $(\log_4\iota_j,\ \log_4\iota_j+1.5]$.  This interval has length $1.5$, so it
contains at most two integers.  Distinct bad offsets give distinct integers $c$, since $c$ determines $o$ as its residue modulo $K'$; hence
there are at most two bad offsets.

(b) \emph{The job sets.}  For $a\in\mathcal A$ let $J_a$ be the job set hosted by the node with address $a$, and $J_a=\varnothing$ if there is no
such node.  We show by induction on the length of $a$ that $J_a$ is a function of the entries $\xi_{a'}$ at the proper prefixes $a'$ of $a$.
The root hosts $J_\varnothing=\J_{\mathrm{nr}}$.  Let $a\lambda$ be $a$ followed by a label $\lambda=(m,\pm)$.  If $J_a=\varnothing$, there is no node with
address $a$, hence none with address $a\lambda$, and $J_{a\lambda}=\varnothing$.  Otherwise $J_{a\lambda}$ is the cell set $J_{m,\pm}$ of the node with address $a$, which is determined by the job set $J_a$ of that node, its separator,
which is determined by $J_a$, and its offset $\xi_a$.  The proper prefixes of $a\lambda$ are those of $a$ and $a$ itself.  In particular,
$J_a$ does not depend on $\xi_a$.

\noindent\emph{The events.}  Fix $j\in\J_{\mathrm{nr}}$.  For $a\in\mathcal A$ let $E_a$ be the event that the node with address $a$ exists, hosts $j$, and has $j$ as
an uncovered chain job.  If $E_a$ occurs, $j$ is in the chain of that node, which is then its home, by \Cref{lem:shape}(d), and $j$ is
uncovered at its home.  Conversely, if $j$ is uncovered at its home, $E_a$ occurs for the address $a$ of its home.  Since $j$ has only one
home, the event that $j$ is uncovered at its home is the disjoint union of the events $E_a$, $a\in\mathcal A$, and
\[
 \Pr[j\text{ is uncovered at its home}]=\sum_{a\in\mathcal A}\Pr[E_a].
\]

\noindent\emph{One address.}  Fix $a\in\mathcal A$.  Let $\psi$ be the vector of the entries of $\xi$ at the proper prefixes of $a$, and write $J_a(\pi)$ for
the value of $J_a$ when $\psi=\pi$.  For a value $\pi$ of $\psi$ with $j\notin J_a(\pi)$, the events $E_a$ and $\psi=\pi$ do not occur
together.  Consider a value $\pi$ with $j\in J_a(\pi)$.  Given $\psi=\pi$, the node with address $a$ exists and hosts $J_a(\pi)$, its
separator $t_\pi$ is determined by $J_a(\pi)$.  Since a job is an uncovered chain job of a node exactly when the offset
of the node is bad for the separator and the job, $E_a$ occurs exactly when $\xi_a\in B_\pi$, where
$B_\pi=\mathrm{Bad}(t_\pi,j)$ is a set of at most two offsets, by (a), determined by $\pi$.  The entries of $\xi$ are independent, and $a$ is not a
proper prefix of itself, so $\xi_a$ is independent of $\psi$; it is chosen uniformly, so $\Pr[\xi_a\in B_\pi\mid\psi=\pi]=|B_\pi|/K'\le2/K'$.  Hence, with the sums over the values $\pi$ with $j\in J_a(\pi)$,
\[
 \Pr[E_a]=\sum_{\pi}\Pr[\psi=\pi]\,\Pr[\xi_a\in B_\pi\mid\psi=\pi]\ \le\ \frac2{K'}\sum_{\pi}\Pr[\psi=\pi]\ =\ \frac2{K'}\Pr[j\in J_a].
\]

\noindent\emph{One depth.}  Fix a length $k$.  In every outcome, distinct addresses of length $k$ belong to distinct nodes of depth $k$, if any,
and these host pairwise disjoint job sets, by \Cref{lem:shape}(c); so at most one address $a$ of length $k$ has $j\in J_a$.  The events
$j\in J_a$ with $|a|=k$ are therefore pairwise disjoint, and their probabilities add up to at most $1$.

\noindent\emph{Conclusion.}  Every address in $\mathcal A$ has length at most $\ell-1$, so
\begin{align*}
 \Pr[j\text{ is uncovered at its home}]&=\sum_{k=0}^{\ell-1}\ \sum_{a\in\mathcal A,\ |a|=k}\Pr[E_a]\ \le\ \sum_{k=0}^{\ell-1}\frac2{K'}\sum_{a\in\mathcal A,\ |a|=k}\Pr[j\in J_a]\\
 &\ \le\ \frac{2\ell}{K'}\ =\ \frac{2\ell}{4\ell+4}\ <\ \frac12 .
\end{align*}

(c) For a set $S$ of jobs let $\overline W(S)=\sum_{j\in S}w_jz_j$, so that $W_{\mathrm{nr}}=\overline W(\J_{\mathrm{nr}})$.  Step~1 of \DPNR{} builds $R$ from the root down, so $J_v$ and $t_v$
are known when $o_v$ is chosen.  For an offset $o$, a job $j\in J_v$ is in the chain of $v$ exactly when the grid of $o$ meets
$(\iota_j,\rho_j)$; so the quantity minimized in Step~1 is
\[
 \mathrm{Loss}_v(o)=\sum\bigl\{w_jz_j:\ j\in J_v,\ o\in\mathrm{Bad}(t_v,j)\bigr\},
\]
the LP weight of the jobs that $o$ would make uncovered chain jobs of $v$.  Exchanging the order of summation and using (a) and
$w_jz_j\ge0$, we obtain
\[
 \sum_{o=0}^{K'-1}\mathrm{Loss}_v(o)=\sum_{j\in J_v}w_jz_j\,|\mathrm{Bad}(t_v,j)|\ \le\ 2\overline W(J_v).
\]
Since $o_v$ minimizes $\mathrm{Loss}_v$, $K'\,\mathrm{Loss}_v(o_v)\le\sum_o\mathrm{Loss}_v(o)$, so $\mathrm{Loss}_v(o_v)\le\frac2{K'}\overline W(J_v)$.  By
\Cref{lem:shape}(d), a job is a chain job of exactly one node, its home.  Consequently, the set of jobs that are uncovered at their homes is the
disjoint union, over all nodes $v$ of $R$, of the sets of uncovered chain jobs of $v$, and its LP weight is
$\sum_v\mathrm{Loss}_v(o_v)\le\frac2{K'}\sum_v\overline W(J_v)$.  By \Cref{lem:shape}(c), the nodes of one depth host pairwise disjoint subsets of
$\J_{\mathrm{nr}}$, and there are at most $\ell$ depths, so $\sum_v\overline W(J_v)\le\ell\,W_{\mathrm{nr}}$.  Hence the jobs uncovered at their homes have LP weight at most
$\frac{2\ell}{K'}W_{\mathrm{nr}}\le\frac12W_{\mathrm{nr}}$.  Every job of $\J_{\mathrm{nr}}$ has a home, by \Cref{lem:shape}(d), and is covered or uncovered there, so the jobs
covered at their homes have LP weight at least $\frac12W_{\mathrm{nr}}$.
\end{proof}
%---------------------------------------------------------------------
\subsection{The dynamic program}\label{sec:dpalg}

\begin{figure}[t]
\centering
\begin{tikzpicture}[x=1.1cm,y=0.55cm,font=\small]
 % cells between consecutive grid points; cells of Occ(v) in orange
 \foreach \m in {3,4,8} {\fill[orange!35] (\m,0) rectangle (\m+1,0.5);}
 \foreach \m in {0,...,9} {\draw[gray] (\m,0) rectangle (\m+1,0.5);}
 \node[font=\scriptsize,anchor=east] at (-0.1,0.25) {cells};
 \node[font=\scriptsize,orange!50!black] at (4,-0.35) {children of $v$};
 \node[font=\scriptsize,orange!50!black] at (8.5,-0.35) {children of $v$};
 % points of E_v
 \draw[gray] (-0.3,-0.85) -- (10.3,-0.85);
 \foreach \x/\l in {0/0,1/1,2/2,3/3,4/4,5/5,6/6,8/7,9/8} {\draw[thick] (\x,-0.75) -- (\x,-0.95); \node[font=\scriptsize] at (\x,-1.35) {$e_{\l}$};}
 \node[font=\scriptsize,anchor=east] at (-0.1,-0.85) {$E_v$};
 % busy-distance intervals of the chain jobs; solid = chosen, outlined = left out
 \foreach \a/\b/\y/\lab/\ch in {0/2/1/A/1,1/3/1.6/B/0,0/3/2.2/C/1,2/6/2.8/D/0,5/8/1/E/1,6/8/1.6/F/1} {
   \ifnum\ch=1 \fill[blue!35] (\a,\y) rectangle (\b,\y+0.4); \fi
   \draw[blue!60] (\a,\y) rectangle (\b,\y+0.4);
   \node[font=\scriptsize] at ({(\a+\b)/2},\y+0.2) {$\lab$};
 }
 \node[font=\scriptsize,anchor=east,blue!60!black] at (-0.1,2.0) {chain jobs};
 % the packing taken: each block drawn as a frame with its content
 \foreach \a/\b/\col in {0/3/blue!60,5/8/blue!60,3/4/orange!70!black,4/5/orange!70!black,8/9/orange!70!black}
   {\draw[\col,thick] (\a+0.04,-3.25) rectangle (\b-0.04,-1.85);}
 \foreach \a/\b/\y/\lab in {0/2/-2.35/{A\ \to\ 1},0/3/-2.95/{C\ \to\ 2},5/8/-2.35/{E\ \to\ 1},6/8/-2.95/{F\ \to\ 2}} {
   \fill[blue!35] (\a+0.1,\y-0.2) rectangle (\b-0.1,\y+0.2);
   \node[font=\scriptsize] at ({(\a+\b)/2},\y) {$\lab$};
 }
 \foreach \m in {3,4,8} {\node[font=\scriptsize,orange!50!black,align=center] at (\m+0.5,-2.55) {$S$};}
 \node[font=\scriptsize,anchor=east] at (-0.1,-2.55) {packing};
 % a block that is not taken
 \draw[dashed,blue!60,thick] (2,-4.25) rectangle (6,-3.55);
 \node[font=\scriptsize] at (4,-3.9) {block $(2,6)$: only $D$};
 \node[font=\scriptsize,anchor=east] at (-0.1,-3.9) {not taken};
\end{tikzpicture}
\caption{The dynamic program at a node, on the line of distances from $t_v$ (logarithmic scale; a distance stands for the two points at
that distance on either side of $t_v$); the cells of $\operatorname{Occ}(v)$ are orange.  Above: the busy-distance intervals of the chain
jobs.  Below: the points $e_0<\dots<e_8$ of $E_v$; not every grid point is one of them.  The packing taken consists of the blocks $(0,3)$
and $(5,7)$ and the one-cell blocks of the three occupied cells, which contribute the schedules $S$ of the children.  Each chain block carries a
maximum-weight independent set of the chain jobs inside it, matched to distinct machines (the numbers); outlined jobs are left out.  Different blocks may
use the same machine.  The block $(2,6)$ contains the occupied cells $3$ and $4$ but would carry only chain jobs, here only $D$; it is not
taken.}
\label{fig:dp}
\end{figure}
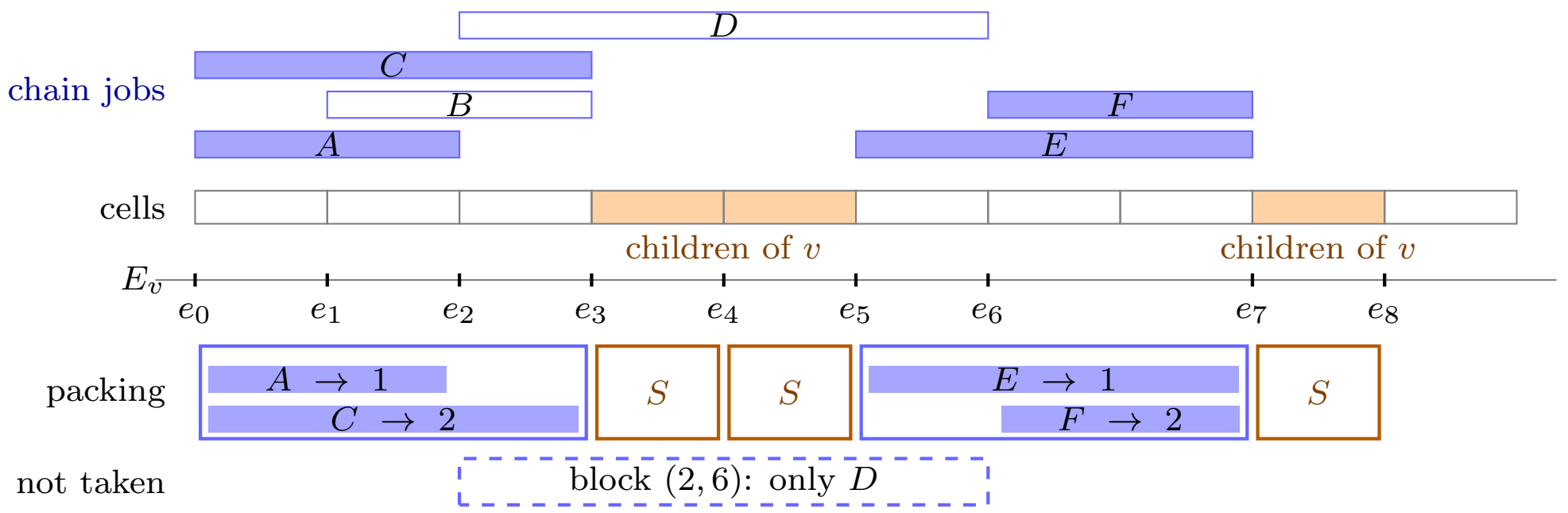

This subsection gives \DPNR{}, see \Cref{fig:dp}.  At every node $v$, from the leaves of $R$ up, it chooses a packing, that is, a set of
blocks, which are intervals of distances from $t_v$ that meet at most in endpoints.  Each block contributes to the
schedule either an independent set of chain jobs of $v$ or the schedules already computed for the children of $v$ in
one cell.
\Cref{lem:recursion} shows that, given the values already computed for the children of $v$, the dynamic program finds a packing at $v$
of maximum value, and \Cref{lem:dp-feasible} that the result is a nonmigratory schedule.

Throughout, $v$ is a node of $R$ with separator $t=t_v$ and grid $\mathcal G_v=\{\gamma_m\}$, and all distances and cells are taken
from $t_v$.  Blocks and packings describe each job only by the distances from $t_v$ at which it can be busy, not by its side or sides.  This suffices, since jobs busy at disjoint sets of distances are never busy at the same time, and it covers the jobs whose windows contain
$t_v$, which are busy on both sides at the same distances.  It is not tight for one-sided jobs on opposite sides of $t_v$: they never
run at the same time, but if their busy-distance intervals overlap in more than one point, only the same chain block can carry them, and
then they need distinct machines.  Here the analysis incurs a factor $2$, in Step~2
of the proof of \Cref{lem:local}, which counts the groups of both sides together.

\begin{definition}[Busy-distance intervals]\label{def:cellint}
Let $v$ be a node of $R$ and $j\in C_v$.  Let $b_j=\mathrm{cell}^-(\rho_j)$, and, if $\iota_j>0$, let $a_j=\mathrm{cell}(\iota_j)$.  The \emph{busy-distance
interval} of $j$ is the interval of distances
\[
 I_j=[\gamma_{a_j},\ \gamma_{b_j+1}]\ \ \text{if }\iota_j>0,\qquad\quad I_j=[0,\ \gamma_{b_j+1}]\ \ \text{if }\iota_j=0 .
\]
\end{definition}

Hence $I_j$ is the interval $[\iota_j,\rho_j]$ with its left end rounded down to a grid point, or to $0$, and its right end rounded up to a
grid point: $\gamma_{a_j}\le\iota_j<\gamma_{a_j+1}$ and $\gamma_{b_j}<\rho_j\le\gamma_{b_j+1}$.  The algorithm uses $I_j$ only to decide
which chain jobs may be scheduled together; this suffices because, by \Cref{lem:radii}(a), the support of $j$ lies at
distances in $[\iota_j,\rho_j]$ from $t_v$, and $[\iota_j,\rho_j]\subseteq I_j$, since $\gamma_{a_j}\le\iota_j$ when $\iota_j>0$ and
$\rho_j\le\gamma_{b_j+1}$.  A chain job has a grid point strictly between $\iota_j$ and $\rho_j$, so its busy-distance interval has a grid point in its
interior; in particular, it does not lie in $[\gamma_m,\gamma_{m+1}]$ for any cell $m$.

\begin{definition}[Blocks and packings]\label{def:recursion}
Let $v$ be a node of $R$.
\begin{enumerate}[label=\arabic*.,nosep]
\item \emph{Points.}  Let $e_0<e_1<\dots<e_\kappa$ be the elements of the set
\[
 E_v=\{\text{the two endpoints of }I_j:\ j\in C_v\}\ \cup\ \{\gamma_m,\gamma_{m+1}:\ m\in\operatorname{Occ}(v)\}.
\]
Each $e_l$ is a grid point or $0$, but not every grid point is in $E_v$.
\item \emph{Blocks.}  A \emph{block} at $v$ is a pair $(i,k)$ with $0\le i<k\le\kappa$; its \emph{interval} is $[e_i,e_k]$.  Since
$e_0<\dots<e_\kappa$, a block is determined by its interval.
\item \emph{One-cell blocks.}  If $[e_i,e_k]=[\gamma_m,\gamma_{m+1}]$ for a cell $m$, then $(i,k)$ is a \emph{one-cell block}, the one-cell
block of $m$, and $\mathrm{val}_v(i,k)=f(v_{m,+})+f(v_{m,-})$, where an absent child contributes $0$ and $f(u)$ is the value computed
for a child $u$ in Step~2 of \DPNR{}, which treats the children of $v$ before $v$.
\item \emph{Chain blocks.}  Every other block $(i,k)$ is a \emph{chain block}.  Let $C_v(i,k)$ be the set of chain jobs $j$ with
$I_j\subseteq[e_i,e_k]$, and let $T_v(i,k)$ be the independent subset of $C_v(i,k)$ of maximum weight found by the greedy
algorithm, with ties broken by a fixed order of the jobs.  The block \emph{carries} the jobs of $T_v(i,k)$, and
\[
 \mathrm{val}_v(i,k)=\sum_{j\in T_v(i,k)}w_j=\max\Bigl\{\sum_{j\in T}w_j:\ T\subseteq C_v(i,k)\text{ independent}\Bigr\}.
\]
For every block we also write $\mathrm{val}_v([e_i,e_k])$ for $\mathrm{val}_v(i,k)$.
\item \emph{Packings.}  A \emph{packing} at $v$ is a set of blocks whose intervals pairwise meet in at most one point.  Its \emph{value} is the
sum of the values $\mathrm{val}_v$ of its blocks.
\end{enumerate}
\end{definition}

The set $E_v$ has at least two elements, so $\kappa\ge1$: since $J_v$ is nonempty, there is a chain job, whose busy-distance interval has two
distinct endpoints, or a child of $v$ in some cell $m$, which contributes $\gamma_m$ and $\gamma_{m+1}$.  A one-cell block $(i,k)$ has $k=i+1$,
since no grid point lies strictly between $\gamma_m$ and $\gamma_{m+1}$.  A busy-distance interval has a grid point in its interior,
so it lies in no one-cell block; as a chain block, a one-cell block would therefore carry no chain job; if $m\notin\operatorname{Occ}(v)$, the one-cell block of $m$, if
it exists, has value $0$.  The interval of a chain block may contain cells of $\operatorname{Occ}(v)$; the children in such a cell $m$ are used
only through the one-cell block of $m$, and the one-cell blocks of adjacent cells meet only at a grid point, so a packing can use them side by
side.  The
two kinds of block are kept separate because chain jobs and the jobs of the children may be busy at the same distances.

The sets $C_v(i,k)$ include the uncovered chain jobs.  The algorithm uses coverage only in Step~1 of \DPNR{}, to choose the
offsets.

\begin{center}
\fbox{\begin{minipage}{0.94\textwidth}
\smallskip
\noindent\phantomsection\label{alg:dpnr}\textbf{Algorithm} \textnormal{\textsc{DP-NR}}
\begin{enumerate}[label=\arabic*.,nosep]
\item \emph{Tree and offsets.}  If $\J_{\mathrm{nr}}=\varnothing$, output the empty schedule and stop.  Otherwise build $R$ from the root down.  At each node $v$, choose the offset
\[
 o_v\in\arg\min_{o\in\{0,\dots,K'-1\}}\ \sum\Bigl\{w_jz_j:\ j\in C_v,\ \iota_j>0,\ \mathcal G_v\cap(8\iota_j,\rho_j)=\varnothing\Bigr\},
\]
where $C_v$ and $\mathcal G_v$ are those of the offset $o$; that is, minimize the LP weight of the uncovered chain jobs of $v$.
\item \emph{Values}, from the leaves of $R$ up.  For each node $v$, compute $\mathrm{val}_v(i,k)$ for every block $(i,k)$, and $T_v(i,k)$ for
every chain block; then
\begin{equation*}
 \mathrm{DP}_v(0)=0,\qquad \mathrm{DP}_v(k)=\max\Bigl\{\mathrm{DP}_v(k-1),\ \max_{0\le i<k}\bigl[\mathrm{DP}_v(i)+\mathrm{val}_v(i,k)\bigr]\Bigr\}\quad(1\le k\le\kappa),
\tag{DP}\label{eq:DP}
\end{equation*}
and set $f(v)=\mathrm{DP}_v(\kappa)$.
\item \emph{Schedules}, from the leaves of $R$ up.  For each node $v$, trace back the maximizing choices in~\eqref{eq:DP} from $k=\kappa$ to
obtain a packing $P_v$ at $v$.  The schedule $S(v)$ is the set of the following pieces.
\begin{itemize}[nosep]
\item For the one-cell block of a cell $m$ in $P_v$: the pieces of the schedules $S(v_{m,+})$ and $S(v_{m,-})$, already computed, of
those of the two children that exist.
\item For a chain block $(i,k)$ in $P_v$: match the jobs it carries, $T_v(i,k)$, to distinct eligible machines, and run each
$j\in T_v(i,k)$ on its machine during the earliest $p_j$ units of its support $Y_j$ seen from $t_v$.
\end{itemize}
\item \emph{Output} $S(\mathrm{root})$.
\end{enumerate}
\smallskip
\end{minipage}}
\end{center}

\begin{lemma}[Packings of maximum value]\label{lem:recursion}
Let $v$ be a node of $R$, and let the values of the one-cell blocks at $v$ be given.  Then the number $f(v)$ computed in
Step~2 of \DPNR{} is the maximum value of a packing at $v$, and the packing $P_v$ obtained in Step~3 of \DPNR{} has value
$f(v)$.
\end{lemma}

\begin{proof}
For $0\le k\le\kappa$ let $M(k)$ be the maximum value of a packing whose intervals lie in $[e_0,e_k]$; there are finitely many such
packings, and the empty packing has value $0$.  We show $\mathrm{DP}_v(k)=M(k)$ by induction on $k$.  For $k=0$, no interval of a block lies in
$[e_0,e_0]$, since a block $(i,k')$ has $e_i<e_{k'}$; so $M(0)=0=\mathrm{DP}_v(0)$.  Let $k\ge1$.

\emph{$M(k)\le \mathrm{DP}_v(k)$.}  Let $P$ be a packing whose intervals lie in $[e_0,e_k]$.  If no interval of $P$ ends at $e_k$, all of them lie
in $[e_0,e_{k-1}]$, and the value of $P$ is at most $M(k-1)=\mathrm{DP}_v(k-1)$.  Otherwise some block $(i,k)$ of $P$ has its interval ending at
$e_k$.  Let $(i',k')$ be another block of $P$.  Then $k'<k$, since two intervals ending at $e_k$ both contain $[e_{k-1},e_k]$.  If $i'\ge i$,
the interval $[e_{i'},e_{k'}]$ would lie in $[e_i,e_k]$ and meet it in more than one point; so $i'<i$, and as $[e_{i'},e_{k'}]$ meets
$[e_i,e_k]$ in at most one point, $k'\le i$.  Therefore the other blocks of $P$ form a packing whose intervals lie in $[e_0,e_i]$, of value at most
$M(i)=\mathrm{DP}_v(i)$, and the value of $P$ is at most $\mathrm{DP}_v(i)+\mathrm{val}_v(i,k)$.  In both cases it is at most $\mathrm{DP}_v(k)$.

\emph{$M(k)\ge \mathrm{DP}_v(k)$.}  A packing whose intervals lie in $[e_0,e_{k-1}]$ also has them in $[e_0,e_k]$, so $M(k)\ge M(k-1)=\mathrm{DP}_v(k-1)$.  For
$0\le i<k$, adding the block $(i,k)$ to a packing of value $M(i)$ whose intervals lie in $[e_0,e_i]$ gives a packing, since $[e_i,e_k]$ meets
each of these intervals in at most the point $e_i$; so $M(k)\ge \mathrm{DP}_v(i)+\mathrm{val}_v(i,k)$.  Hence $M(k)\ge \mathrm{DP}_v(k)$.

Finally, every interval of a block lies in $[e_0,e_\kappa]$, so $f(v)=\mathrm{DP}_v(\kappa)=M(\kappa)$ is the maximum value of a packing at $v$.  Tracing
back the maximizing choices in~\eqref{eq:DP} repeats the constructions of the second part of the proof, so it yields a packing of value
$\mathrm{DP}_v(\kappa)=f(v)$.
\end{proof}

\begin{lemma}[Feasibility]\label{lem:dp-feasible}
For every node $v$ of $R$, the set $S(v)$ computed in Step~3 of \DPNR{} is a nonmigratory schedule of weight $f(v)$.  Every job $j$ that
has a piece in $S(v)$ is a chain job of a node $u$ in the subtree of $v$, and $S(v)$ runs $j$ on one machine during the earliest $p_j$
units of its support $Y_j$ seen from $t_u$.
\end{lemma}

\begin{proof}
By induction from the leaves of $R$.  Let $P_v$ be the packing obtained at $v$ in Step~3 of \DPNR{}; its value is $f(v)$ by
\Cref{lem:recursion}.  A job $j$ with a piece in $S(v)$ is \emph{taken from} a chain block $(i,k)$ of $P_v$ if $j\in T_v(i,k)$, and from
the one-cell block of a cell $m$ in $P_v$ if $j$ has a piece in $S(v_{m,+})$ or in $S(v_{m,-})$.  For an interval $D$ of distances, let
$\widehat D=\{x\in\mathbb R:\ |x-t_v|\in D\}$ be the set of times at distance in $D$ from $t_v$.

\noindent\emph{Step 1: every job is taken from exactly one block.}  Let $j$ have a piece in $S(v)$.  If $j$ is taken from a chain block, then
$j\in C_v$.  If $j$ is taken from the one-cell block of a cell $m$, then, by induction, $j$ is a chain job of a node in the subtree of
$v_{m,+}$ or of $v_{m,-}$, so it lies in the cell set $J_{m,+}$ or $J_{m,-}$ of $v$, since every node of that subtree hosts a subset of
it.  The chain $C_v$ and the cell sets of $v$ are pairwise disjoint.  Thus $j$ is taken either only from chain
blocks, or only from the one-cell block of a single cell $m$ and then from one of the two children in cell $m$.  Finally, no two chain
blocks of $P_v$ carry $j$: both would have $I_j$ in their intervals, which is impossible, since these intervals meet in at most one point,
as $P_v$ is a packing, while $I_j$ has positive length.

\noindent\emph{Step 2: the pieces of one job.}  Let $j$ have a piece in $S(v)$.  If $j$ is taken from a chain block, then $S(v)$ runs $j$ on the
eligible machine to which it is matched, during the earliest $p_j$ units of $Y_j$ seen from $t_v$; these are defined because $Y_j$ is a
support of $j$, by \Cref{lem:radii}(a).  If $j$ is taken from a one-cell block, then the pieces of $j$ in $S(v)$ are, by Step~1, its pieces in the
schedule of one child, and by induction that schedule runs $j$ on one machine during the earliest $p_j$ units of its support seen
from $t_u$, where $u$ is the node whose chain contains $j$.  In both cases, by \Cref{def:run}, the pieces of $j$ in $S(v)$ are on
one machine, eligible for $j$; their intervals are pairwise disjoint, lie in a support of $j$, hence in $K_j$, and have total length $p_j$.

\noindent\emph{Step 3: the pieces of a job lie at distances in the interval of its block.}  Let $j$ be taken from a block of $P_v$ with interval $D$.
We show that the interval of every piece of $j$ lies in $\widehat D$.
\begin{itemize}[nosep]
\item If the block is a chain block $(i,k)$, then $D=[e_i,e_k]$, and, by Step~2, the pieces of $j$ lie in $Y_j$, whose points are at distance in
$[\iota_j,\rho_j]$ from $t_v$, by \Cref{lem:radii}(a).  The left end of $I_j$ is at most $\iota_j$: it is $\gamma_{a_j}\le\iota_j$ if
$\iota_j>0$, since $a_j=\mathrm{cell}(\iota_j)$, and it is $0$ if $\iota_j=0$.  Also $\rho_j\le\gamma_{b_j+1}$, since $b_j=\mathrm{cell}^-(\rho_j)$.  Consequently,
$[\iota_j,\rho_j]\subseteq I_j$, and $I_j\subseteq D$ because $j\in C_v(i,k)$.
\item If the block is the one-cell block of a cell $m$, then $D=[\gamma_m,\gamma_{m+1}]$, and, by Step~1, $j$ lies in $J_{m,+}$ or in
$J_{m,-}$.  Hence, seen from $t_v$, no grid point lies in $(\iota_j,\rho_j)$ and $m=\mathrm{cell}(\iota_j)$, by the definition of the cell sets, and
$\delta_j=\iota_j$, by \Cref{lem:radii}(c).  Hence $\gamma_m\le\delta_j<\gamma_{m+1}$, and, since $\gamma_{m+1}$ does not lie in
$(\iota_j,\rho_j)$, also $\rho_j\le\gamma_{m+1}$.  Every point of $K_j$ is at distance in $[\delta_j,\rho_j]\subseteq D$ from $t_v$, and, by Step~2, the pieces of $j$ lie in $K_j$.
\end{itemize}

\noindent\emph{Step 4: pieces of different jobs on the same machine have disjoint intervals.}  Let $j\ne j'$ have pieces on the same machine,
with intervals $U$ and $U'$.
\begin{itemize}[nosep]
\item \emph{Both jobs are taken from the same chain block.}  Impossible: the jobs carried by a chain block are matched to distinct machines.
\item \emph{Both jobs are taken from the one-cell block of a cell $m$, both in $J_{m,+}$ or both in $J_{m,-}$.}  Both pieces belong to the schedule of
the same child, which is a schedule by induction, so $U$ and $U'$ are disjoint.
\item \emph{Both jobs are taken from the one-cell block of a cell $m$, with $j\in J_{m,+}$ and $j'\in J_{m,-}$.}  By the definition of the
cell sets, $K_j$ lies on side $+$ of $t_v$ and $K_{j'}$ on side $-$.  By Step~2, the pieces lie in the windows, so $U$ and
$U'$ are disjoint.
\item \emph{The two jobs are taken from different blocks, with intervals $D$ and $D'$.}  By Step~3, $U\cap U'\subseteq\widehat D\cap\widehat{D'}
=\widehat{D\cap D'}$.  Since $P_v$ is a packing, $D\cap D'$ is empty or a single distance $e$, so $\widehat{D\cap D'}$ is empty or
contained in $\{t_v-e,\,t_v+e\}$.  If $U$ and $U'$ met, their intersection would be a half-open
interval of positive length, which has more than two points.  Therefore $U$ and $U'$ are disjoint.
\end{itemize}

\noindent\emph{Step 5: conclusion.}  $S(v)$ is a finite set of pieces.  Two of its pieces with the same job have disjoint intervals by Step~2, and
two with the same machine but different jobs by Step~4; so $S(v)$ is a schedule.  By Step~2, it is nonmigratory, every job with a piece in
$S(v)$ is completed, and these jobs are run as the lemma states.  By Step~1, the weight of $S(v)$ is the sum, over the blocks of $P_v$, of
the weights of the jobs taken from them: $\mathrm{val}_v(i,k)$ for a chain block $(i,k)$ and, by induction, $f(v_{m,+})+f(v_{m,-})$ for the
one-cell block of a cell $m$, where an absent child contributes $0$.  This sum is the value of $P_v$, which is $f(v)$.
\end{proof}

%---------------------------------------------------------------------
\subsection{The weight of the schedule}\label{sec:guarantee}

This subsection proves \Cref{thm:near}.  For every choice of the offsets, the schedule $S(\mathrm{root})$ has weight at least
the LP weight of the jobs covered at their homes divided by $\Psi(n)$, where $\Psi$ is defined below, and
hence at least $W_{\mathrm{nr}}/(62\log n+38)$ with the offsets chosen in Step~1 of \DPNR{}.
The proof shows $f(v)\ge W_{\mathrm{cov}}(v)/\Psi(|J_v|)$ for every node $v$ by induction from the leaves of $R$ up.  The induction step,
\Cref{lem:local}, constructs a fractional packing at $v$ of value at least $W_{\mathrm{cov}}(v)/\Psi(|J_v|)$, and \Cref{lem:tu} shows that $f(v)$ is at least that value.
The fractional packing uses groups of chain jobs with common witnesses, which \Cref{def:anchor,def:groups} and
\Cref{lem:anchor,lem:groups} provide.

For a node $u$ of $R$, let $W_{\mathrm{cov}}(u)$ be the LP weight $\sum w_jz_j$ of the jobs whose home lies in the subtree of $u$
and that are covered there, and let $C^{\mathrm{cov}}_u$ be the covered chain jobs of $u$.  Then
\begin{equation}\label{eq:wcov}
 W_{\mathrm{cov}}(u)=\sum_{j\in C^{\mathrm{cov}}_u}w_jz_j+\sum_{u'\text{ child of }u}W_{\mathrm{cov}}(u'),
\end{equation}
since the subtree of $u$ consists of $u$ and the subtrees of its children, which are pairwise disjoint, and every job has exactly one
home, by \Cref{lem:shape}(d).
For $N\ge1$ let
\[
 \Psi(N)=31\log N+19 .
\]

Let $v$ be a node of $R$, and let $\omega$ assign a real number $\omega_{ik}\ge0$, the \emph{multiplicity} of the block $(i,k)$, to every block
$(i,k)$ at $v$.  The \emph{load} of $\omega$
at a distance $r$ is the sum of $\omega_{ik}$ over the blocks $(i,k)$ whose interval contains $r$.  Each distance has capacity $1$: $\omega$ is a
\emph{fractional packing} at $v$ if its load at every distance $r>0$ that is not a grid point is at most $1$.  A packing gives multiplicity $1$
to its blocks and $0$ to the others; this is a fractional packing, since the intervals of its blocks meet at most in endpoints, which are grid
points or $0$.

The grid points and $0$ are left out of this condition because the intervals of two blocks of a packing may share an endpoint, and the
load there is then $2$.

\begin{lemma}[Fractional packings]\label{lem:tu}
Let $v$ be a node of $R$.  For every fractional packing $\omega$ at $v$, $f(v)\ge\sum_{(i,k)}\omega_{ik}\,\mathrm{val}_v(i,k)$, the sum over all blocks at $v$.
\end{lemma}

\begin{proof}
Let $A$ be the matrix with a row for each $l=0,\dots,\kappa-1$ and a column for each block $(i,k)$, whose entry is $1$ if $i\le l<k$, i.e.,
if the interval of the block contains $(e_l,e_{l+1})$, and $0$ otherwise.  The open interval $(e_l,e_{l+1})$ contains a distance $r>0$ that is
not a grid point, and the blocks whose interval contains $r$ are exactly those with $i\le l<k$, since all endpoints of intervals of blocks lie in
$E_v$.  Thus the load of $\omega$ at $r$ is the $l$-th entry of $A\omega$, and $A\omega\le1$.  The ones of every column of $A$ lie in consecutive rows, so $A$
is an interval matrix and totally unimodular.  Hence the polytope $\{\omega\ge0:A\omega\le1\}$ has integral vertices.  Every column has an entry $1$,
since $i<k$, so every point of the polytope has coordinates at most $1$, and an integral point is the indicator vector of a set of blocks
no two of which have an entry $1$ in a common row.  The intervals of two such blocks meet in at most one point: two intervals with endpoints
in $E_v$ that overlap in more than one point both contain some $(e_l,e_{l+1})$.  Consequently, the vertices of the polytope are indicator vectors of
packings.  The polytope is bounded, so the linear function $\sum_{(i,k)}\omega_{ik}\,\mathrm{val}_v(i,k)$ attains its maximum over it at a vertex,
and that maximum is at most $f(v)$ by \Cref{lem:recursion}.
\end{proof}

To partition the covered chain jobs into groups, the induction step, \Cref{lem:local}, places the items defined below in $[0,1)$ and gives each covered chain job
with $\iota_j>0$ a subinterval and a dyadic point in it; the algorithm does not compute these objects.

\begin{definition}[Positions and split points]\label{def:anchor}
Let $v$ be a node of $R$.
\begin{enumerate}[label=\arabic*.,nosep]
\item \emph{The cell $a'_j$.}  For a covered chain job $j\in C_v$ with $\iota_j>0$ let $a'_j=\mathrm{cell}(8\iota_j)$.  A grid point lies in $(8\iota_j,\rho_j)$, so
$a'_j<b_j$, where $b_j=\mathrm{cell}^-(\rho_j)$.
\item \emph{Items.}  The \emph{items} of $v$ are the following cells, in general not consecutive ones: the cells $a'_j$ and $b_j$ of every
covered chain job $j\in C_v$ with $\iota_j>0$, and the cells of $\operatorname{Occ}(v)$; a cell is a single item even if it arises in several of these ways.  Each item $m$ has \emph{frequency} $\mu(m)=1+\sum_\pm\bigl(|J_{m,\pm}|+|J_{m-1,\pm}|\bigr)$, where
$|J_{m,\pm}|=0$ if there is no such child; the cell $m-1$ need not be an item.  Let $\mu_v=\sum_m\mu(m)$, the sum over the items $m$.
\item \emph{Positions.}  Item $m$ has \emph{position} $c_m=\bigl(\sum_{m'<m}\mu(m')+\mu(m)/2\bigr)/\mu_v\in(0,1)$, the sum over the items $m'<m$.
Cells that are not items have no frequency and no position.
\item \emph{Levels and split points.}  A \emph{dyadic point of level $k\ge1$} is a number $i/2^k$ with $i$ odd.  A multiple of $2^{-k}$
in $(0,1)$ is a dyadic point of level at most $k$.  For a covered chain job $j\in C_v$ with $\iota_j>0$ let $P_j=(c_{a'_j},\,c_{b_j}]$, a
subinterval of $(0,1)$ of positive length, since $a'_j<b_j$ and positions increase with the cell.  The \emph{level} of $j$ is
\[
 k_j=\min\{k\ge1:\ P_j\text{ contains a multiple of }2^{-k}\},
\]
which exists because $P_j$ has positive length.  The \emph{split point} $q_j$ of $j$ is the largest multiple of $2^{-k_j}$ in $P_j$.
For a split point $q$ of some covered chain job, let $m(q)$ be the least item $m$ with $c_m\ge q$.
\end{enumerate}
\end{definition}

\begin{lemma}[The split point of a job]\label{lem:split}
Let $v$ be a node of $R$, and let $j\in C_v$ be a covered chain job with $\iota_j>0$.  The split point $q_j$ is a dyadic point of level
$k_j$, and it is the only dyadic point of level at most $k_j$ in $P_j$.
\end{lemma}

\begin{proof}
Every dyadic point of level at most $k_j$ is a multiple of $2^{-k_j}$, so it suffices to show that $q_j$ is the only multiple of
$2^{-k_j}$ in $P_j$ and that its numerator is odd.  If $k_j=1$, then $q_j=\frac12$, since $\frac12$ is the only multiple of $\frac12$ in
$(0,1)$, and its numerator is $1$.  Let $k_j\ge2$.  If $P_j$ contained two multiples of $2^{-k_j}$, it would contain two consecutive ones,
since $P_j$ is an interval; one of these has an even numerator, so it would be a multiple of $2^{-(k_j-1)}$ in $P_j$, contrary to the choice
of $k_j$.  For the same reason, the numerator of $q_j$ is odd.
\end{proof}

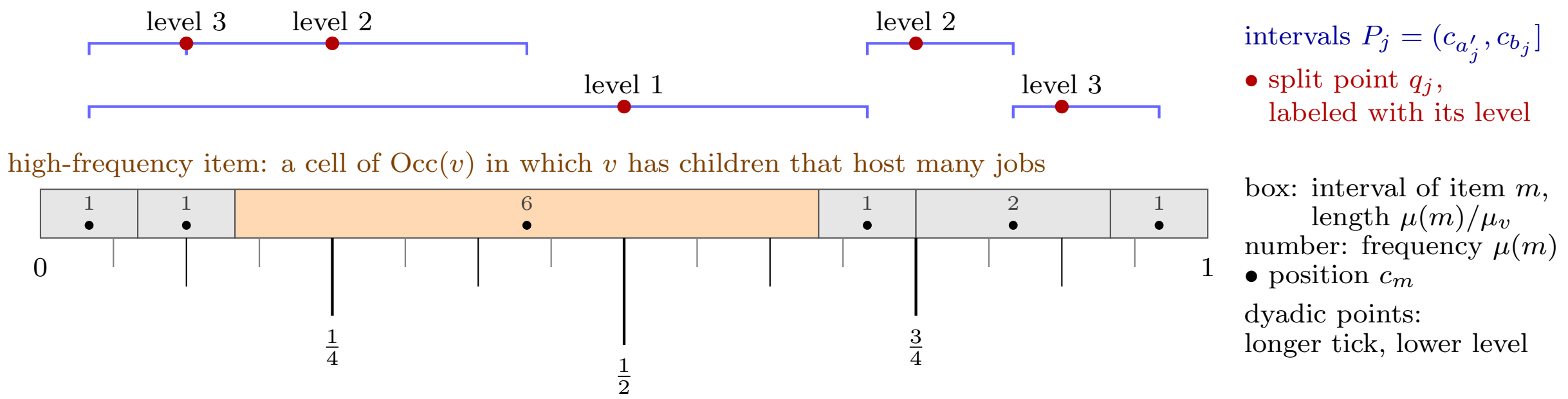
\begin{figure}[t]
\centering
\begin{tikzpicture}[x=12cm,y=1cm,font=\small]
 % items as boxes of widths proportional to frequencies 1,1,6,1,2,1 (total 12)
 \foreach \a/\b/\col/\w in {0/1/gray!20/1,1/2/gray!20/1,2/8/orange!30/6,8/9/gray!20/1,9/11/gray!20/2,11/12/gray!20/1} {
   \fill[\col] (\a/12,0) rectangle (\b/12,0.5);
   \draw[gray!70!black] (\a/12,0) rectangle (\b/12,0.5);
   \node[font=\tiny,gray!50!black] at ({(\a+\b)/24},0.36) {$\w$};
 }
 \foreach \c in {0.5,1.5,5,8.5,10,11.5} {\fill (\c/12,0.13) circle (1.3pt);}
 \node[font=\scriptsize,anchor=west] at (1.02,0.5) {box: interval of item $m$,};
 \node[font=\scriptsize,anchor=west] at (1.02,0.2) {\phantom{box: }length $\mu(m)/\mu_v$};
 \node[font=\scriptsize,anchor=west] at (1.02,-0.1) {number: frequency $\mu(m)$};
 \node[font=\scriptsize,anchor=west] at (1.02,-0.4) {$\bullet$ position $c_m$};
 \node[font=\scriptsize,orange!50!black] at (5/12,0.75) {high-frequency item: a cell of $\operatorname{Occ}(v)$ in which $v$ has children that host many jobs};
 \node[font=\scriptsize] at (0,-0.3) {$0$}; \node[font=\scriptsize] at (1,-0.3) {$1$};
 % dyadic ticks: height decreasing with level
 \draw[thick] (0.5,0) -- (0.5,-1.1) node[below,font=\scriptsize] {$\frac12$};
 \foreach \d/\lab in {0.25/{\frac14},0.75/{\frac34}} {\draw[thick] (\d,0) -- (\d,-0.8) node[below,font=\scriptsize] {$\lab$};}
 \foreach \d in {0.125,0.375,0.625,0.875} {\draw (\d,0) -- (\d,-0.5);}
 \foreach \d in {0.0625,0.1875,0.3125,0.4375,0.5625,0.6875,0.8125,0.9375} {\draw[gray] (\d,0) -- (\d,-0.3);}
 \node[font=\scriptsize,anchor=west] at (1.02,-0.8) {dyadic points:};
 \node[font=\scriptsize,anchor=west] at (1.02,-1.1) {longer tick, lower level};
 % jobs P_j as brackets with split points
 \foreach \a/\b/\y/\q/\lab in {0.5/8.5/1.35/0.5/{level 1},1.5/5/2.0/0.25/{level 2},8.5/10/2.0/0.75/{level 2},10/11.5/1.35/0.875/{level 3},0.5/1.5/2.0/0.125/{level 3}} {
   \draw[blue!60,thick] (\a/12,\y-0.12) -- (\a/12,\y) -- (\b/12,\y) -- (\b/12,\y-0.12);
   \fill[red!70!black] (\q,\y) circle (2pt);
   \node[font=\scriptsize,above] at (\q,\y) {\lab};
 }
 \node[font=\scriptsize,anchor=west,blue!60!black] at (1.02,2.0) {intervals $P_j=(c_{a'_j},c_{b_j}]$};
 \node[font=\scriptsize,anchor=west,red!70!black] at (1.02,1.6) {$\bullet$ split point $q_j$,};
 \node[font=\scriptsize,anchor=west,red!70!black] at (1.02,1.3) {\phantom{$\bullet$} labeled with its level};
\end{tikzpicture}
\caption{Positions and split points at one node.  The items are laid out on $[0,1)$ in the order of the cells, with lengths
proportional to their frequencies.  A job with a high-frequency item between $a'_j$ and $b_j$ has a long interval $P_j$ and therefore a small level,
so few groups have $I_G$ meeting the distances covered by the cell of a large child.  Split points of the same level are separated: if $q_j<q_{j'}$ and $k_j=k_{j'}$, then
$b_j<a'_{j'}$, by \Cref{lem:anchor}(c).}
\label{fig:anchors}
\end{figure}

Positions increase with $m$.  If $m'<m''$ are items with no item between them (as cells they need not be adjacent), then
$c_{m''}-c_{m'}=(\mu(m')+\mu(m''))/(2\mu_v)$.  Hence for all items $m'<m''$,
\begin{equation}\label{eq:gap}
 c_{m''}-c_{m'}\ \ge\ \frac{\mu(m')+\mu(m'')}{2\mu_v},\qquad\text{and}\qquad c_{m''}-c_{m'}\ \ge\ \frac{\mu(m^*)}{2\mu_v}\ \text{ for every item }m'\le m^*\le m'' .
\end{equation}
The positions are those of the alphabetic code of Gilbert and Moore \cite{GM}, in which the \emph{code word} of an item $m$ consists of
the first $\lceil\log (\mu_v/\mu(m))\rceil+1$ digits after the binary point of $c_m$.  The analysis uses code words only through their lengths.  The shortest
binary expansion of a dyadic point of level $k$ has exactly $k$ digits after the binary point, so, by \Cref{lem:split}, the split point of $j$
is the point of $P_j$ with the shortest binary expansion, see \Cref{fig:anchors}.  By \Cref{lem:anchor}(d), its level $k_j$ is at most the length of the code word of any item
$m^*$ with $a'_j\le m^*\le b_j$, plus $1$.  The frequency of an item $m$ also counts the children of $v$ in the cell $m-1$, because a
covered chain job with $a'_j=m$ can be busy in the cell $m-1$, by \Cref{lem:anchor}(a); Case~2 of the proof of \Cref{lem:local}
uses this.

\begin{lemma}[Split points]\label{lem:anchor}
Let $v$ be a node of $R$ with at least one item.  Below, $j$ and $j'$ denote covered chain jobs of $v$ with $\iota_j>0$ and $\iota_{j'}>0$.
\begin{enumerate}[label=(\alph*),nosep]
\item $\mu_v\le3|J_v|$.  Every covered chain job $j\in C_v$ with $\iota_j>0$ has
$a'_j-1\le a_j$, so its busy-distance interval satisfies $I_j\subseteq[\gamma_{a'_j-1},\gamma_{b_j+1}]$.
\item \emph{(A common witness.)}  Every covered chain job $j\in C_v$ with $\iota_j>0$ satisfies $8\iota_j<\gamma_{m(q_j)}<\rho_j$, and its
busy-distance interval contains $[\gamma_{m(q_j)-1},\gamma_{m(q_j)+1}]$.
\item \emph{(Separation.)}  If $q_j<q_{j'}$ and $k_j=k_{j'}$, then $b_j<a'_{j'}$.
\item \emph{(Level.)}  If $m^*$ is an item with $a'_j\le m^*\le b_j$, then $k_j\le\log (2\mu_v/\mu(m^*))+1$.  In all cases $k_j\le\log \mu_v+1$.
\end{enumerate}
\end{lemma}

\begin{proof}
(a) By the definition of the frequencies, $\mu_v$ is the number of items plus the sum, over the items $m$, of
$\sum_\pm(|J_{m,\pm}|+|J_{m-1,\pm}|)$.  The chain and the cell sets partition $J_v$, so the children of $v$ host $|J_v|-|C_v|$ jobs together,
and each child hosts at least one job.  Each covered chain job with $\iota_j>0$ contributes the two items $a'_j$ and $b_j$, possibly shared
with other jobs, and each child contributes at most one item, its cell; so there are at most $2|C_v|+(|J_v|-|C_v|)$ items.  The jobs of
a child $v_{m,\pm}$ are counted in the sum only for the items $m$ and $m+1$, so every job not in the chain is counted at most twice, and
the sum is at most $2(|J_v|-|C_v|)$.  Hence
$\mu_v\le2|C_v|+3(|J_v|-|C_v|)\le3|J_v|$.  Finally $\iota_j=8\iota_j/8\ge\gamma_{a'_j}/Q=\gamma_{a'_j-1}$, since
$Q\ge8$, so $a_j=\mathrm{cell}(\iota_j)\ge a'_j-1$.

(b) Let $q=q_j$.  Since $q\in P_j$, we have $c_{a'_j}<q\le c_{b_j}$, so $a'_j<m(q)\le b_j$.  Hence $\gamma_{m(q)}\ge\gamma_{a'_j+1}>8\iota_j$ and
$\gamma_{m(q)}\le\gamma_{b_j}<\rho_j$.  Also $a_j\le a'_j\le m(q)-1$ and $b_j\ge m(q)$, so $\gamma_{a_j}\le\gamma_{m(q)-1}$ and
$\gamma_{b_j+1}\ge\gamma_{m(q)+1}$.

(c) Let $k=k_j$.  By \Cref{lem:split}, $q_j$ and $q_{j'}$ are dyadic points of level $k$.  The points $q_j\pm2^{-k}$ are therefore
multiples of $2^{-(k-1)}$, so they are not in $P_j$, by the choice of
$k_j$ if $k\ge2$, and since $P_j\subseteq(0,1)$ if $k=1$.  Since $P_j$ is an interval containing $q_j$, its right end satisfies
$c_{b_j}<q_j+2^{-k}$.  Likewise $q_{j'}-2^{-k}\notin P_{j'}$, so the left end of $P_{j'}$ satisfies $c_{a'_{j'}}\ge q_{j'}-2^{-k}$.
Distinct dyadic points of level $k$ differ by at least $2^{1-k}$, so $c_{a'_{j'}}\ge q_j+2^{-k}>c_{b_j}$, and $a'_{j'}>b_j$.

(d) By \eqref{eq:gap}, $P_j$ has length $L\ge\mu(m^*)/(2\mu_v)$.  Let $k\ge1$ be least with $2^{-k}\le L$; then
$k\le\log (1/L)+1$.  Writing $P_j=(c',c'']$, the point $\lfloor2^kc''\rfloor/2^k$ is greater than $c''-2^{-k}\ge c'$ and at most
$c''$, so it lies in $P_j$.  It is a multiple of $2^{-k}$, so $k_j\le k$ by the definition of $k_j$.  For the last statement use
the first inequality of \eqref{eq:gap} instead, which gives $L\ge1/\mu_v$.
\end{proof}

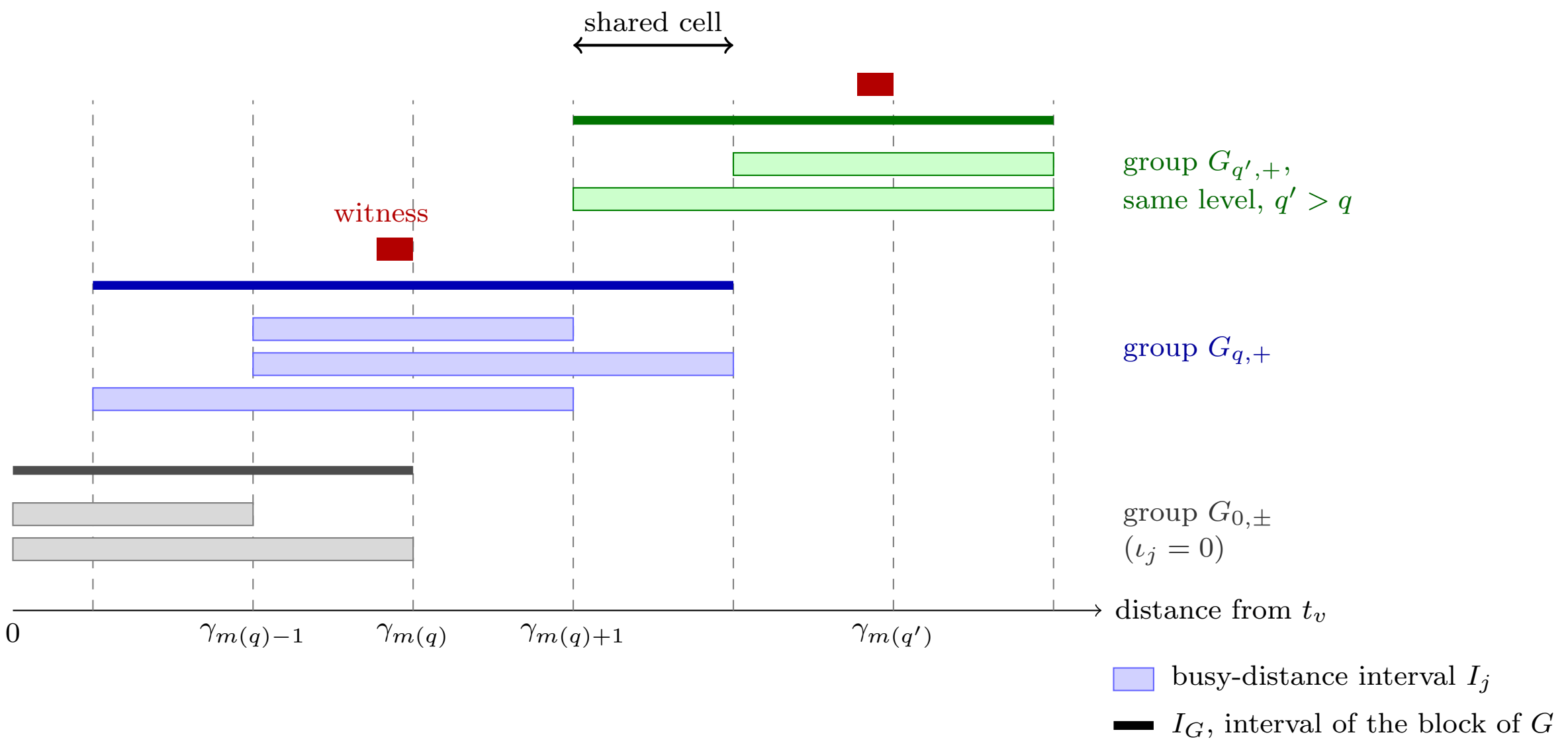
\begin{figure}[t]
\centering
\begin{tikzpicture}[x=0.8cm,y=0.5cm,font=\small]
 % distance axis with grid points
 \draw[->] (0,0) -- (13.6,0) node[right,font=\scriptsize] {distance from $t_v$};
 \node[font=\scriptsize,below] at (0,0) {$0$};
 \foreach \x/\lab in {1/{},3/{\gamma_{m(q)-1}},5/{\gamma_{m(q)}},7/{\gamma_{m(q)+1}},9/{},11/{\gamma_{m(q')}},13/{}} {
   \draw[dashed,gray] (\x,0) -- (\x,10.2);
   \node[font=\scriptsize,below] at (\x,0) {$\lab$};
 }
 % G_0 group (gray): rigid jobs at t_v
 \foreach \b/\y in {5/1,3/1.7} {\fill[gray!30] (0,\y) rectangle (\b,\y+0.45); \draw[gray] (0,\y) rectangle (\b,\y+0.45);}
 \draw[gray!60!black,line width=2.5pt] (0,2.8) -- (5,2.8);
 \node[font=\scriptsize,anchor=west,gray!40!black] at (13.7,1.9) {group $G_{0,\pm}$};
 \node[font=\scriptsize,anchor=west,gray!40!black] at (13.7,1.2) {($\iota_j=0$)};
 % blue group G_{q,+}
 \foreach \a/\b/\y in {1/7/4,3/9/4.7,3/7/5.4} {\fill[blue!18] (\a,\y) rectangle (\b,\y+0.45); \draw[blue!60] (\a,\y) rectangle (\b,\y+0.45);}
 \draw[blue!70!black,line width=2.5pt] (1,6.5) -- (9,6.5);
 \fill[red!70!black] (4.55,7.0) rectangle (5,7.45);
 \node[font=\scriptsize,red!70!black,anchor=south] at (4.6,7.5) {witness};
 \node[font=\scriptsize,anchor=west,blue!60!black] at (13.7,5.2) {group $G_{q,+}$};
 % green group G_{q',+}, same level
 \foreach \a/\b/\y in {7/13/8,9/13/8.7} {\fill[green!20] (\a,\y) rectangle (\b,\y+0.45); \draw[green!50!black] (\a,\y) rectangle (\b,\y+0.45);}
 \draw[green!45!black,line width=2.5pt] (7,9.8) -- (13,9.8);
 \fill[red!70!black] (10.55,10.3) rectangle (11,10.75);
 \node[font=\scriptsize,anchor=west,green!40!black] at (13.7,8.9) {group $G_{q',+}$,};
 \node[font=\scriptsize,anchor=west,green!40!black] at (13.7,8.2) {same level, $q'>q$};
 % legend for bars
 \fill[blue!18] (13.75,-1.6) rectangle (14.25,-1.15); \draw[blue!60] (13.75,-1.6) rectangle (14.25,-1.15);
 \node[font=\scriptsize,anchor=west] at (14.3,-1.37) {busy-distance interval $I_j$};
 \draw[line width=2.5pt] (13.75,-2.3) -- (14.25,-2.3);
 \node[font=\scriptsize,anchor=west] at (14.3,-2.3) {$I_G$, interval of the block of $G$};
 % shared cell brace
 \draw[<->,thick] (7,11.3) -- (9,11.3) node[midway,above,font=\scriptsize] {shared cell};
\end{tikzpicture}
\caption{Groups at one node, on the line of distances from $t_v$.  The busy-distance intervals of the members of a group
$G_{q,\pm}$ all contain $[\gamma_{m(q)-1},\gamma_{m(q)+1}]$, and the distances in $[B_G/4,B_G)$ on the common long side form a common witness, where $B_G=\gamma_{m(q)}$ is the
common distance.
Two groups of the same level and side are separated, by \Cref{lem:anchor}(c), yet their intervals $I_G$ can share one cell; three
cannot meet at one distance.}
\label{fig:groups}
\end{figure}

The proof of \Cref{lem:local} charges the LP weight of the covered chain jobs to blocks.  For this they are partitioned into
groups with a common witness, so that the block of a group has value at least $\frac23$ of its LP weight, by \Cref{lem:groups};
grouping by split points keeps the number of groups that meet at any distance small, see \Cref{fig:groups}.

\begin{definition}[Groups]\label{def:groups}
Let $v$ be a node of $R$.
\begin{enumerate}[label=\arabic*.,nosep]
\item \emph{Groups.}  For a dyadic point $q$ and a side $\pm$, let $G_{q,\pm}$ be the set of covered chain jobs $j\in C_v$ with $\iota_j>0$,
split point $q$ and long side $\pm$.  For a side $\pm$, let $G_{0,\pm}$ be the set of chain jobs $j\in C_v$ with $\iota_j=0$ and long side $\pm$;
they are covered.  No dyadic point is $0$, so the two kinds of set are distinct.  The nonempty sets among these are the \emph{groups} of $v$,
and every covered chain job of $v$ lies in exactly one group, since the split point of a covered chain job with $\iota_j>0$ is a dyadic
point, by \Cref{lem:split}.
\item \emph{Level.}  The \emph{level} of a group $G_{q,\pm}$ with $q$ a dyadic point is the level of $q$; it is the level $k_j$ of each
member $j$, since $q_j=q$ and $q_j$ is a dyadic point of level $k_j$, by \Cref{lem:split}.  The groups $G_{0,\pm}$ have no level.
\item \emph{Common distance.}  The \emph{common distance} of a group $G$ is $B_G=\gamma_{m(q)}$ if $G=G_{q,\pm}$ with $q$ a dyadic point,
and $B_G=\frac12\min_{j\in G}\rho_j$ if $G=G_{0,\pm}$.
\item \emph{Block.}  Let $e_i$ be the smallest left end and $e_k$ the largest right end of the busy-distance intervals of the members of a
group $G$; both lie in $E_v$, and $i<k$, since busy-distance intervals have positive length.  We write $I_G=[e_i,e_k]$ and call $(i,k)$
the \emph{block of} $G$.
\end{enumerate}
\end{definition}

Several groups can have the same block: for instance, $G_{q,+}$ and $G_{q,-}$ have the same block when the busy-distance intervals of
their members have the same union.

\begin{lemma}[Properties of groups]\label{lem:groups}
Let $G$ be a group of a node $v$ of $R$.
\begin{enumerate}[label=(\alph*),nosep]
\item $8\iota_j<B_G<\rho_j$ for every member $j$ of $G$, and $G$ has an independent subset of weight at least $\frac23\sum_{j\in G}w_jz_j$.
\item $I_G$ is the union of the busy-distance intervals of the members of $G$.
\item The block of $G$ is a chain block, and $\mathrm{val}_v(I_G)\ge\frac23\sum_{j\in G}w_jz_j$.
\end{enumerate}
\end{lemma}

\begin{proof}
(a) For $G=G_{q,\pm}$ with $q$ a dyadic point this is \Cref{lem:anchor}(b), since every member $j$ has $q_j=q$.  For $G=G_{0,\pm}$, every member has $\iota_j=0$ and
$0<B_G<\rho_j$.  In both cases the members of $G$ have the common long side $\pm$, so \Cref{lem:witness}, with $B=B_G$, gives the
independent subset.

(b) The busy-distance intervals of the members of $G$ have a common point: $\gamma_{m(q)}$ if $G=G_{q,\pm}$, by \Cref{lem:anchor}(b),
and $0$ if $G=G_{0,\pm}$.  Hence their union is an interval, from the smallest left end to the largest right end, which is $I_G$.

(c) Let $j$ be a member of $G$.  As a chain job, $j$ has a grid point $g$ with $\iota_j<g<\rho_j$, by the definition of the chain.
By (b), $I_G$ contains $I_j$, and $I_j$ contains $[\iota_j,\rho_j]$, since $\gamma_{a_j}\le\iota_j$ if $\iota_j>0$ and
$\rho_j\le\gamma_{b_j+1}$.  Therefore the grid point $g$ lies in the interior of $I_G$, $I_G$ is not of the form
$[\gamma_m,\gamma_{m+1}]$, and the block of $G$ is not a one-cell block.  Hence, with $(i,k)$ the block of $G$, $\mathrm{val}_v(I_G)=\mathrm{val}_v(i,k)$ is the largest weight of an independent subset of $C_v(i,k)$.
Every member of $G$ lies in $C_v(i,k)$, so (a) gives the bound.
\end{proof}

\begin{lemma}[Packing at one node]\label{lem:local}
Let $v$ be a node of $R$.  If $f(u)\ge W_{\mathrm{cov}}(u)/\Psi(|J_u|)$ for every child $u$ of $v$, then $f(v)\ge W_{\mathrm{cov}}(v)/\Psi(|J_v|)$.
\end{lemma}

\begin{proof}
Throughout the proof, $v$ is the node of the lemma, with job set $J_v$, separator $t_v$, grid $\mathcal G_v$ and chain $C_v$; blocks,
items, split points, groups and cells are those of $v$, and distances are measured from $t_v$.  Write $\Psi_v=\Psi(|J_v|)$.  Using the groups of $v$ and \Cref{lem:groups}, the paragraph ``The fractional multiplicities'' assigns a fractional
multiplicity $\omega_{ik}\ge0$ to every block $(i,k)$ at $v$.  The paragraph ``Value'' shows that
$\sum_{(i,k)}\omega_{ik}\,\mathrm{val}_v(i,k)\ge W_{\mathrm{cov}}(v)/\Psi_v$, and the paragraph ``Load'' shows that the
load of $\omega$ is at most $1$ at every distance $r>0$ that is not a grid point, so that $\omega$ is a fractional
packing.
\Cref{lem:tu} then gives $f(v)\ge W_{\mathrm{cov}}(v)/\Psi_v$.  The hypothesis of the lemma, which is the induction hypothesis of the
proof of \Cref{thm:near}, is used only in the paragraph ``Value''.

\medskip
\noindent\emph{The fractional multiplicities.}  For $m\in\operatorname{Occ}(v)$ let $N_m=\max\{|J_{m,+}|,|J_{m,-}|\}\ge1$, where an absent child counts
$0$.  The multiplicity $\omega_{ik}$ of a block $(i,k)$ at $v$ is defined as follows.
\begin{itemize}[nosep]
\item If $(i,k)$ is the one-cell block of a cell $m\in\operatorname{Occ}(v)$, then $\omega_{ik}=\omega_m$, where $\omega_m=\Psi(N_m)/\Psi_v$.  By
\Cref{lem:shape}(b), $N_m\le|J_v|/2$, so $\omega_m<1$.
\item Otherwise, $\omega_{ik}$ is $3/(2\Psi_v)$ times the number of groups whose block is $(i,k)$; this number can exceed $1$, and it is $0$ if there
is no such group.
\end{itemize}

\medskip
\noindent\emph{Value.}  We bound the sum of $\omega_{ik}\,\mathrm{val}_v(i,k)$ over all blocks $(i,k)$, the right-hand side of \Cref{lem:tu}.
Only two kinds of block have $\omega_{ik}>0$: the one-cell blocks of the cells of $\operatorname{Occ}(v)$, and the blocks of the groups.  No block is of both kinds, by \Cref{lem:groups}(c).  Thus the sum is $S_1+S_2$, where $S_1$ is the sum over the blocks of the first kind and $S_2$
the sum over the blocks of the second kind.
\begin{itemize}[nosep]
\item \emph{The children, where the hypothesis is used.}  Let $m\in\operatorname{Occ}(v)$, and write $\mathrm{val}_v(m)$ for the value of the
one-cell block of $m$, whose interval is $[\gamma_m,\gamma_{m+1}]$.  By \Cref{def:recursion}, it is the sum of $f(u)$ over the at most two children $u$ of $v$ in
cell $m$, namely those of $v_{m,+}$ and $v_{m,-}$ that exist.  Each such child has $|J_u|\le N_m$, so $\Psi(|J_u|)\le\Psi(N_m)$, since $\Psi$ is
increasing, and the hypothesis gives $f(u)\ge W_{\mathrm{cov}}(u)/\Psi(|J_u|)$.  Hence
\[
 \omega_m\,\mathrm{val}_v(m)\ =\ \frac{\Psi(N_m)}{\Psi_v}\sum_uf(u)\ \ge\ \frac{\Psi(N_m)}{\Psi_v}\sum_u\frac{W_{\mathrm{cov}}(u)}{\Psi(|J_u|)}\ \ge\
 \frac1{\Psi_v}\sum_uW_{\mathrm{cov}}(u),
\]
the sums over these at most two children.  Every child of $v$ is $v_{m,\pm}$ for exactly one cell $m$ and one side $\pm$, so
$S_1\ge\frac1{\Psi_v}\sum_uW_{\mathrm{cov}}(u)$, the sum over all children $u$ of $v$.
\item \emph{The chain.}  By the definition of the multiplicities, each group $G$ contributes $3/(2\Psi_v)$ to the multiplicity of its own
block, so $S_2=\sum_G\frac{3}{2\Psi_v}\mathrm{val}_v(I_G)$, the sum over all groups.  By \Cref{lem:groups}(c), $\mathrm{val}_v(I_G)\ge\frac23\sum_{j\in G}w_jz_j$
for every group $G$.  Every covered chain job lies in exactly one group, so $S_2\ge\frac1{\Psi_v}\sum_{j\in C^{\mathrm{cov}}_v}w_jz_j$.
\end{itemize}
By \eqref{eq:wcov}, $S_1+S_2\ge W_{\mathrm{cov}}(v)/\Psi_v$.

\medskip
\noindent\emph{Load.}  We show that the load of $\omega$ is at most $1$ at every distance $r>0$ from $t_v$ that is not a grid point of $\mathcal G_v$; by definition, this makes $\omega$ a
fractional packing.  Fix such an $r$, and let $m=\mathrm{cell}(r)$, so that $\gamma_m<r<\gamma_{m+1}$.  The one-cell block of a cell $m'\in\operatorname{Occ}(v)$
has interval $[\gamma_{m'},\gamma_{m'+1}]$, which contains $r$ only if $m'=m$, and the block of a group is never a one-cell block, by \Cref{lem:groups}(c).  Consequently, the load of $\omega$ at $r$ is
\[
 \lambda(r)\ =\ \omega_m\,[m\in\operatorname{Occ}(v)]\ +\ \frac{3}{2\Psi_v}\,N(r),
\]
where $[m\in\operatorname{Occ}(v)]$ is $1$ if $m\in\operatorname{Occ}(v)$ and $0$ otherwise, and $N(r)$ is the number of groups $G$ with $r\in I_G$.  These
groups are of two kinds: the groups $G_{0,+}$ and $G_{0,-}$, at most two, and groups $G_{q,\pm}$ with $q$ a dyadic point, possibly of many
different levels.  We first count the second kind level by level, and then bound the levels that occur.

\noindent\emph{Step 1: at most two groups per level and side.}  Fix a level $h$ and a side $\pm$.  We claim that at most two groups $G_{q,\pm}$ of
level $h$ have $r\in I_G$.  Suppose that three do.  They are distinct groups on the same side, so their split points are distinct, say
$q_1<q_2<q_3$.  By \Cref{lem:groups}(b), we can pick $j_1$ in the first group with $r\in I_{j_1}$ and $j_3$ in the third group with
$r\in I_{j_3}$; let $j_2$ be any member of the second.  By \Cref{lem:anchor}(c), twice, and $a'_{j_2}<b_{j_2}$, $b_{j_1}<a'_{j_2}<b_{j_2}<a'_{j_3}$, so
$b_{j_1}+1<a'_{j_3}-1$.  By \Cref{lem:anchor}(a), $I_{j_1}$ ends at $\gamma_{b_{j_1}+1}$ and $I_{j_3}$ starts at $\gamma_{a'_{j_3}-1}$ or later, which
is larger; so $I_{j_1}\cap I_{j_3}=\varnothing$, a contradiction.

\noindent\emph{Step 2: summing over levels.}  Let $h(r)$ be the largest level of a group $G_{q,\pm}$, with $q$ a dyadic point, such that
$r\in I_G$, and $h(r)=0$ if there is none.  Each such group has one of the $h(r)$ levels $1,\dots,h(r)$ and one of the two sides.  For each
level and each side, Step~1 allows at most two such groups, so there are at most $2\cdot2\cdot h(r)=4h(r)$ of them.  Adding the at most two groups $G_{0,+}$ and $G_{0,-}$, whose members have $\iota_j=0$ and therefore no split point
and no level,
\[
 N(r)\ \le\ 4h(r)+2 .
\]
It remains to bound $h(r)$.  There are two cases.

\noindent\emph{Case 1: $m\notin\operatorname{Occ}(v)$.}  The first term of $\lambda(r)$ is $0$, and $\lambda(r)=\frac{3}{2\Psi_v}N(r)$.  Every covered
chain job $j\in C_v$ with $\iota_j>0$ gives $v$ the items $a'_j$ and $b_j$, so \Cref{lem:anchor} applies to $v$, and by its parts (d)
and (a), $k_j\le\log \mu_v+1\le\log (3|J_v|)+1$.  The level of a group $G_{q,\pm}$ with $q$ a dyadic point is the level $k_j$ of each of
its members, so $h(r)\le\log (3|J_v|)+1$; this also holds when $h(r)=0$, in particular when $v$ has no item, since $|J_v|\ge1$.  Hence $N(r)\le4\log |J_v|+4\log 3+6$ and
\[
 \lambda(r)\ \le\ \frac{3}{2\Psi_v}\bigl(4\log |J_v|+4\log 3+6\bigr)\ <\ \frac{6\log |J_v|+19}{\Psi_v}\ \le\ 1,
\]
using $6\log 3+9<19$.

\noindent\emph{Case 2: $m\in\operatorname{Occ}(v)$.}  Here the first term of $\lambda(r)$ is $\omega_m$, and the load of the groups must be
at most $1-\omega_m=(\Psi_v-\Psi(N_m))/\Psi_v$.  Let $G=G_{q,\pm}$, with $q$ a dyadic point, be a group with $r\in I_G$.  Since $I_G$ is the union of the busy-distance intervals of the members, by \Cref{lem:groups}(b), some member $j$ has $r\in I_j$; fix one.  The level of $G$ is $k_j$.  By \Cref{lem:anchor}(a), $\gamma_{a'_j-1}\le r\le\gamma_{b_j+1}$, and $r$ is
not a grid point, so $a'_j-1\le m\le b_j$.  Let $m^*=\max\{m,a'_j\}$, which is $m$ or $m+1$.  It is an item, since $m\in\operatorname{Occ}(v)$ and
$a'_j$ are items, and $a'_j\le m^*\le b_j$.  Moreover $\mu(m^*)\ge|J_{m,+}|+|J_{m,-}|\ge N_m$: by \Cref{def:anchor}, item~2, the frequency $\mu(m')$ of an item $m'$
counts the jobs of the children of $v$ in the cells $m'$ and $m'-1$, and $m=m^*$ or $m=m^*-1$.  By \Cref{lem:anchor}(d) and (a), with $\beta=|J_v|/N_m\ge2$, by \Cref{lem:shape}(b),
\[
 k_j\ \le\ \log \frac{2\mu_v}{\mu(m^*)}+1\ \le\ \log \frac{2\mu_v}{N_m}+1\ \le\ \log (6\beta)+1 .
\]
This holds for every such group, so $h(r)\le\log (6\beta)+1$, also when $h(r)=0$, and $N(r)\le4\log \beta+4\log 6+6$.  Hence
\begin{align*}
 \lambda(r)\ \le\ \omega_m+\frac{3}{2\Psi_v}\bigl(4\log \beta+4\log 6+6\bigr)\ &<\ \omega_m+\frac{6\log \beta+25}{\Psi_v}\\
 &\le\ \omega_m+\frac{31\log \beta}{\Psi_v}\ =\ \omega_m+\frac{\Psi_v-\Psi(N_m)}{\Psi_v}\ =\ 1,
\end{align*}
using $6\log 6+9<25$ and $\log \beta\ge1$.

In both cases $\lambda(r)\le1$, so $\omega$ is a fractional packing, and \Cref{lem:tu}, with the paragraph ``Value'', gives
$f(v)\ge W_{\mathrm{cov}}(v)/\Psi_v$.
\end{proof}

\begin{theorem}[Nearly rigid jobs]\label{thm:near}
Assume that $\J_{\mathrm{nr}}\ne\varnothing$.  Then \DPNR{} computes in deterministic polynomial time a nonmigratory
schedule of weight at least
\[
 \frac{W_{\mathrm{nr}}}{62\log n+38}.
\]
\end{theorem}

\begin{proof}
For every choice of the offsets, $S(\mathrm{root})$ is a nonmigratory schedule of weight $f(\mathrm{root})$ by
\Cref{lem:dp-feasible}.  We show $f(v)\ge W_{\mathrm{cov}}(v)/\Psi(|J_v|)$ for every node $v$ by induction from the leaves of $R$ up.  At a
leaf of $R$ the hypothesis of \Cref{lem:local} is empty, and at any other node it is the induction hypothesis for the
children; in both cases \Cref{lem:local} gives the claim for $v$.  At the root,
$|J_{\mathrm{root}}|=|\J_{\mathrm{nr}}|\le n$ and $\Psi$ is increasing, so, for every choice of the offsets,
\[
 f(\mathrm{root})\ \ge\ \frac{W_{\mathrm{cov}}(\mathrm{root})}{\Psi(n)}\ =\ \frac{W_{\mathrm{cov}}(\mathrm{root})}{31\log n+19},
\]
where $W_{\mathrm{cov}}(\mathrm{root})$ is the LP weight of all jobs covered at their homes, since every home lies in the subtree of the
root.  With the offsets of Step~1 of \DPNR{}, $W_{\mathrm{cov}}(\mathrm{root})\ge W_{\mathrm{nr}}/2$ by \Cref{lem:coverage}(c), and
$2\Psi(n)=62\log n+38$.

\noindent\emph{Running time.}  Release dates, due dates and processing times are rationals given in binary, and every separator is an endpoint of
a window, by the definition of $t(J)$.  Hence there is a number $b$, polynomial in the input size, such that every radius $\iota_j>0$
or $\rho_j$ of a job seen from a separator, and every number $8\iota_j$ with $\iota_j>0$, can be written as $P/R$ with integers
$1\le P,R\le2^b$, and therefore lies between $2^{-b}$ and $2^b$.  Since $\gamma_m=4^{o+K'm}$ with $0\le o<K'$, the integers
$\mathrm{cell}(x)$ and $\mathrm{cell}^-(x)$ of such a number $x$ have absolute value at most $b/(2K')+3$, and they are found by exact
comparisons of $x$ with powers of $4$.  Consequently, every point of $E_v$ is $0$ or a power of $4$ of polynomial bit length.  The
tree $R$ has at most $n\ell$
nodes, by \Cref{lem:shape}(c), and Step~1 of \DPNR{} tries $K'$ offsets at each node.  The set $E_v$ of a node $v$ has at most $2|J_v|$
elements: the busy-distance intervals of the chain jobs contribute at most $2|C_v|$, and the cells of $\operatorname{Occ}(v)$ at most
$2|\operatorname{Occ}(v)|\le2(|J_v|-|C_v|)$, since the children of $v$ host the $|J_v|-|C_v|$ jobs not in the chain, at least one
each.  Therefore each node $v$ has $O(|J_v|^2)$ blocks, and the value of a chain block is found by the
greedy algorithm, whose independence tests are bipartite matching computations, by \Cref{lem:greedy}.  Hence Steps~1--3 of \DPNR{} run in
polynomial time.
\end{proof}

\begin{remark}[Dependence on the LP]\label{rem:lp}
The dynamic program uses windows, slacks, weights and eligibility sets; the LP solution enters only
through the choice of offsets.  With random offsets the nearly rigid part does not use the
LP at all.  Every job of $\J_{\mathrm{nr}}$ is then covered at its home with probability more than $1/2$, by
\Cref{lem:coverage}(b), and the expected weight of $S(\mathrm{root})$ is at least $W_{\mathrm{nr}}/(62\log n+38)$
simultaneously for every \StrongLP{} solution: by the proof of \Cref{thm:near}, $f(\mathrm{root})\ge
W_{\mathrm{cov}}(\mathrm{root})/\Psi(n)$ for every outcome, and by \Cref{lem:coverage}(b) and linearity of expectation, the expected value of
$W_{\mathrm{cov}}(\mathrm{root})$ is at least $W_{\mathrm{nr}}/2$.
In $S(\mathrm{root})$, every support consists of at most two intervals, so each completed job is
preempted at most once.  This does not hold for the other two schedules of \Cref{sec:assembly}: a moderate-slack
support is a window minus shorter windows, a big-slack job runs in the idle time left by the jobs admitted before it, and both
kinds of set can consist of many intervals.
\end{remark}

%=====================================================================
\section{Proof of the restricted-assignment theorem}\label{sec:assembly}
%=====================================================================

\begin{proof}[Proof of \Cref{thm:main-ra}]
We use the classes $\J_{\mathrm{big}}$, $\J_{\mathrm{mod}}$ and $\J_{\mathrm{nr}}$ of \Cref{def:ra-classes} and their LP weights, so that
$W=W_{\mathrm{big}}+W_{\mathrm{mod}}+W_{\mathrm{nr}}$.  Compute three schedules; for an empty class, take the empty
schedule.
\begin{itemize}[nosep]
\item \emph{Big slack.}  By \Cref{lem:ra-split} with $\delta=1/8$, $z_j^{\mathrm B}=z_j$ for every $j\in\J_{\mathrm{big}}$, so \Cref{thm:big-rounding} with $\delta=1/8$ and $\J'=\J_{\mathrm{big}}$ gives
a nonmigratory schedule of weight at least $W_{\mathrm{big}}/96$.
\item \emph{Moderate slack.}  \Cref{thm:moderate} gives a $1$-witness system for $\J_{\mathrm{mod}}$ of size at most
$54\log n+138$, and \Cref{thm:ws} gives a nonmigratory schedule of weight at least
$W_{\mathrm{mod}}/(108\log n+276)$.
\item \emph{Nearly rigid.}  \Cref{thm:near} gives a nonmigratory schedule of weight at least
$W_{\mathrm{nr}}/(62\log n+38)$.
\end{itemize}

\noindent\emph{Combining.}  Output the heaviest of the three schedules.  Since $\max\{a/\alpha,b/\beta,c/\gamma\}\ge(a+b+c)/(\alpha+\beta+\gamma)$
for nonnegative $a,b,c$ and positive $\alpha,\beta,\gamma$, its weight is at least
\[
 \frac{W_{\mathrm{big}}+W_{\mathrm{mod}}+W_{\mathrm{nr}}}{96+(108\log n+276)+(62\log n+38)}\ =\ \frac{W}{170\log n+410}.
\]
Every step is deterministic and runs in polynomial time; for the moderate-slack schedule this uses that the supports
of \Cref{thm:moderate} consist of at most $n+1$ intervals each.
\end{proof}

\begin{proof}[Proof of \Cref{cor:ra}]
As in the proof of \Cref{cor:unrelated}, compute an optimal $\StrongLP$ solution, so that $W=\OPTLP$, and apply
\Cref{thm:main-ra}: $\OPTLP\le(170\log n+410)\,W(A)\le(170\log n+410)\,\OPTnm$, and $\OPT\le\OPTLP$ by
\Cref{lem:strong-valid}.
\end{proof}

%=====================================================================
\section{A matching logarithmic lower bound}\label{sec:restricted-log-gap}
%=====================================================================

The lower bound of \Cref{sec:tightness} uses machine-dependent processing times.  Under restricted assignment, the ratio $\OPT/\OPTnm$
can still be logarithmic: the construction below gives instances with $\OPT/\OPTnm=\Omega(\log n)$,
even with unit weights, rigid jobs and exactly two eligible machines per job.  The edge-isoperimetric problem on the hypercube, and in
particular the role of initial segments of the binary order, is classical; see
\cite{Harper1964,Lindsey1964,Bernstein1967,Hart1976}.  We need the following additional elementary structural property
of these edge boundaries, which we prove in \Cref{app:hypercube-initial-segment-proof}.

For a graph $G=(V,E)$ and $S\subseteq V$, the \emph{edge boundary} of $S$ is
\begin{equation}\label{eq:edge-boundary}
 \partial_G(S)
 :=\bigl\{\{x,y\}\in E:\ |\{x,y\}\cap S|=1\bigr\}.
\end{equation}

\begin{lemma}[Boundaries of initial segments of the hypercube]
\label{lem:hypercube-initial-segment-forest}
Let $Q_q$ be the $q$-dimensional hypercube.  Number its vertices by $0,1,\ldots,2^q-1$,
identifying each integer with its $q$-bit binary representation.  For $0\le k\le2^q-2$, let
\[
 I_k:=\{0,1,\ldots,k\},
\]
the initial segment ending at $k$ in the binary order.  Then the graph with vertex set $V(Q_q)$ and edge set
$\partial_{Q_q}(I_k)$ is a forest.
\end{lemma}

\begin{theorem}[A logarithmic lower bound under restricted assignment]
\label{thm:restricted-log-gap}
For every power of two $m\ge4$ there is a unit-weight restricted-assignment instance with $m$ machines and
$n=\frac12m\log m$ jobs, every job rigid with exactly two eligible machines, such that
\[
 \OPTLP\ =\ \OPT\ \ge\ \frac{\log m}4\,\OPTnm\ \ge\ \frac{\log n}8\,\OPTnm .
\]
\end{theorem}

\begin{proof}
Let $q=\log m\ge2$, so that $n=mq/2=2^{q-1}q$.  Identify the $m=2^q$ machines with the vertices of $Q_q$.  For every hypercube edge $e=\{u,v\}$, with $u<v$, create one
job $j_e$ with
\begin{equation}\label{eq:restricted-job}
 K_e=[u,v),
 \qquad
 L_e=v-u,
 \qquad
 p_e=v-u,
 \qquad
 \Gamma(j_e)=\{u,v\}.
\end{equation}
This is a restricted-assignment instance, and every job is rigid.  The hypercube has $2^{q-1}q=mq/2$ edges, so there
are exactly $n=mq/2$ jobs.

\noindent\emph{A migratory schedule completing every job.}
Every window lies in $[0,m-1)$, which is the disjoint union of the unit intervals $C_k=[k,k+1)$,
$k=0,\dots,m-2$.  (Since $q\ge2$, every vertex is an endpoint of an edge, so these are exactly the elementary intervals
of the instance.)  We construct the schedule separately on each unit interval.  Fix $C_k$.  The window of a job
$j_{\{u,v\}}$ with $u<v$ contains $C_k$ exactly when $u\le k<v$, i.e., when $u\in I_k$ and $v\notin I_k$, and it is disjoint
from $C_k$ otherwise.  Therefore the jobs whose windows contain $C_k$ are exactly the edges of the boundary
$\partial_{Q_q}(I_k)$.  By \Cref{lem:hypercube-initial-segment-forest}, these edges form a forest.

We use the following elementary property of a forest.  Root every tree component that has an edge, and assign each
edge to its endpoint farther from the root.  Each non-root vertex is assigned exactly one edge, the edge to its parent,
and each root is assigned no edge.  Hence this assignment of edges to incident vertices is injective.

Apply this assignment to the forest $\partial_{Q_q}(I_k)$, and for every edge $e$ of it use the piece $(i,j_e,C_k)$, where
$i$ is the vertex to which $e$ is assigned.  This machine is one of the two eligible machines of $j_e$.  By injectivity,
the pieces in $C_k$ have distinct machines, and each job has at most one piece in $C_k$.  Different unit intervals
may use different assignments, so a job may migrate between its two machines.

The pieces of different unit intervals have disjoint intervals, so all pieces together form a schedule.  Fix a job
$j_e$ with $e=\{u,v\}$ and $u<v$.  Its window is the disjoint union $C_u\cup C_{u+1}\cup\cdots\cup C_{v-1}$, and $e$ lies in
$\partial_{Q_q}(I_k)$ for every $k=u,\dots,v-1$.  Consequently, $j_e$ has one piece in each of these unit intervals, the union of the
intervals of its pieces is $K_e$, and the completion condition of \Cref{def:schedule} holds with equality.  Every job is
completed, so $\OPT=n$.  By \Cref{lem:strong-valid}, $\OPTLP\ge n$, and no $\StrongLP$ solution has value more than the total
weight $n$, since $z_j\le1$.  Hence $\OPTLP=\OPT=n$.

\noindent\emph{Every nonmigratory schedule completes at most $2m$ jobs.}
Fix a machine $v$.  An incident job $j_{\{u,v\}}$ with $u<v$ has window $[u,v)$; call it \emph{incoming} at $v$.  An
incident job $j_{\{v,w\}}$ with $v<w$ has window $[v,w)$; call it \emph{outgoing} at $v$.

Let a nonmigratory schedule complete a job $j_e$ on machine $v$.  Its pieces on $v$ are pairwise
disjoint subintervals of $K_e$ of total length at least $p_e=|K_e|$, so they cover $K_e$.  If
$v\ge1$, every incoming job at $v$ has the unit interval $C_{v-1}=[v-1,v)$ in its window.  Two incoming jobs completed on
$v$ would both have pieces on $v$ covering $C_{v-1}$, which is impossible, since pieces with the same machine have
disjoint intervals.  Hence at most one incoming job is completed on $v$; for $v=0$ there are no incoming jobs.  Similarly,
if $v\le m-2$, every outgoing job at $v$ has $C_v=[v,v+1)$ in its window, so at most one outgoing job is completed on
$v$; for $v=m-1$ there are no outgoing jobs.  Consequently a nonmigratory schedule completes at most two jobs on each
machine, and
\[
 \OPTnm\le2m.
\]
Combining this with $\OPTLP=\OPT=n=mq/2$ gives $\OPT\ge(q/4)\,\OPTnm$.  Finally, $\log m=q$, and
$\log n\le2\log m$, since $n=mq/2\le m^2$.
\end{proof}

\begin{remark}[The instances of Megow and Zondervan]\label{rem:tower}
The instances of Megow and Zondervan \cite[Proposition~19]{MZ} form another family of rigid restricted-assignment
instances on which the ratio $\OPT/\OPTnm$ is logarithmic.  Such an instance has $n=(k+2)2^{k-1}$ unit-weight rigid jobs on
$m=2^k$ machines, and every job is eligible on at most two machines.  All jobs can be scheduled if migration is allowed, so
$\OPTLP=\OPT=n$.  Their argument shows that every nonmigratory schedule uses a separate machine copy for
each job; in particular, on the original machines every nonmigratory schedule completes at most one job on each machine,
hence at most $2^k$ jobs.  The ratio is $(k+2)/2=\Omega(\log n)$.
\end{remark}

%=====================================================================
\section{Conclusion}\label{sec:summary}
%=====================================================================

We showed that the migratory optimum is at most $O(\sqrt n)$ times the nonmigratory optimum on unrelated machines,
when every eligible machine can complete a job by itself, and at most $O(\log n)$ times the nonmigratory optimum
under restricted assignment.  Both bounds remain valid when $\OPT$ is replaced by the optimum of the $\StrongLP$.
Since our proofs are constructive, we obtain the first nontrivial approximation algorithms for migratory throughput
maximization in these settings.  The instances of \Cref{thm:sqrt-tightness,thm:restricted-log-gap} show that the
bounds are best possible up to constant factors.  They also show that approximation algorithms that output
nonmigratory schedules cannot do better: on these instances, the weight of every nonmigratory schedule is smaller
than $\OPT$ by these factors.

The main open problem is whether migratory throughput maximization admits an approximation ratio $o(\sqrt n)$ on
unrelated machines or $o(\log n)$ under restricted assignment.  Such an algorithm would have to output schedules in
which some jobs do migrate.  Even a constant-factor approximation under restricted assignment is open, while for
identical machines one is known; see, e.g., \cite{KP,MZ}.  A natural approach is to round the $\StrongLP$ to migratory schedules.
For a constant factor, this approach requires the integrality gap $\OPTLP/\OPT$ to be bounded by a constant, which
is open.

Under restricted assignment, the largest possible ratio $\OPT/\OPTnm$ as a function of $m$ alone is not known.  It
lies between $\Omega(\log m)$, by \cite{MZ} and \Cref{thm:restricted-log-gap}, and $2m-1$, by \Cref{prop:machines}.
By \Cref{thm:main-ra}, a family of instances with ratio $\omega(\log m)$ must have a number of jobs superpolynomial
in $m$.
For the special case of rigid unit-weight jobs we showed in \cref{prop:rigid-unit} that
$O(\log m)$ holds, which is best possible.  
A related question concerns the number of copies of each machine that suffice to eliminate migration.
Under restricted assignment, this number lies between $\Omega(\log m)$ \cite{MZ} and $O(\log^2n)$, by
\Cref{cor:copies}.  A positive answer to the question of \cite{MZ} whether $O(\log m)$ copies suffice would also
give an $O(\log m)$ bound on $\OPT/\OPTnm$.

\section*{AI statement}

We used ChatGPT5.6 and Claude Opus 5.5 (and earlier versions thereof as well as Gemini at
the beginning of the project)
as an interactive research assistant for drafting, editing, verification and
figure production. Based on the unrelated machines result, we produced
an $O(\log^2 n)$-approximation algorithm for restricted
assignment. The idea was the following.  For every node $v$ of the
midpoint tree, organize the jobs $j$ in $F_v,$ by creating a balanced
tree, called the \emph{distance tree}, on the midpoints of their
usable-distance intervals $U_j$.  At a distance-tree node, all
intervals crossing the node separator have a common usable distance
and therefore give one matching. At one fixed distance-tree depth,
intervals belonging to different nodes are pairwise disjoint and  we
showed that different node matchings can safely reuse the same machines.
Since the recursion is balanced, only $O(\log n)$ distance
depths are needed.  To remove the second $\log n$ we identified the
nearly rigid case as the difficult one.  We came up with the idea of using a
single tree, the recursion tree, in order to compute bottom-up by dynamic
programming the value at every node. AI provided the idea to use
fractional packings at each node and  the Gilbert-Moore codes in order
to lower bound the values. This enabled the crucial \cref{lem:local}.

We verified all AI-assisted contributions and take full responsibility for the content and correctness
of the paper.

%=====================================================================
\appendix
\section{Instances that violate \texorpdfstring{\Cref{asm:valid}}{Assumption 2.2}}\label{app:fractional-progress}

This appendix drops \Cref{asm:valid}: the processing time $p_{ij}$ of an
eligible pair may exceed the length $L_j$ of the window, so slacks may be negative; \Cref{def:schedule,def:run}, the
$\StrongLP$ and $\OPT$, $\OPTnm$ and $\OPTLP$ are unchanged.  A machine $i$ with $p_{ij}>L_j$ cannot complete $j$ by
itself, but in a migratory schedule it can contribute partial progress toward the completion of $j$.  This is the
standard model of preemptive scheduling on unrelated machines \cite{LL}, which places no restriction on $p_{ij}$
relative to the window.  An eligible pair $(i,j)$ is \emph{valid} if $p_{ij}\le L_j$, so \Cref{asm:valid} says that every
eligible pair is valid, and a schedule \emph{uses only valid pairs} if every piece $(i,j,[a,b))$ of it has a valid pair
$(i,j)$.  The rest of the paper uses only \Cref{lem:fp-invalid}, in \Cref{sec:model}.

\begin{lemma}[Jobs without a valid pair]\label{lem:fp-invalid}
Let $j$ be a job with $p_{ij}>L_j$ for every $i\in\Gamma(j)$.  Then no schedule completes $j$,
and every $\StrongLP$ solution has $z_j=0$.
\end{lemma}

\begin{proof}
Let $p=\min_{i\in\Gamma(j)}p_{ij}$, so that $p>L_j$.  The pieces of $j$ in a schedule have pairwise disjoint intervals in $K_j$,
so $\sum_i|\Pi_{ij}|\le L_j$ and
\[
 \sum_{i\in\Gamma(j)}\frac{|\Pi_{ij}|}{p_{ij}}\ \le\ \frac{L_j}{p}\ <\ 1 .
\]
Therefore no schedule completes $j$.

In a $\StrongLP$ solution, $y_{ijk}>0$ only if $\Delta_k\subseteq K_j$, and $K_j$ is the disjoint union of the elementary intervals it
contains, since its endpoints are among the $\tau_k$; so summing non-parallelism \eqref{eq:lp-job} over $k$ gives
$\sum_i\sum_ky_{ijk}\le z_jL_j$.  Coverage \eqref{eq:lp-coverage} gives $z_j=\sum_i\sum_ky_{ijk}/p_{ij}\le z_jL_j/p$.  Since
$L_j/p<1$, this forces $z_j=0$.
\end{proof}

\begin{proposition}[Upper bounds without \Cref{asm:valid}]\label{prop:fp-upper}
Every instance with $n$ jobs and $m$ machines, whether or not it satisfies \Cref{asm:valid}, satisfies the following.
\begin{enumerate}[label=(\alph*),nosep]
\item $\OPT\le\min\{n,\,2m-1\}\,\OPTnm$ and $\OPTLP\le n\,\OPTnm$.
\item Running a heaviest job that has a valid pair $(i,j)$ alone on machine $i$ is a nonmigratory schedule of weight at least
$\OPT/n$ and at least $\OPTLP/n$.  For every $\varepsilon>0$, a nonmigratory schedule of weight at least
$(1-\varepsilon)\OPT/(2m-1)$ can be computed in time polynomial in the input size and $1/\varepsilon$, and with unit
weights one of weight at least $\OPT/(2m-1)$ can be computed in polynomial time.
\end{enumerate}
\end{proposition}

\begin{proof}
\emph{The factor $n$.}  Let $V$ be the set of jobs that have a valid pair.  If $V=\varnothing$, then $\OPT=\OPTLP=0$ by
\Cref{lem:fp-invalid}, and there is nothing to show.  For $j\in V$ with valid pair $(i,j)$, running $j$ alone on $i$ during the
earliest $p_{ij}$ units of $K_j$ is a nonmigratory schedule that completes $j$, so $\max_{j\in V}w_j\le\OPTnm$.
By \Cref{lem:fp-invalid}, a schedule completes only jobs of $V$, and every $\StrongLP$ solution has $z_j=0$ for $j\notin V$; with
$0\le z_j\le1$ this gives
\[
 \OPT\ \le\ \sum_{j\in V}w_j\ \le\ n\max_{j\in V}w_j,
 \qquad
 \OPTLP\ \le\ \sum_{j\in V}w_j\ \le\ n\max_{j\in V}w_j .
\]

\noindent\emph{The factor $2m-1$.}  Run the algorithm in the proof of \Cref{prop:machines}: assign every job to a fastest machine and
compute a schedule of each single-machine instance by \cite{La90} or \cite{PW}.  The output is a nonmigratory schedule, as
there.  The analysis in that proof uses \Cref{asm:valid} only to obtain $p^*_j\le L_j$ for the jobs $j$ that the
schedule $A$ completes.  Without \Cref{asm:valid}, these jobs have a valid pair by \Cref{lem:fp-invalid}, so
$p^*_j\le L_j$ holds for them, and the rest of the analysis applies without change.  It gives $\OPT\le(2m-1)\,\OPTnm$ and the
bounds on the weight of the output.
\end{proof}

\begin{proposition}[A matching lower bound]\label{prop:fp-lower}
For every $n\ge2$ there is a unit-weight instance with $n$ jobs and $m=n+1$ machines that violates \Cref{asm:valid}, in
which all jobs have the same window, every job is eligible on every machine, and the processing time of a job on a machine
depends only on the machine, such that
\[
 \OPTLP=\OPT=n,\qquad \OPTnm=1,
\]
and every schedule that uses only valid pairs completes at most one job.
\end{proposition}

\noindent In particular, the factor $n$ of \Cref{prop:fp-upper} is attained, and the factor $2m-1$ is best possible up
to a factor $3$.

\begin{proof}
\emph{The instance.}  The machines are $F,S_1,\dots,S_n$.  Every job $j\in\{1,\dots,n\}$ has window $K_j=[0,1)$, so $L_j=1$,
weight $1$, $\Gamma(j)=\M$, and
\[
 p_{Fj}=\frac23,\qquad p_{S_kj}=\frac{2n-2}{2n-3}>1\quad(1\le k\le n).
\]
Consequently, the valid pairs are the pairs $(F,j)$, each with slack $1/3$.  This instance is similar to the one in the proof of
\cite[Theorem~1.2]{IM}: in the schedule below, the machine $F$ is shared in short slices by all jobs, and each job
spends the rest of its window on a machine on which it cannot complete by itself.

\noindent\emph{A schedule that completes every job.}  For each $j$ take the piece $(F,j,[\tfrac{j-1}n,\tfrac jn))$ and the pieces
$(S_j,j,[0,\tfrac{j-1}n))$ and $(S_j,j,[\tfrac jn,1))$, omitting an empty one.  Pieces on $F$ have pairwise disjoint intervals,
$S_j$ processes only $j$, and the pieces of $j$ have pairwise disjoint intervals, so this is a schedule.  Job $j$ receives $1/n$
on $F$ and $(n-1)/n$ on $S_j$, and
\[
 \frac{1/n}{2/3}+\frac{(n-1)/n}{(2n-2)/(2n-3)}\ =\ \frac3{2n}+\frac{2n-3}{2n}\ =\ 1,
\]
so every job is completed, and $\OPT=n$.  The only elementary interval is $\Delta_1=[0,1)$.  Set $z_j=1$, $y_{Fj1}=1/n$ and
$y_{S_jj1}=(n-1)/n$ for every $j$, and all other $y_{ijk}=0$.  Coverage holds by the computation above; machine capacity holds,
since $\sum_jy_{Fj1}=n\cdot\frac1n=1$ and $y_{S_jj1}=\frac{n-1}n$; and non-parallelism holds, since $\frac1n+\frac{n-1}n=1$.  This
solution has value $n$, and no $\StrongLP$ solution has larger value, since $z_j\le1$; so $\OPTLP=n$.

\noindent\emph{Schedules that complete jobs on $F$ alone.}  Let a schedule complete a set $C$ of jobs, each with
$|\Pi_{Fj}|\ge p_{Fj}=\frac23$.  These sets are pairwise disjoint subsets of $[0,1)$, since pieces with the same machine
have disjoint intervals, so $\frac23|C|\le1$ and $|C|\le1$.  This applies to every schedule that uses only valid pairs, since all
its pieces have machine $F$.  It also applies to every nonmigratory schedule: a job $j$ that it completes has all its pieces on
one machine $i$, with $|\Pi_{ij}|\ge p_{ij}$, and $i=S_k$ is impossible, since $p_{S_kj}>1=|K_j|$.  Running job $1$ on $F$
during $[0,\frac23)$ shows $\OPTnm=1$.

Finally, $\OPT/\OPTnm=n=m-1$, and $2m-1\le3(m-1)$ since $m\ge2$.
\end{proof}

By \Cref{prop:fp-upper}, without \Cref{asm:valid}, $\OPT$ can be approximated in polynomial time within a
factor $\min\{n,(2m-1)/(1-\varepsilon)\}$.  We leave open whether a substantially better
approximation exists; by \Cref{prop:fp-lower}, on some instances such an algorithm must output a schedule that
migrates and uses invalid pairs.

\section{Machine copies}\label{app:copies}

This appendix states and proves the bounds on machine copies mentioned in \Cref{sec:results}.  For an integer $\mu\ge1$, let $\M\times[\mu]$ be the set
of machines $(i,c)$ with $i\in\M$ and $c\in[\mu]=\{1,\dots,\mu\}$, where job $j$ is eligible on $(i,c)$ if and only if
$i\in\Gamma(j)$, with processing time $p_{ij}$.  We call $(i,c)$ the $c$-th \emph{copy} of $i$ and identify $(i,1)$
with $i$, so $\M\times[\mu]$ consists of $\mu$ copies of each machine of $\M$, the machine itself included.

\begin{corollary}[Machine copies]\label{cor:copies}
Suppose that a migratory schedule on $\M$ completes every job of $\J$.  Then one can compute in deterministic
polynomial time an integer $\mu$ and a nonmigratory schedule on $\M\times[\mu]$ that completes every job of $\J$, where
$\mu\le\min\{2m-1,\,O(\sqrt n\log n)\}$ on unrelated machines and $\mu\le\min\{2m-1,\,O(\log^2n)\}$ under restricted
assignment.
\end{corollary}

\begin{proof}
Let $S^*$ be a migratory schedule on $\M$ that completes every job of $\J$.  Let $\rho=246\sqrt n+49152$ on unrelated
machines and $\rho=170\log n+410$ under restricted assignment.  The statement does not involve weights, and in all
instances below every job has weight one, so the weight of a schedule is the number of jobs it completes.

\noindent\emph{Rounds.}  Let $R_0=\J$.  For $c=1,2,\dots$, as long as $R_{c-1}\ne\emptyset$, round $c$ computes a nonempty set
$J_c\subseteq R_{c-1}$ and a nonmigratory schedule $S_c$ on $\M$ that processes only jobs of $J_c$ and completes each of
them, and sets $R_c=R_{c-1}\setminus J_c$.  Let $I_c$ be the instance with machine set $\M$, job set $R_{c-1}$, unit
weights, and all other data as in the given instance.  It satisfies \Cref{asm:valid}, and it is a
restricted-assignment instance if the given one is.  Deleting from $S^*$ all processing of the jobs outside $R_{c-1}$
gives a migratory schedule that completes every job of $R_{c-1}$, so $\OPT=|R_{c-1}|$ for $I_c$.
\begin{itemize}[nosep]
\item If $|R_{c-1}|\ge2$, apply \Cref{cor:unrelated} or \Cref{cor:ra} to $I_c$.  Since $|R_{c-1}|\le n$ and the
factors $246\sqrt{n'}+49152$ and $170\log n'+410$ are nondecreasing in the number $n'$ of jobs, this computes a
nonmigratory schedule on $\M$ that completes a set $J_c\subseteq R_{c-1}$ with $|J_c|\ge|R_{c-1}|/\rho>0$.  The
schedule $S_c$ is obtained from it by deleting all processing of the jobs outside $J_c$.
\item If $R_{c-1}=\{j\}$, let $J_c=\{j\}$, choose $i\in\Gamma(j)$, and let $S_c$ process $j$ on $i$ during
$[r_j,r_j+p_{ij})$, which lies in $K_j$ by \Cref{asm:valid}.
\end{itemize}
Since every $J_c$ is nonempty, $R_c=\emptyset$ for some $c$; let $\mu$ be the least such $c$.  Then the sets
$J_1,\dots,J_\mu$ are pairwise disjoint and their union is $\J$.

\noindent\emph{The schedule.}  Let $S$ be the schedule on $\M\times[\mu]$ that, for every $c\in[\mu]$, every $i\in\M$ and every
$j\in J_c$, processes $j$ on $(i,c)$ at exactly the times at which $S_c$ processes $j$ on $i$.  At every time, machine
$(i,c)$ processes at most the one job that $S_c$ processes on $i$ at that time.  Let $j\in\J$, and let $c$ be the
index with $j\in J_c$.  For $c'\ne c$, the schedule $S_{c'}$ processes only jobs of $J_{c'}$, so it does not process
$j$.  The schedule $S_c$ is nonmigratory and completes $j$, so it processes $j$ on a single machine $i\in\Gamma(j)$, only
inside $K_j$, and for a total time of at least $p_{ij}$.  Hence $S$ processes $j$ only on $(i,c)$, which is eligible
for $j$ with processing time $p_{ij}$, only inside $K_j$, and for a total time of at least $p_{ij}$.  Hence $S$ is a
nonmigratory schedule on $\M\times[\mu]$ that completes every job of $\J$.

\noindent\emph{The number of copies.}  The sizes $|R_0|>|R_1|>\dots>|R_\mu|=0$ decrease, so the rounds $c$ with
$|R_{c-1}|\ge2$ are the rounds $1,\dots,c_0$ for some $c_0\ge0$, and at most one round follows them.  Each of the
rounds $1,\dots,c_0$ gives $|R_c|\le(1-1/\rho)|R_{c-1}|$.  If $c_0\ge1$, then
$2\le|R_{c_0-1}|\le n(1-1/\rho)^{c_0-1}\le n\,e^{-(c_0-1)/\rho}$, so $c_0\le1+\rho\ln(n/2)$ and
$\mu\le c_0+1\le\rho\ln n+2$.  If $c_0=0$, then $\mu=1$.  With the two values of $\rho$ this gives
$\mu=O(\sqrt n\log n)$ and $\mu=O(\log^2n)$.  There are $\mu\le\rho\ln n+2$ rounds, a number polynomial in $n$, and
each round and the construction of $S$ take polynomial time.

\noindent\emph{The bound $2m-1$.}  For every job $j$ fix a fastest machine $i^*(j)$, with $p^*_j=p_{i^*(j)j}$, as in the
proof of \Cref{prop:machines}, and let $I'$ be the instance with the machines of $\M$ and the jobs of $\J$, in which every
job $j$ is eligible on every machine with processing time $p^*_j$.  As in that proof, the pieces of $S^*$ form a schedule
of $I'$ that completes every job, and $I'$ has identical machines.  By \cite{MZ}, one can compute in strongly polynomial
time a nonmigratory schedule $B$ on $2m-1$ identical machines $1,\dots,2m-1$, with the windows and processing times of
$I'$, that completes every job; let $a(j)$ be the machine of $j$ in $B$.  Replace every piece $(a(j),j,[x,y))$ of $B$ by
the piece $((i^*(j),a(j)),j,[x,y))$.  The pieces on a copy $(i,a)$ come from pieces of $B$ on the machine $a$, so their
intervals are pairwise disjoint.  Every job $j$ runs only on $(i^*(j),a(j))$, which is eligible for $j$ with processing
time $p^*_j$, and it receives at least $p^*_j$ units of $K_j$ there.  Hence this is a nonmigratory schedule on
$\M\times[2m-1]$ that completes every job.  The algorithm outputs it if $2m-1<\mu$, and the schedule $S$ otherwise.
\end{proof}

For rigid jobs, a single tree of separators gives a better bound.

\begin{proposition}[Machine copies for rigid jobs]\label{prop:rigid-copies}
Consider a restricted-assignment instance in which every job is rigid, and suppose that a migratory schedule on $\M$
completes every job of $\J$.  Then one can compute in polynomial time a nonmigratory schedule on
$\M\times[\mu]$ that completes every job of $\J$, where $\mu=\min\{m,\lfloor\log n\rfloor+1\}$.
\end{proposition}

\begin{proof}
\emph{Step 1: at every time, completed jobs use distinct machines.}  Let $A$ be any schedule, and let $C$ be the set of
jobs that $A$ completes.  Let $j\in C$.  Since $j$ is rigid, $p_j=L_j$.  Since $A$ completes $j$, the intervals of
the pieces of $j$ are pairwise disjoint subintervals of $K_j$ of total length at least $L_j$, so they cover $K_j$.
Hence every time $t\in K_j$ lies in the interval of exactly one piece of $j$; let $\mu_t(j)$ be the machine of this
piece.  It is eligible for $j$, since every piece has an eligible machine.  Now let $j$ and $j'$ be distinct jobs of $C$, and let $t\in K_j\cap K_{j'}$.
Then $\mu_t(j)\ne\mu_t(j')$.  Indeed, otherwise the piece of $j$ and the piece of $j'$ that contain $t$ would have the
same machine and intersecting intervals, while pieces with the same machine have disjoint intervals.

\noindent\emph{The bound $m$.}  By Step~1, applied to the given migratory schedule, every time lies in the windows of at
most $m$ jobs.  Consider the jobs in nondecreasing order of release date, and give each job $j$ one of the colors
$1,\dots,m$ that is not carried by an earlier job whose window contains $r_j$.  Together with $K_j$, at most $m$ windows
contain $r_j$, so such a color exists.  Two jobs $h$ and $j$ of the same color, with $h$ considered first, have disjoint
windows: otherwise $r_h\le r_j<d_h$, so $K_h$ contains $r_j$, and $j$ would not have received the color of $h$.  For a
job $j$ of color $c$, choose $i\in\Gamma(j)$ and use the piece $((i,c),j,K_j)$.  The pieces on a copy $(i,c)$ belong to
jobs of color $c$ and have pairwise disjoint intervals, and every job runs on one eligible copy throughout its window,
for a total time $|K_j|=p_j$.  Hence this is a nonmigratory schedule on $\M\times[m]$ that completes every job, computed
in polynomial time.  If $m\le\lfloor\log n\rfloor+1$, the algorithm outputs it; otherwise it proceeds as follows.

\noindent\emph{Computing a migratory schedule.}  By Step~1, applied to the given migratory schedule, for every elementary
interval $\Delta_k$ the jobs whose windows contain $\Delta_k$ can be assigned to distinct eligible machines.  For each
$\Delta_k$, compute such an assignment by bipartite matching, and run each of these jobs on its machine during
$\Delta_k$.  Every window is the union of the elementary intervals it contains, so the result is a migratory schedule
$A$ that completes every job, computed in polynomial time.  We apply Step~1 to $A$, with $C=\J$.

\noindent\emph{Step 2: one nonmigratory schedule for each depth.}  Consider the midpoint tree on $\J$, as in
\Cref{def:midpoint-tree}.  Let $v$ be a node.  Every job $j\in F_v$ has $t_v\in K_j$, by the definition of the
crossing set, so by Step~1 the machines $\mu_{t_v}(j)$, for $j\in F_v$, are eligible and pairwise distinct.  In
particular, $|F_v|\le m$.  Fix a depth $d$, and let $B_d$ be the set of the pieces $(\mu_{t_v}(j),\,j,\,K_j)$ over all
nodes $v$ of depth $d$ and all jobs $j\in F_v$.  By \Cref{lem:midpoint-tree}(e), every job lies in the crossing set of
exactly one node, so every job has at most one piece in $B_d$.  Consider two distinct pieces of $B_d$ with the same
machine, of jobs $j\in F_v$ and $j'\in F_{v'}$.  Then $j\ne j'$.  If $v=v'$, then the two machines $\mu_{t_v}(j)$ and
$\mu_{t_v}(j')$ would be distinct, contrary to the assumption; so $v\ne v'$.  By \Cref{lem:midpoint-tree}(c), $K_j$
lies in the region of $v$ and $K_{j'}$ in the region of $v'$, and these regions are disjoint by
\Cref{lem:midpoint-tree}(d).  So the two pieces have disjoint intervals, and $B_d$ is a schedule.  The piece of a job
$j$ runs $j$ on an eligible machine throughout $K_j$, for a total time $|K_j|=p_j$, so it completes $j$.  Every job of
$B_d$ has a single piece, so $B_d$ is nonmigratory, and it completes exactly the jobs that lie in the crossing sets of
the nodes of depth $d$.

\noindent\emph{The schedule.}  By \Cref{lem:midpoint-tree}(b), a node of depth $d$ hosts at most $n/2^d$ jobs.  Every node
hosts at least one job, so $2^d\le n$, and every node has depth at most $\lfloor\log n\rfloor$.  Let $S$ be the
schedule on $\M\times[\lfloor\log n\rfloor+1]$ that, for every depth $d$ and every piece $(i,j,K_j)$ of $B_d$,
contains the piece $((i,d+1),j,K_j)$.  Pieces on different copies have different machines, and the pieces on the
copies $(i,d+1)$, $i\in\M$, are those of $B_d$.  So $S$ is a schedule.  Every job lies in exactly one crossing set,
so it has exactly one piece in $S$, which completes it.  Hence $S$ is a nonmigratory schedule that completes every
job.  The midpoint tree and the matchings are computed in polynomial time.
\end{proof}

\noindent \Cref{prop:rigid-copies} improves \Cref{cor:copies} for rigid jobs.  It is tight up to a constant factor on
the instances of \Cref{rem:tower}: there $n=(k+2)2^{k-1}$ and $m=2^k$, so $\log n=O(\log m)$, and Megow and
Zondervan \cite{MZ} show that $\Omega(\log m)$ copies of some machine are needed.

\section{Derandomization of the big-slack rounding}\label{app:big-derandomization}

This appendix proves \Cref{thm:big-rounding}.  We use the notation of \Cref{sec:big-covered}: the random choices $R$,
the schedule $A(R)$ built by the greedy algorithm, and the quantities $\mathcal I_{ij}$ and $Z_{ij}$.  The weight
$W(A(R))$ is not a simple function of $R$, so we derandomize a lower bound on it that is quadratic in $R$:
\begin{equation}\label{eq:pessimistic-F}
 F(R):=
 \sum_{j\in\J'}\sum_{i:(i,j)\in\EB}
 w_jR_{ij}\left(1-\frac{Z_{ij}}{\delta L_j}\right).
\end{equation}

\begin{lemma}[Pointwise lower bound]\label{lem:pointwise-certificate}
For every outcome $R$ of the choices, $W(A(R))\ge F(R)$.
\end{lemma}

\begin{proof}
Fix a job $j$.  If $j$ chooses no machine, its terms in \eqref{eq:pessimistic-F} vanish.  Otherwise $j$ chooses a
unique machine $i$, and its terms reduce to $w_j(1-Z_{ij}/(\delta L_j))$.  If the greedy algorithm admits $j$, then $j$
contributes $w_j$ to $W(A(R))$, and $1-Z_{ij}/(\delta L_j)\le1$.  If the algorithm discards $j$, the proof of
\Cref{lem:big-expected} shows that $Z_{ij}>\delta L_j$, so $1-Z_{ij}/(\delta L_j)<0$, and $j$ contributes $0$.  In both
cases the contribution of $j$ to $W(A(R))$ is at least its terms in \eqref{eq:pessimistic-F}.  Summing over jobs proves
the claim.
\end{proof}

\begin{lemma}[Expected lower bound]\label{lem:expected-certificate}
\[
 \E[F(R)]\ \ge\ \frac\delta{12}\sum_{j\in\J'}w_jz_j^{\mathrm B}.
\]
\end{lemma}

\begin{proof}
For a fixed pair $(i,j)$, the variable $R_{ij}$ depends only on the choice of $j$, and $Z_{ij}$ only on the choices of
other jobs, so they are independent.  By \Cref{lem:big-volume},
$\E[R_{ij}Z_{ij}]=\E[R_{ij}]\,\E[Z_{ij}]\le\gamma x_{ij}\cdot3\gamma L_j$.  Taking expectations in
\eqref{eq:pessimistic-F},
\[
 \E[F(R)]
 \ \ge\ \sum_{j\in\J'}\sum_{i:(i,j)\in\EB}w_j\gamma x_{ij}\left(1-\frac{3\gamma}{\delta}\right)
 \ =\ \frac{\gamma}{2}\sum_{j\in\J'}w_jz_j^{\mathrm B}
 \ =\ \frac\delta{12}\sum_{j\in\J'}w_jz_j^{\mathrm B},
\]
because $\gamma=\delta/6$.
\end{proof}

\begin{lemma}[Conditional expectations]\label{lem:derandomization}
One can compute in deterministic polynomial time a choice $R^*$ of at most one big-slack machine for every job of
$\J'$, with $R^*_{ij}=1$ only if $x_{ij}>0$, such that $F(R^*)\ge\E[F(R)]$.
\end{lemma}

\begin{proof}
The procedure fixes the choices of the jobs one at a time.  Suppose that some job choices have
already been fixed and the others retain their original independent distributions.  For the next job $j$, the
procedure computes the conditional expectation of $F(R)$ under each possible outcome for $j$: one of its big-slack
machines $i$, or no machine.  The current conditional expectation is the probability-weighted average of these next
conditional expectations, so at least one positive-probability outcome has value at least the current one.  The
procedure selects such an outcome and continues.  Choosing machine $i$ has positive probability only if $x_{ij}>0$.

These conditional expectations are computable in polynomial time.  Indeed,
\eqref{eq:pessimistic-F} is a sum of linear terms $R_{ij}$ and bilinear terms
$R_{ij}R_{ih}$ involving two distinct jobs $j\ne h$.  After conditioning, variables of a fixed
job are known zero--one constants; for an unfixed job their one-variable marginals remain
$\gamma x_{ij}$.  Every bilinear term involves two distinct jobs, so if both are still unfixed
its expectation factors into the product of these marginals.  Hence, for each candidate outcome
of the next job, the conditional expectation is a sum of polynomially many terms, each computable in
constant time, so the whole procedure runs in polynomial time.  After all jobs have been fixed, the resulting
choice $R^*$ satisfies $F(R^*)\ge\E[F(R)]$.
\end{proof}

\begin{proof}[Proof of \Cref{thm:big-rounding}]
By \Cref{lem:derandomization}, one can compute in deterministic polynomial time a choice $R^*$, in which $R^*_{ij}=1$
only if $x_{ij}>0$, with $F(R^*)\ge\E[F(R)]$, and $\E[F(R)]\ge\frac\delta{12}\sum_{j\in\J'}w_jz_j^{\mathrm B}$ by
\Cref{lem:expected-certificate}.  The greedy algorithm applied to $R^*$ runs in polynomial time, and the schedule
$A_{\J'}=A(R^*)$ satisfies $W(A(R^*))\ge F(R^*)$ by \Cref{lem:pointwise-certificate}, which gives
\eqref{eq:big-rounding}.  Every job $j$ that $A(R^*)$ completes runs on the machine $i$ it chose, with $(i,j)\in\EB$
and $x_{ij}>0$.
\end{proof}

\section{Proof of \texorpdfstring{\Cref{lem:hypercube-initial-segment-forest}}{Lemma 17.1}}
\label{app:hypercube-initial-segment-proof}

This appendix proves \Cref{lem:hypercube-initial-segment-forest}.

\begin{proof}[Proof of \Cref{lem:hypercube-initial-segment-forest}]
We induct on $q$.  For $q=1$, the only allowed initial segment is $I_0=\{0\}$, and its boundary consists
of the single edge of $Q_1$.

Assume the statement for $Q_{q-1}$ and set
\[
 h:=2^{q-1}.
\]
Split $Q_q$ according to the most significant bit into the two half-cubes
\[
 V_0:=\{0,1,\ldots,h-1\},
 \qquad
 V_1:=\{h,h+1,\ldots,2h-1\}.
\]
The induced graphs $Q_q[V_0]$ and $Q_q[V_1]$ are both copies of $Q_{q-1}$.  The natural
isomorphism from $Q_{q-1}$ to the upper copy is
\[
 \varphi_1(x):=x+h.
\]
The only edges with one endpoint in $V_0$ and the other in $V_1$ are
\[
 \{x,x+h\},\qquad 0\le x<h.
\]
We now determine the boundary $\partial_{Q_q}(I_k)$ exactly.

\noindent\emph{Case 1: $0\le k\le h-2$.}
Here $I_k$ is a proper initial segment of $V_0$.  A boundary edge can have one of only two forms:
\begin{enumerate}[label=(\roman*),leftmargin=2em]
 \item both endpoints lie in $V_0$, with one endpoint in $I_k$ and the other in
       $V_0\setminus I_k$; these edges form $\partial_{Q_q[V_0]}(I_k)$;
 \item the edge is $\{x,x+h\}$ for some $x\in I_k$.
\end{enumerate}
No edge contained in $V_1$ belongs to the boundary, because every vertex of $V_1$ lies outside
$I_k$.  Therefore
\begin{equation}\label{eq:hypercube-boundary-lower}
 \partial_{Q_q}(I_k)
 =\partial_{Q_q[V_0]}(I_k)
  \;\cup\;
  \bigl\{\{x,x+h\}:x\in I_k\bigr\}.
\end{equation}
Under the identification $Q_q[V_0]\cong Q_{q-1}$, the first edge set is exactly the initial segment
boundary of $I_k$ in $Q_{q-1}$, and hence is a forest by induction.

For every edge $\{x,x+h\}$ in the second family, the upper endpoint $x+h$ has degree one in the
full boundary graph.  Indeed, every neighbor of $x+h$ inside $V_1$ is also outside $I_k$, so no
edge of $Q_q[V_1]$ incident with $x+h$ lies in the boundary.  Thus the edges of the second family
attach distinct leaves to the forest of the first family, so the union is a forest.

\noindent\emph{Case 2: $k=h-1$.}
Now $I_k=V_0$.  Every edge contained in one half-cube has both endpoints on the same side of the
cut, while every edge $\{x,x+h\}$ between the two halves has one endpoint on each side.  Hence
\[
 \partial_{Q_q}(I_{h-1})
 =\bigl\{\{x,x+h\}:0\le x<h\bigr\},
\]
which is a matching and therefore a forest.

\noindent\emph{Case 3: $h\le k\le2h-2$.}
Write
\[
 k=h+k',\qquad 0\le k'\le h-2,
\]
and denote by
\[
 I^{(1)}_{k'}:=\{h,h+1,\ldots,h+k'\}\subset V_1
\]
the part of the initial segment lying in the upper half-cube.  Thus
\[
 I_k=V_0\cup I^{(1)}_{k'}.
\]
To identify the boundary, consider the three possible locations of a hypercube edge.
\begin{enumerate}[label=(\roman*),leftmargin=2em]
 \item \emph{Both endpoints in $V_0$.}  Such an edge never lies in the boundary, because all
       of $V_0$ lies in $I_k$.

 \item \emph{Both endpoints in $V_1$.}  Such an edge lies in the boundary exactly when one endpoint lies
       in $I^{(1)}_{k'}$ and the other in $V_1\setminus I^{(1)}_{k'}$.  These edges form
       \[
        \partial_{Q_q[V_1]}\!\left(I^{(1)}_{k'}\right).
       \]
       Under the isomorphism $\varphi_1^{-1}:Q_q[V_1]\to Q_{q-1}$, this boundary corresponds
       exactly to $\partial_{Q_{q-1}}(I_{k'})$.  It is therefore a forest by induction.

 \item \emph{One endpoint in each half-cube.}  Such an edge has the form $\{x,x+h\}$ with
       $0\le x<h$.  Its lower endpoint $x$ always belongs to $I_k$.  Its upper endpoint belongs
       to $I_k$ exactly when
       \[
        x+h\le h+k',
       \]
       equivalently when $x\le k'$.  Hence this edge lies in the boundary exactly for
       $k'<x<h$.
\end{enumerate}
Consequently
\begin{equation}\label{eq:hypercube-boundary-upper}
 \partial_{Q_q}(I_k)
 =\partial_{Q_q[V_1]}\!\left(I^{(1)}_{k'}\right)
  \;\cup\;
  \bigl\{\{x,x+h\}:k'<x<h\bigr\}.
\end{equation}

It remains to show that the union is a forest.  Fix
$\{x,x+h\}$ with $x>k'$.  The lower endpoint $x\in V_0$ is not incident with any other boundary
edge: every neighbor of $x$ inside $V_0$ also lies in $I_k$, because all of $V_0\subseteq I_k$.
Thus $x$ has degree one in the full boundary graph.  Thus the edges of the second family
attach distinct leaves to the forest of the first family, so the union is a forest.

The boundary is a forest in all three cases, completing the induction.
\end{proof}

%=====================================================================
\section{Proof of the feasibility criterion on one machine}\label{app:one-machine}
%=====================================================================

This appendix proves \Cref{lem:processor-demand}.

\begin{proof}[Proof of \Cref{lem:processor-demand}]
\emph{Necessity.}  In such a schedule, the pieces of a job $j\in \J'$ have total length at
least $p_{ij}$.  For the jobs with $K_j\subseteq I$ these pieces lie in $I$, and they are pairwise disjoint, since all
pieces have the same machine.  Thus their total length, which is at least the left side of
\eqref{eq:processor-demand}, is at most $|I|$.

\noindent\emph{Earliest due date.}  Assume \eqref{eq:processor-demand}.  A job $j\in\J'$ is \emph{pending} at a time $x$ if
$r_j\le x<d_j$ and its pieces before $x$ have total length less than $p_{ij}$.  The \emph{earliest-due-date algorithm} (EDD)
processes on machine $i$, at every time, a pending job with the smallest due date, ties broken by a fixed order, and leaves
the machine idle when no job is pending.  The pending jobs, and hence the choice of EDD, change only at release dates, at
due dates, and at times at which the processed job receives its $p_{ij}$-th unit.  There are at most $3|\J'|$ such
times; EDD finds them from left to right in polynomial time and uses one piece between two consecutive ones.  Its pieces
lie in the windows of their jobs, and no job receives more than $p_{ij}$ units.  By construction, EDD has two
properties.
\begin{enumerate}[label=(E\arabic*),nosep]
\item If a job $h$ is pending at a time at which EDD processes a job $j$, then $d_j\le d_h$.
\item If some job is pending at a time, then EDD processes a job at that time.
\end{enumerate}

\noindent\emph{EDD completes every job.}  Suppose not, and let $d$ be the smallest due date of a job that EDD does not
complete, attained by a job $j^\star$.  Then $j^\star$ is pending at every time of $[r_{j^\star},d)$.  By (E2) and (E1), at
every such time EDD processes a job with due date at most $d$.  Let $a$ be the least time such that the machine processes
a job with due date at most $d$ at every time of $[a,d)$; then $a\le r_{j^\star}$.

We claim that every job $h$ with due date at most $d$ that has a piece meeting $[a,d)$, and also $j^\star$, has
$r_h\ge a$.  Suppose that $r_h<a$.  By the choice of $a$, at some time $x'\in[r_h,a)$ the machine is idle or processes a
job with due date larger than $d$.  The pieces of $h$ before $x'$ have total length less than $p_{ih}$: if $h\ne j^\star$,
a piece of $h$ meeting $[a,d)$ gives $h$ positive processing after $a$, and $h$ receives at most $p_{ih}$ units in total;
and $j^\star$ receives less than $p_{ij^\star}$ units in total.  The due date of $h$ exceeds $a$: a piece of $h$ meeting
$[a,d)$ lies in $K_h$, and $d>a$.  Hence $h$ is pending at $x'$.  By (E2) and (E1), EDD processes at $x'$ a job with due
date at most $d_h\le d$, contradicting the choice of $x'$.

Let $G$ be the set of jobs $h\in \J'$ with $r_h\ge a$ and $d_h\le d$.  By the claim, $[a,d)$ is covered by the pieces of
jobs of $G$, so the jobs of $G$ receive at least $d-a$ units in total.  Each $h\in G$ receives at most $p_{ih}$ units,
and $j^\star\in G$ receives less than $p_{ij^\star}$.  Hence $\sum_{h\in G}p_{ih}>d-a$, while $K_h\subseteq[a,d)$ for every
$h\in G$.  This contradicts \eqref{eq:processor-demand} for $I=[a,d)$.
\end{proof}

\section{Rigid jobs with unit weights}\label{app:rigid-unit}

This appendix bounds $\OPT/\OPTnm$ in terms of the number of machines alone, for rigid jobs with unit weights under
restricted assignment.  The bound compares with $\OPT$, not with $\OPTLP$.
The bound is existential; a technique of \cite{KK} makes it constructive.

\begin{proposition}[Rigid jobs with unit weights]\label{prop:rigid-unit}
Consider a restricted-assignment instance on $m$ machines in which every job is rigid and has weight one.  Then
\[
 \OPT\ \le\ (2\log m+6)\,\OPTnm .
 \]
 A nonmigratory schedule of weight at least $(1-1/e)\OPT / (2\log m+6)  $  can be computed in
 deterministic polynomial time. 
\end{proposition}

\begin{proof}
Let $A$ be a schedule of weight $\OPT$, and let $C$ be the set of jobs that $A$ completes.  All weights are one, so
$|C|=\OPT$.  If $C$ is empty, there is nothing to show.  Otherwise let $N=|C|\ge1$.

\noindent\emph{Step 1: one nonmigratory schedule for each depth.}  Consider the instance with job set $C$ and the same
machines.  Deleting from $A$ the pieces of the jobs outside $C$ gives a schedule of this instance that completes all $N$
jobs, so by \Cref{lem:strong-valid} this instance has a $\StrongLP$ solution of value $N$.  Since all weights are one and
$z_j\le1$, this solution has $z_j=1$ for every $j\in C$.  Every job of $C$ is rigid, so \Cref{lem:rigid-depth} applies to
this solution and to the midpoint tree on $C$.  The set $I_v$ of a node $v$ satisfies $|I_v|\ge|F_v|$ and $I_v\subseteq F_v$,
so $I_v=F_v$.  Hence the jobs of every crossing set are on pairwise distinct machines, so every crossing set has at
most $m$ jobs, and for every depth $d$ there is a nonmigratory schedule $B_d$ that completes exactly the jobs of $C$
that lie in the crossing sets of the nodes of depth $d$.  This schedule uses only jobs of $C$ and the given machines,
so it is also a schedule of the given instance.  For $d\ge0$ let $c_d$ be the number of these jobs.  Then
\[
 \OPTnm\ \ge\ c_d\qquad\text{for every }d\ge0 .
\]
By \Cref{lem:midpoint-tree}(b), a node of depth $d$ hosts at most $\lfloor N/2^d\rfloor$ jobs, and every node hosts at least
one, so every node has depth at most $h=\lfloor\log N\rfloor$.  By \Cref{lem:midpoint-tree}(e), every job of $C$ lies in
the crossing set of exactly one node, so $c_d=0$ for $d>h$ and
\[
 c_0+c_1+\dots+c_h\ =\ N .
\]

\noindent\emph{Step 2: counting.}  Every node has at most two children, so there are at most $2^d$ nodes of depth $d$, each
with at most $m$ jobs in its crossing set; hence $c_d\le m2^d$.  Let $s$ be the least integer $s\ge0$ with
$m2^s>N/4$.

We first show that the depths below $s$ hold fewer than half of the jobs.  If $s=0$, there are no such depths, and
their number of jobs is $0<N/2$.  If $s\ge1$, then $m2^{s-1}\le N/4$ by the choice of $s$, so $m2^s\le N/2$, and
\[
 \sum_{d=0}^{s-1}c_d\ \le\ \sum_{d=0}^{s-1}m2^d\ =\ m\,(2^s-1)\ <\ m2^s\ \le\ \frac N2 .
\]
In both cases, the depths $d\ge s$ hold more than half of the jobs:
\[
 \sum_{d=s}^{h}c_d\ >\ \frac N2 .
\]
In particular $s\le h$, since otherwise this sum would be empty.

Next we count these depths.  From $m2^s>N/4$ we get $s>\log N-\log m-2$, and $h\le\log N$.  Hence the number of
depths from $s$ to $h$ satisfies
\[
 h-s+1\ <\ \log N-(\log N-\log m-2)+1\ =\ \log m+3 .
\]
The largest of the $h-s+1$ numbers $c_s,\dots,c_h$ is at least their sum divided by $h-s+1$.  So some depth $d$ has
\[
 c_d\ >\ \frac{N/2}{\log m+3}\ =\ \frac{N}{2\log m+6} .
\]
By Step~1, $\OPTnm\ge c_d>N/(2\log m+6)=\OPT/(2\log m+6)$.
Karakostas and Kolliopoulos \cite{KK} have shown that a $(1-1/e - \varepsilon)$-approximation to
$\OPTnm$ can be computed in polynomial time. It is not difficult to see that in the case
of rigid jobs the bound can be improved to $(1-1/e)$ by solving exactly the Configuration LP
relaxation. 
\end{proof}

\noindent The instances of \Cref{thm:restricted-log-gap} and of \Cref{rem:tower} consist of rigid jobs with unit
weights and have ratio $\OPT/\OPTnm=\Omega(\log m)$.  Hence, for rigid jobs with unit weights under restricted
assignment, the largest ratio $\OPT/\OPTnm$ on $m$ machines is $\Theta(\log m)$.

\end{document}